\documentclass[a4paper]{article}[11pts]

\usepackage[english]{babel}
\usepackage[utf8x]{inputenc}
\usepackage[T1]{fontenc}

\normalsize

\usepackage[a4paper,top=1in,bottom=1in,left=1in,right=1in,marginparwidth=1in]{geometry}
\usepackage{setspace}

\usepackage{appendix}
\usepackage{color}
\definecolor{strcolor}{rgb}{0.6, 0.2, 0.6}
\definecolor{commentcolor}{rgb}{0.3125, 0.5, 0.3125}
\definecolor{keycol}{rgb}{0, 0, 1}

\usepackage{amssymb}
\usepackage{amsmath}
\usepackage{amsthm}
\usepackage{amsfonts, mathtools}
\usepackage{bbm}
\usepackage{tablefootnote}
\usepackage{booktabs}
\usepackage{multirow}

\newtheorem{proposition}{Proposition}
\newtheorem{lemma}{Lemma}

\newtheorem{theorem}{Theorem}

\usepackage{listings}
\usepackage{url}
\usepackage{hyperref}
\hypersetup{
    hidelinks,
    colorlinks=true,
    linkcolor=red,
    citecolor=magenta
} 

\newcommand {\bea}{\begin{eqnarray}}
	\newcommand {\eea}{\end{eqnarray}}

\newtheorem{algorithm}{Algorithm}
\newtheorem{remark}{Remark}

\def\blot{\quad \mbox{$\vcenter{ \vbox{ \hrule height.4pt
				\hbox{\vrule width.4pt height.9ex \kern.9ex \vrule width.4pt}
				\hrule height.4pt}}$}}

\usepackage{natbib}
 \bibpunct[, ]{(}{)}{,}{a}{}{,}%

 \makeatletter
\newenvironment{breakablealgorithm}
  {
   \begin{center}
     \refstepcounter{algorithm}
     \hrule height.8pt depth0pt \kern2pt
     \renewcommand{\caption}[2][\relax]{
       {\raggedright\textbf{\fname@algorithm~\thealgorithm} ##2\par}%
       \ifx\relax##1\relax 
         \addcontentsline{loa}{algorithm}{\protect\numberline{\thealgorithm}##2}%
       \else 
         \addcontentsline{loa}{algorithm}{\protect\numberline{\thealgorithm}##1}%
       \fi
       \kern2pt\hrule\kern2pt
     }
  }{
     \kern2pt\hrule\relax
   \end{center}
  }
\makeatother
\allowdisplaybreaks
\usepackage{diagbox}

\usepackage{xspace}

\newcommand{\mE}{\mathbb{E}}
\newcommand{\mP}{\mathbb{P}}

\newcommand{\mcO}{\mathcal{O}}
\newcommand{\mcOd}{\mathcal{O}_{\delta}}
\newcommand{\mfS}{\mathfrak{S}}
\newcommand{\mcT}{\mathcal{T}}
\newcommand{\mcP}{\mathcal{P}}
\newcommand{\mcA}{\mathcal{A}}

\newcommand{\mcK}{\mathcal{K}}

\newcommand{\tb}[1]{\textbf{#1} \xspace}

\newcommand{\bs}[1]{\boldsymbol{#1}}

\usepackage{subfig}
\usepackage{adjustbox}
\usepackage{algorithm}
\usepackage{algpseudocode}
\usepackage{tikz}
\usepackage{pgfplots}
\pgfplotsset{compat=1.5}

\begin{document}

	\title{Infrequent Resolving Algorithm for Online Linear Programming}

	\author{Guokai Li\textsuperscript{1} \and Zizhuo Wang\textsuperscript{2} \and Jingwei Zhang\textsuperscript{2}} %
    \date{\small
        1. Desautels Faculty of Management, McGill University, Montreal, Quebec H3A 1G5, Canada\\
        2. School of Data Science, The Chinese University of Hong Kong, Shenzhen, Guangdong, 518172, P.R. China\\
        guokai.li@mcgill.ca, wangzizhuo@cuhk.edu.cn, zhangjingwei@cuhk.edu.cn}

	
     \maketitle

     \onehalfspacing

    \begin{abstract}

        Online linear programming (OLP) has gained significant attention from both researchers and practitioners due to its extensive applications such as online auctions, network revenue management, order fulfillment and advertising. Existing OLP algorithms fall into two categories: LP-based algorithms and LP-free algorithms. The former typically guarantees better performance but requires solving a large number of LPs, which could be computationally expensive. In contrast, LP-free algorithms only require first-order computations but induce a worse performance. In this work, we bridge the gap between these two extremes by proposing a well-performing algorithm that solves LPs at a few selected time points and conducts first-order computations at other time points. Specifically, for the case where the inputs are drawn from an unknown finite-support distribution, the proposed algorithm achieves a constant regret (even for the hard ``degenerate'' case) while solving LPs only $\mcO(\log\log T)$ times over the time horizon $T$. Moreover, when we are allowed to solve LPs only $M$ times, we design the corresponding schedule such that the proposed algorithm can guarantee a nearly $\mcO\left(T^{(1/2)^{M-1}}\right)$ regret. Our work highlights the value of resolving both at the beginning and the end of the selling horizon, and provides a novel framework to prove the performance guarantee of the proposed policy under different infrequent resolving schedules. Numerical experiments are conducted to demonstrate the efficiency of the proposed algorithms. The code is available at \url{https://github.com/guokai-li/infrequent_resolving_olp}.
    
    \end{abstract}

    \section{Introduction}
\noindent Online linear programming (OLP) is a classical problem in online decision making. In this problem, a decision maker manages multiple types of resources, such as airplane seats or products, with limited inventory. Customers arrive sequentially, each requesting a combination of resources (e.g., multi-leg flights or bundled products) and offering a bid price. Upon observing an arriving customer's request and bid price, the decision maker needs to make an irrevocable decision to accept or reject the request in order to maximize the total expected revenue without violating resource constraints. OLP has applications in various areas, including online auctions (\citealt{buchbinder2007online}), advertisement (\citealt{mehta2005adwords}), covering and packing (\citealt{buchbinder2009online}), e-commerce fulfillment (\citealt{jasin2015lp}), and network revenue management (\citealt{jasin2015performance}), among many others. To facilitate our discussion, we first provide the formal formulation of the OLP problem. 

Consider a decision maker with $m$ types of resources facing sequentially arriving customers over $T$ discrete time periods. We assume the initial inventory is $T\bs{\rho}$, where $\bs{\rho}=(\rho_1,\dots,\rho_m)\in\mathbb{R}_{+}^m$ is given and denotes the vector of average inventories per period. Each customer arriving at period $t$ is characterized by a reward $r_t\in \mathbb{R}_+$ and a consumption vector $\bs{A}_t\in \mathbb{R}_+^m$. The time periods are indexed forward, starting from period $t=1$ and ending at period $t=T$. In each time period $t$, after observing the arriving customer characterized by $(r_t,\boldsymbol{A}_t)$, the decision maker must irrevocably decide whether to accept or reject the customer ($x_t\in \{0,1\}$) without knowing future information. 
Given full information $\{(r_t, \bs{A}_t)\}_{t=1}^T$, we can solve the following (offline) integer linear program:
\begin{align}\label{eq:exact_formulation}
    \max_{\bs{x}}\quad & \sum_{t=1}^T r_t x_t \nonumber\\
    \text{s.t.}\quad & \sum_{t=1}^T \bs{A}_t x_t \le T\bs{\rho}, \\
    & x_t \in \{0, 1\}, \quad \forall \; t.\nonumber
\end{align}
In the corresponding online setting, the coefficients $(r_t, \bs{A}_t)$ in \eqref{eq:exact_formulation} are revealed in each period $t$, and then the decision maker needs to determine the decision $x_t$ without knowing future information $\{(r_\ell, \bs{A}_\ell)\}_{\ell=t+1}^T$. 
The goal is to find a nonanticipative policy to maximize the expected revenue over the entire time horizon. Such a problem is commonly referred to as \textit{online linear programming (OLP)}. 

In this paper, we assume that the customers' features $\left\{(r_t, \bs{A}_t)\right\}_{t=1}^T$ are generated i.i.d. from an \textit{unknown} distribution $\mathcal{P}$ with $n\in \mathbb{Z}_+$ supports. Such an assumption is widely adopted in the online resource allocation literature (e.g., \citealt{jasin2015performance}, \citealt{gupta2024greedy}, \citealt{xie2025benefits}). It is also motivated by practical considerations. In many industries including airline and e-commerce, the resource combinations and the prices are designed by the seller, and only customers interested in one of these options are considered. Thus, the number of customer types is the same as the number of provided options, which is typically finite. Moreover, we assume that the time horizon $T$ is known in advance. For example, in the airline industry, the time horizon is determined by the known departure dates of the flights (see, e.g., \citealt{talluri1998analysis}, \citealt{jasin2015performance}, \citealt{chen2024improved}). Without the knowledge of the time horizon, it is hard to derive a policy with a good performance guarantee because the remaining time plays a crucial role in the decision process (see, e.g., \citealt{jasin2012re}, \citealt{agrawal2014dynamic}, \citealt{bumpensanti2020re}). 
With the above assumptions, the OLP problem can be simplified as follows:

There are $n$ types of customers with type-$j$ customers characterized by the coefficients $(r_j, \bs{A}_j)$, and each arriving customer's type is drawn according to a distribution $\{p_j\}_{j=1}^n$ (with $\sum_{j=1}^n p_j=1$). Slightly abusing notation, we define $\bs{r}\in \mathbb{R}_+^n$ and $\bs{A}\in \mathbb{R}_+^{m\times n}$ as the reward vector and the consumption matrix for all types of customers, respectively.  The decision maker's problem is to select a nonanticipative policy $\mu$ to maximize the total expected reward while satisfying the inventory constraints. The OLP problem can be written as follows:
\begin{equation}\label{eq:exact_formula_finite_type}
\begin{split}
    V^*(T) := \max_{\mu\in \Pi} \quad & \mE\left[ \sum_{t=1}^T \sum_{j=1}^n r_j x^t_{\mu, j}\right]\\
    \text{s.t.} \quad & \sum_{t=1}^T\sum_{j=1}^n \bs{A}_j x^t_{\mu, j} \le T\bs{\rho}, \ (a.s.), \\
    & x^t_{\mu, j} \le Y_j^t, \   \forall j, t, \ (a.s.),\\
    & x^t_{\mu, j} \in \{0, 1\}, \   \forall  j, t,
\end{split}
\end{equation}
where $\Pi$ is the set of all nonanticipative policies, $x^t_{\mu, j}$ denotes whether the policy $\mu$ accepts a type-$j$ customer at time $t$, and $Y_j^t \in \{0, 1\}$ is the random indicator whether a type-$j$ customer arrives at time $t$. 

Solving the optimal policy $\mu^*$ for \eqref{eq:exact_formula_finite_type} is computationally intractable even if the underlying distribution $\mathcal{P}$ is known. To address such challenges, researchers have proposed various heuristic policies with performance guarantees. A widely adopted performance measure is \textit{regret}, which quantifies the optimality gap of a policy. Specifically, let $V^{\mu}(T)$ denote the expected revenue accumulated over the time horizon $T$ under a given policy $\mu$. We define the regret formally as follows:
\begin{align}\label{eq:regret_define}
    \text{Reg}^{\mu}(T)  = V^*(T) - V^{\mu}(T) = V^*(T) - \mE\left[\sum_{t=1}^T\sum_{j=1}^n r_j x_{\mu, j}^{t} \right].
\end{align}
In an asymptotic regime where $T$ scales up, a policy $\mu$ is asymptotically optimal if $\text{Reg}^{\mu}(T)=o(T)$, implying the competitive ratio $V^{\mu}(T)/V^*(T)$ converges to one as the time horizon $T$ goes to infinity. Thus, a lower-order regret typically implies a better performance, and designing a policy with constant regret, i.e., $\text{Reg}^{\mu}(T)=\mcO(1)$, is highly desirable. Furthermore, since $V^*(T)$ is hard to analyze, researchers typically replace it with an upper bound. In our work, we consider the ``hindsight'' upper bound $V^H(T)$, which knows the types of all future arrivals, and utilize the fact that $\text{Reg}^{\mu}(T) \le V^H(T) - V^{\mu}(T)$ in the proof. The details are provided in Section~\ref{subsec:regret_proof}.

Since \eqref{eq:exact_formula_finite_type} is computationally intractable, researchers typically design heuristic policies based on a relaxed problem. Specifically, if we relax the integer constraints in \eqref{eq:exact_formula_finite_type} and replace all random variables with their (estimated) expectations, then we can derive the following linear program (let $y_j$ denote the number of accepted type-$j$ customers, and $\hat{\bs{p}}^t$ denote the estimated probabilities at period $t$)
\begin{equation}\label{eq:fluid_LP_intro}
\begin{split}
    \bar{V}(T) := \max_{\bs{y}\ge \bs{0}} \quad & \bs{r}^{\mathsf{T}} \bs{y}\\
    \text{s.t.} \quad & \bs{A}\bs{y} \le T\bs{\rho}, \\
    & \bs{y} \le \hat{\bs{p}}^1\cdot  T, \   \forall j,
\end{split}
\end{equation}
which is referred to as the ``fluid model'' (at period 1). In some periods, we may update the remaining inventory and the estimated future arrivals in \eqref{eq:fluid_LP_intro} and resolve the (updated) fluid model. 

Broadly speaking, based on the number of LP resolvings, OLP algorithms in prior studies can be categorized into two types: \textit{LP-based algorithms} and \textit{LP-free algorithms}. LP-based algorithms typically make decisions based on the optimal primal/dual solution to the updated fluid LP \eqref{eq:fluid_LP_intro} in each period, offering strong performance guarantees but requiring solving a large number of LPs (e.g., \citealt{li2022online}, \citealt{chen2024improved}, \citealt{xie2025benefits}). Although commercial solvers can efficiently solve LPs, frequent LP resolving remains computationally expensive especially for large-scale or time-sensitive problems. For example, the leading hotel booking platform Booking.com received about 560 million visits per month during 2022-2024, implying an average interarrival time of about five milliseconds (\citealt{statista2024most}). In contrast, it typically takes from seconds to minutes to solve linear programs of practical problems (see, e.g., \citealt{mittelmann2024benchmark}). Moreover, the optimal basis may change during the time horizon, and hence for each period we cannot simply use the last-period optimal solution to significantly speed up LP solving. In response, researchers have recently proposed LP-free algorithms that typically use gradient descent methods to derive an approximate solution converging to the optimal dual solution to the fluid LP, and make decisions based on the approximate solution. Thus, such algorithms rely only on first-order computations and avoid solving any LP (e.g., \citealt{li2020simple}, \citealt{gao2024decoupling}, \citealt{ma2025optimal}).  However, these LP-free approaches usually lead to weaker performance bounds than the LP-based algorithms.

In this paper, we aim to strike a balance between the performance guarantee of LP-based algorithms and the computational efficiency of LP-free algorithms. Specifically, we propose an algorithm which achieves an $\mcO(1)$ regret for the OLP problem while requiring solving LPs only $\mcO(\log\log T)$ times. In our algorithm, we concentrate the resolving (of the fluid LP) periods 1)  on the beginning of the time horizon and 2) toward the end of the time horizon. The first set of resolvings aims to update the policy when the data is scarce so that correcting learning errors is important, while the latter set of resolvings aims to update the policy when the inventory is running out so that the optimal policy varies drastically. Between two resolvings, we use the optimal fluid solution of the latest resolving plus some first-order computations to guide the allocation. Moreover, when we are only allowed to solve LPs for up to $M$ times, we propose an algorithm that can achieve an $\mcO\left(T^{ (1/2+\epsilon)^{M-1}}\right)$ regret, where $\epsilon$ can be any positive constant. If we take $\epsilon\to 0$, the regret bound is about $\mcO\left(T^{ (1/2)^{M-1}}\right)$.
In addition, we show that the proposed algorithms can be easily adapted to the settings where the arrival probabilities of each type of request are known, which also lead to near-optimal regrets in those settings. In this case, a modified algorithm can achieve an $\mcO\left(T^{ (1/2+\epsilon)^M}\right)$ regret by solving LPs only $M$ times.
Therefore, our results demonstrate that one can achieve near-optimal regrets only with a few resolvings, and depict the precise tradeoff between the frequency of resolving and the performance of the proposed algorithm.

In addition to the strong performance of our proposed algorithm, we would like to highlight one especially significant feature of our algorithm. In many previous studies, an important ``\textit{non-degeneracy}'' assumption is imposed on the underlying input data and such an assumption would greatly affect the performance of proposed algorithms. Particularly, the ``non-degeneracy'' assumption refers to the assumption that the fluid model \eqref{eq:fluid_LP_intro} with the estimation $\hat{\bs{p}}^1$ replaced by the true value $\bs{p}$ (called the ``no-learning fluid model'') is non-degenerate. In some works (e.g., \citealt{wei2023constant}, \citealt{gupta2024greedy}), a $\delta$ is defined to be a measure of the distance between the current inventory configuration $T\bs{\rho}$ and the nearest inventory configuration under which the no-learning fluid model is degenerate, and the derived regret bounds contain $1/\delta$ terms (thus their regret bounds drastically increase and tend to infinity as $\delta$ tends to zero). Importantly, our results do not rely on the non-degeneracy assumption. In fact, to the best of our knowledge, our algorithm is the first that achieves a constant regret with such few resolvings for the case without the non-degeneracy assumption and distribution knowledge. To distinguish different results, we use $\mcOd(\cdot)$ ($\mcO(\cdot)$, resp.) to denote regret bounds containing (without, resp.) $1/\delta$.

Finally, we would like to highlight the technical contribution of this work.
In the literature on LP-based algorithms for solving OLP, the algorithm usually solves an updated fluid model in each time period (e.g., \citealt{vera2019bayesian}, \citealt{chen2024improved}, \citealt{xie2025benefits}). This approach allows updating the probability estimation frequently and bridging their policies with the optimal hindsight policy through two LPs: the fluid model and the hindsight benchmark. 
In contrast, our approach solves the updated fluid model only at a few selected periods ($\mcO(\log\log T)$ periods), and hence cannot update the probability estimation and access the optimal solution of the updated fluid model in most periods, posing challenges for the analysis. In order to overcome these technical challenges, we approximate the optimal solution of the updated fluid model based on the latest obtained fluid solution and some first-order computations. While this solution is sub-optimal for the updated fluid model for the corresponding period, we prove that it is optimal to a surrogate LP with high probability.
Therefore, we can bridge our policy with the optimal hindsight policy through the surrogate LP and the hindsight LP, from which we can obtain the desired result.

The remainder of this paper is organized as follows. In the rest of this section, we review literature related to our work. In Section~\ref{sec:main_result}, we propose the main algorithm and prove the regret bounds under the infrequent resolving schedule and the finite-resolving schedule. In Section~\ref{sec:nrm}, we study the case with known arrival probabilities, which is referred to as the known-probability case. 
In Section~\ref{sec:numerical}, we compare our policy with several existing policies and provide additional insights through numerical experiments. 
We conclude the paper in Section~\ref{sec:conclusion}. All proofs are relegated to the appendix. 
 
\subsection{Literature Review}\label{subsec:literature}

Online decision making has a rich history within operations research and theoretical computer science, and remains a vibrant and flourishing area. Academic studies in this field typically focus on designing online
algorithms that make real-time decisions based on limited information and adapt their strategies as new data becomes available. For a comprehensive review, we refer readers to \cite{borodin2005online}, \cite{buchbinder2009design} and \cite{hazan2016introduction}. \textit{Online linear programming} (OLP) is a classical problem in online decision making. In the OLP problem, in order to maximize the expected revenue under resource constraints, the decision maker needs to dynamically make irrevocable decisions to accept or reject the arriving customers' requests. There are two streams of research categorized by whether the arrival probabilities are known at the beginning. In the following, we review the literature of these two streams separately.

\textit{Unknown Distribution.} 
We start with the stream assuming unknown type distribution. Our work closely relates to studies under the random input assumption, where the coefficients $\left\{(r_t, \bs{A}_t)\right\}_{t=1}^T$ are generated i.i.d. from an unknown distribution $\mcP$. For the underlying distribution, there are two distinct assumptions: finite-support distribution and continuous-support distribution. 

We start with papers under the finite-support distribution assumption, which is the same as our setting. Under such an assumption, the distribution $\mcP$ is supported by finite bounded points such that arrivals can be categorized into finite types. We first review the so-called \textit{LP-based algorithms} which require solving many LPs but guarantee good performance. For example, under the non-degeneracy assumption (i.e., $\delta>0$), \cite{jasin2015performance} proposes an $\mcOd(\log^2 T)$-regret algorithm that requires solving LPs $\mcO(\log T)$ times. Subsequently, \cite{chen2024improved} consider a similar algorithm that requires solving LPs $T$ times, achieving an $\mcOd(1)$ regret under the non-degeneracy assumption and $\mcO(\sqrt{T}\log T)$ otherwise. \cite{wei2023constant} propose a primal-dual algorithm that solves LPs $T$ times and the corresponding regret is $\mcOd(1)$ under the non-degeneracy assumption and $\mcO(\sqrt{T})$ in general. \cite{xie2025benefits} remove the non-degeneracy assumption and provide an OLP algorithm with $\mcO(1)$ regret but still requiring solving LPs $T$ times. As mentioned, although LP-based algorithms have good performance guarantees, frequent LP solving can be computationally expensive for large-scale or time-sensitive problems. To address these computational challenges, recent studies have developed \emph{LP-free algorithms} without LP resolving. The pioneering works \cite{balseiro2020dual} and \cite{li2020simple} adopt stochastic gradient descent methods to learn the optimal dual prices and provide $\mcO(\sqrt{T})$-regret LP-free algorithms. These algorithms only require first-order computations and never solve any full LP. In this work, we achieve a balance between computational efficiency and algorithm performance by proposing an algorithm that achieves a constant regret bound by solving LPs $\mcO(\log\log T)$ times. Moreover, we also provide regret bounds when the number of resolvings is finite, which to the best of our knowledge have not been provided in previous literature.

Then, we review papers under the continuous-support distribution assumption. In this case, the distribution $\mcP$ is assumed to be supported by a bounded and continuous set and the probability density function is both lower bounded (away from 0) and upper bounded. (Note that finite-support distributions do not satisfy the assumption due to the existence of mass points.) We still start with LP-based algorithms. For example, under the non-degeneracy assumption,  \cite{li2022online} provide an $\mcOd(\log T \log\log T)$-regret algorithm from the dual perspective, and the algorithm requires solving LPs $T$ times. Then, \cite{bray2025logarithmic} shows that the best possible regret bound for this problem is $\Omega(\log T)$. In addition, \cite{bray2025logarithmic}  and \cite{ma2025optimal} prove that the regret bound of \cite{li2022online}'s algorithm is $\mcO_\delta(\log T)$ under the non-degeneracy assumption. Then, we introduce LP-free algorithms for this setting. The $\mcO(\sqrt{T})$ regret bounds established in \cite{balseiro2020dual} and \cite{li2020simple} also hold in this case. Subsequently, researchers have studied variants of these algorithms and tried to derive tighter bounds under the non-degeneracy assumption. For example, \cite{gao2024decoupling} propose variants that improve the bounds to $\mcOd(T^{1/3})$. Then, \cite{ma2025optimal} propose an LP-free algorithm which guarantees an $\mcOd(\log^2 T)$ regret. A summary of the existing results and our result is presented in Table~\ref{tab:OLP_comparison} (where $\epsilon$ can be any positive constant). 

\begin{table}[ht]
    \centering
    \caption{Comparison among OLP algorithms for unknown-probability case. Entries marked ``-'' represent cases where the corresponding bound is not examined or reported in the paper.}
    \label{tab:OLP_comparison}
    \begin{adjustbox}{max width=\textwidth}
     \begin{tabular}{c c c c c c c} 
     \toprule
     Paper  & \multicolumn{1}{c}{\begin{tabular}[c]{@{}c@{}}Regret\\ (Non-degenerate Case)\end{tabular}} &  \multicolumn{1}{c}{\begin{tabular}[c]{@{}c@{}}Regret\\ (Degenerate Case)\end{tabular}} & \# of Resolvings & Distribution Assumption  \\ 
     \midrule
     \cite{jasin2015performance}  & $\mcOd(\log^2 T)$ & -  & $\mcO(\log T)$ & Finite\\
     \cite{chen2024improved}   & $\mcOd(1)$ & $\mcO(\sqrt{T}\log T)$ & $\mcO(T)$ & Finite\\
     \cite{wei2023constant} & $\mcOd(1)$ & $\mcO(\sqrt{T})$ & $\mcO(T)$ & Finite\\
     \cite{xie2025benefits}  & $\mcO(1)$ & $\mcO(1)$ & $\mcO(T)$ & Finite\\
     \cite{li2022online}  & $\mcOd(\log T \log\log T)$ & - & $\mcO(T)$  & Continuous\\
     \cite{bray2025logarithmic} & $\mcOd(\log T)$ & - & $\mcO(T)$  & Continuous\\
     \cite{ma2025optimal} & $\mcOd(\log T)$ & - & $\mcO(T)$ & Continuous\\
     \hline
     \cite{balseiro2020dual} & $\mcO(\sqrt{T})$ & $\mcO(\sqrt{T})$ & $0$ & General\\
     \cite{li2020simple}  & $\mcO(\sqrt{T})$ & $\mcO(\sqrt{T})$ & $0$ & General\\
     \cite{gao2024decoupling}  & $\mcOd(T^{1/3})$ & - & $0$ & Continuous\\
     \cite{ma2025optimal} & $\mcOd(\log^2 T)$ & - & $0$ & Continuous\\
     \hline
     This paper  & $\mcO(1)$ & $\mcO(1)$ & $\mcO(\log \log T)$ & Finite\\
     This paper & $\mcO\left(T^{(1/2+\epsilon)^{M-1}}\right)$ & $\mcO\left(T^{(1/2+\epsilon)^{M-1}}\right)$ & $M$ & Finite\\
     \bottomrule
     \end{tabular}
     \end{adjustbox}
    \end{table}

Moreover, there are some works extending the previous settings. For example, for any bounded distribution (subsuming previous two classes), \cite{balseiro2020dual}, \cite{li2020simple}, \cite{balseiro2022dual} and \cite{gao2023solving} provide LP-free algorithms with $\mcO(\sqrt{T})$ regrets. \cite{balseiro2023best} and \cite{jiang2025online} consider the case when the underlying distribution is non-stationary. \cite{besbes2012blind} and \cite{ferreira2018online} consider the pricing problem without the knowledge of the demand function.

We also note another stream of research under the random permutation assumption. Here, the set $\{(r_t, \bs{A}_t)\}_{t=1}^T$ is adversarially chosen, but the arrival order is uniformly distributed over all the permutations. Researchers aim to derive necessary conditions for the existence of a $(1-\epsilon)$-competitive algorithm, see, e.g., \cite{agrawal2014dynamic}, \cite{gupta2014experts}, \cite{kesselheim2014primal}, \cite{molinaro2014geometry}. Since the concentration bound under the random permutation assumption is weaker, the lower bound in \cite{agrawal2014dynamic} implies that the asymptotic order of the regret is $\Omega(\sqrt{T})$. We also highlight that \cite{agrawal2014dynamic}, \cite{gupta2014experts} and \cite{molinaro2014geometry} solve LPs $\mcO(\log T)$ times and their resolving periods concentrate at the beginning, sharing similarities with the first half of our resolving schedule.

\textit{Known distribution.} We now review the stream assuming the knowledge of the arrival probabilities at the beginning, which is widely studied in the network revenue management (NRM) problem. Originating from the airline industry, NRM has garnered significant attention from both academia and industry. There are two main streams of research, price-based NRM and quantity-based NRM. The former studies the dynamic pricing problem under resource constraints (see \citealt{gallego1994optimal}); the latter studies the dynamic resource allocation problem (see \citealt{talluri1998analysis}). Our work is closely related to the quantity-based NRM, in which most works adopt the finite-support assumption (see \citealt{besbes2025dynamic}, \citealt{chen2025beyond}, \citealt{jiang2025degeneracy} and \citealt{zhang2026online} for some exceptions). Most early works only solve the fluid LP once at the beginning and then design a policy based on the optimal solution. For example, \cite{talluri1998analysis} prove an $\mcO(\sqrt{T})$ regret bound for the bid-price control (BPC) policy, which uses the optimal dual solution to the fluid model as the values of resources and accepts a request if and only if the offered price is larger than the total value of resources. \cite{cooper2002asymptotic} proves an $\mcO(\sqrt{T})$ regret bound for the booking-limit control (BLC) policy, which assigns quotas to each request type according to the fluid model and accepts a request until the corresponding quota is depleted. \cite{reiman2008asymptotically} prove an $\mcO(\sqrt{T})$ regret bound for the probabilistic allocation control (PAC) policy, which probabilistically accepts a request according to the ratio of the fluid model solution to the expected demand.

Observing the potential to resolve the fluid model to reduce the regret, many works consider LP-based algorithms with more resolvings. Although \cite{cooper2002asymptotic} provides a two-period example showing that resolving may increase the value of the regret, subsequent works (e.g., \citealt{jasin2012re}) find that resolving does not worsen (and can even improve) the asymptotic order of the regret (i.e., $O(\sqrt{T})$).
For example, \cite{reiman2008asymptotically} prove that a single resolving can reduce the asymptotic regret of the PAC policy from $\Theta(\sqrt{T})$ to $o(\sqrt{T})$. \cite{jasin2012re} and \cite{jasin2013analysis} show that the PAC policy can significantly benefit from resolving, while neither BPC nor BLC can benefit. Specifically, under the non-degeneracy assumption, the PAC policy with periodic ($\mcO(T)$ times) or midpoint ($\mcO(\log T)$ times) resolving can achieve $\mcOd(1)$ regret. Without the non-degeneracy assumption, \cite{bumpensanti2020re} show that the above regret in general is $\Omega(\sqrt{T})$, and they provide a modified PAC policy with infrequent (i.e., $\mcO(\log \log T)$ times) resolving that guarantees $\mcO(1)$ regret. Note that the resolving schedule in \cite{bumpensanti2020re} shares some similarities with the second half of our schedule, but their proof framework cannot deal with the case without distribution knowledge.  Recently, \cite{arlotto2019uniformly} study an alternative interpretation of the fluid solution, which accepts the request if and only if the ratio of the primal solution to the expected demand is no less than $1/2$. They prove an $\mcO(1)$ regret bound for the multi-secretary problem. In our work, we refer to such a policy as the ``\textit{argmax policy}'' because it takes the action (accept/reject) with the larger value in the primal solution. Then, \cite{vera2019bayesian} and \cite{vera2021online} generalize this idea to the multi-constraint problem, but their policy requires solving LPs in every period. Subsequently, \cite{banerjee2024good} propose a constant-regret algorithm whose \textit{expected} number of resolvings is $\mcO(\log\log T)$. Similarly, the proof in \cite{banerjee2024good} cannot handle the unknown-probability case under infrequent resolving.  In our work, inspired by the argmax policy, we propose a constant-regret policy whose resolving schedule can be determined at the beginning and the number of resolvings is $\mcO(\log\log T)$.

Then, we review some results for the case when the decision maker is only allowed to solve a finite number (i.e., $M$) of LPs under the NRM model. For this problem, \cite{reiman2008asymptotically} show that the regret can be reduced to $\mcO\left(T^{\frac14 + \epsilon}\right)$ if we can solve LPs twice. However, the proof cannot be directly extended to the multiple-resolving case. Under the non-degeneracy assumption, \cite{jasin2012re} prove that solving LPs at $M$ periods can induce an $\left(\rho(M) + \hat{\rho}(M) \cdot T^{(1/2)^M}\right)$ regret bound, where $\rho(M)$ and $\hat{\rho}(M)$ are independent of $T$. Thus, given a finite $M$ independent of $T$, the regret bound in \cite{jasin2012re} can be represented as $\mcOd\left(T^{(1/2)^M}\right)$. Moreover, their finite-resolving schedule is not fixed at the beginning, but is determined adaptively based on the realization of the arrival process. \cite{bumpensanti2020re} prove that the regret bound of their policy is $\mcO\left(T^{5/12}\right)$ given $M=2$. \cite{sun2020near} propose an LP-free algorithm with $\mcOd\left(T^{3/8}(\log T)^{5/4}\right)$ regret under the non-degeneracy assumption. In addition, \cite{gupta2024greedy} and \cite{he2025online} recently propose algorithms that solve the fluid LP only once at the beginning (to find the optimal basis) and then greedily make decisions to minimize some function in each period, and the regret is proved to be $\mcOd(1)$ under the non-degeneracy assumption.
Compared to the above literature, our results can match the single-resolving result of \cite{reiman2008asymptotically} and extend it to the multi-resolving case. We derive regret bounds similar to \cite{jasin2012re} but allow for degenerate cases.  A summary of the existing results and our result is presented in Table~\ref{tab:NRM_comparison} (where $\epsilon$ can be any positive constant).

\begin{table}[ht]
    \centering
    \caption{Comparison among algorithms for known-probability case under finite-support distribution. Entries marked ``-'' represent cases where the corresponding bound is not examined or reported in the paper.}
    \label{tab:NRM_comparison}
    \begin{adjustbox}{max width=\textwidth}
     \begin{tabular}{c c c c c c} 
     \toprule
     Paper  & \multicolumn{1}{c}{\begin{tabular}[c]{@{}c@{}}Regret\\ (Non-degenerate)\end{tabular}} &  \multicolumn{1}{c}{\begin{tabular}[c]{@{}c@{}}Regret\\ (Degenerate)\end{tabular}} & \# of Resolvings\\ 
     \midrule
     \cite{jasin2012re}  & $\mcOd(1)$ & - & $\mcO(\log T)$ \\
     \cite{bumpensanti2020re}  & $\mcO(1)$ & $\mcO(1)$   & $\mcO(\log \log T)$ \\
     \cite{vera2019bayesian}  & $\mcO(1)$ & $\mcO(1)$  & $\mcO(T)$ \\
     \cite{vera2021online}  & $\mcO(1)$ & $\mcO(1)$  & $\mcO(T)$ \\
     \cite{banerjee2024good}   & $\mcO(1)$ & $\mcO(1)$ &  $\mcO(\log \log T)$ (\textbf{In Expectation})  \\
     \midrule 
     \cite{reiman2008asymptotically} & $\mcO(T^{\frac14+\epsilon})$ & $\mcO(T^{\frac14+\epsilon})$ & 2 \\
     \cite{jasin2012re} & $\mcOd(T^{\frac{1}{2^{M}}})$ & -  & $M$ \\
     \cite{bumpensanti2020re} & $\mcO(T^{5/12})$ & $\mcO(T^{5/12})$ & 2 \\
     \cite{sun2020near} & $\mcOd(T^{3/8}(\log T)^{5/4})$ & - & 0 \\
     \cite{gupta2024greedy} & $\mcOd(1)$ & - & 1  \\
     \cite{he2025online} & $\mcOd(1)$ & - & 1  \\
     \midrule
     This paper  & $\mcO(1)$ & $\mcO(1)$  & $\mcO(\log\log T)$ \\
     This paper & $\mcO\left(T^{(1/2+\epsilon)^M}\right)$ & $\mcO\left(T^{(1/2+\epsilon)^M}\right)$ & $M$ \\
     \bottomrule
     \end{tabular}
     \end{adjustbox}
\end{table}

Moreover, our work shares some similar trade-offs with the literature on batched bandits, where the decision maker partitions all individuals into at most $M$ batches and can only observe bandit outcomes batch by batch. For example, under the non-contextual bandit setting with $M$ batch updates, \cite{perchet2016batched} and \cite{gao2019batched} establish lower bounds of regret as $\tilde{\Omega}\left(T^{\frac{1}{2-2^{1-M}}}\right)$, and propose algorithms that match these bounds. Their results imply that achieving the standard regret $\Theta(\sqrt{T})$ under per-period update (i.e., $M=T$) requires only $\Theta(\log\log T)$ batch updates. Then, \cite{han2020sequential}, \cite{ruan2021linear} and \cite{ren2024dynamic} investigate the batched updates under the contextual bandit setting. Although the problem settings of these works significantly differ from our work, their batch update schedules are also concentrated at the beginning of the time horizon, sharing some similarities with the first half of our resolving schedule. Both the above works and our work demonstrate that early-stage learning is crucial and carefully designed infrequent learning can perform comparably to frequent learning. Furthermore, due to the inventory constraints, our work needs to simultaneously learn the arrival rates and allocate inventory, and hence the second half of our resolving schedule is introduced.

\section{Main Results}\label{sec:main_result}
In this section, we propose an algorithm to solve problem \eqref{eq:exact_formula_finite_type} with constant regret. For ease of notation, we define the following LP parameterized by inventory $\bs{b}$ and demand $\bs{d}$:
\begin{equation}\label{eq:hindsight}
\begin{split}
    \phi(\bs{b}, \bs{d}) := \max_{\bs{y}\ge \bs{0}} \quad & \bs{r}^{\mathsf{T}} \bs{y} \\
    \text{s.t.}\quad &  \bs{A}\bs{y} \le \bs{b}\\
    & \bs{y} \le \bs{d}.
    \end{split}
\end{equation}
At time $t$, suppose the remaining inventory is $\bs{b}^t$ and the realized arrivals from period $\ell=1$ to period $\ell=t-1$ are $\{Y^\ell_j: \ell=1, 2,\dots, t-1, j=1, 2, \dots, n\}$. We refer to $\phi(\bs{b}^t, (T-t+1)\hat{\bs{p}}^t)$ as the ``fluid model" in period $t$, where $\hat{p}_j^t = \left(\sum_{\ell=1}^{t-1} Y_j^\ell\right)/(t-1)$ is the empirical estimation of arrival probability $p_j$ at time $t$. We also let $\hat{\bs{p}}^1=\bs{0}$. The fluid model replaces all uncertainties with their expectations. The decision variable $y_j$ represents the expected number of accepted type-$j$ customers. The first constraint ensures that the total resource consumption does not exceed the remaining inventory, and the second constraint ensures that the number of accepted customers does not exceed the demand. For the second constraint, since the arrival probabilities $p_j$ are unknown, we use the empirical estimation $\hat{\bs{p}}^t$, approximating the future demand as $(T-t+1)\hat{\bs{p}}^t$.

\subsection{Argmax with Infrequent Resolving (AIR) Policy}
We now introduce our policy in Algorithm~\ref{alg:period_resolving}, referred to as the Argmax with Infrequent Resolving (AIR) policy. We use $\bs{e}_j$ to denote a vector of zeros except 1 at the $j$-th entry, and $[n]$ to denote the set $\{1, 2, 3, \dots, n\}$. The time set $\mcT$ in Algorithm~\ref{alg:period_resolving} will be specified shortly. Throughout this paper, we adopt the convention that $0/0:=0$. 

\begin{algorithm}
    \caption{Argmax with infrequent resolving (AIR) policy}\label{alg:period_resolving}
    \begin{algorithmic}
    \State Input: Time set $\mathcal{T}= \{T_1, T_2, T_3, \dots, T_{|\mcT|}\}$.
    \State Initialize $\bs{b}^1 \gets T\bs{\rho}$, $\bs{N}^1 \gets \bs{0}$, $\bs{u}^{1} \gets \bs{0}$ and $\bs{d}^1 \gets \bs{0}$.
    \For{$t=1, 2, 3, \dots, T$} 
    \If{$t\in \mathcal{T}$} \Comment{Infrequent resolving}
        \State Update the empirical estimations $\hat{p}_j^t \gets N_j^t/(t-1)$ for each $j$.
        \State Solve the fluid problem $\phi(\bs{b}^t, (T-t+1)\hat{\bs{p}}^t)$ and obtain its optimal solution $\bs{y}^{t, *}$. 
        \State Set $u^t_j\gets y^{t, *}_j$ and $d^t_j\gets \hat{p}^t_j (T-t+1)$ for all $j$.
    \EndIf
        \State Observe arrival type $j$ and set $\bs{N}^{t+1} \gets \bs{N}^t +\bs{e}_j$.
        \If{$\bs{A}_j\le \bs{b}^t$, $u_j^t\ge 1$, and $u_j^t\ge d_j^t-u_j^t$} \Comment{Argmax between $u_j^t$ and $d_j^t-u_j^t$}
            \State Accept the request. 
            \State Set $\bs{b}^{t+1} \gets \bs{b}^t-\bs{A}_j$. \Comment{Update the remaining capacity}
            \State Set $\bs{u}^{t+1} \gets \bs{u}^t-\bs{e}_j$. \Comment{Approximate the optimal solution}
        \Else
        \State Reject the request and set $\bs{b}^{t+1} \gets \bs{b}^t$ and $\bs{u}^{t+1}\gets \bs{u}^t$.
        \EndIf
        \State Set $\bs{d}^{t+1} \gets \bs{d}^{t}-\bs{e}_j$.\Comment{Approximate the future demand}
    \EndFor
    \end{algorithmic}
    \end{algorithm}

We now explain the intuition of Algorithm~\ref{alg:period_resolving}. The algorithm begins with a predetermined time set $\mcT$ specifying the time points where resolving is needed. To implement our policy, we need to maintain two approximations in our algorithm: The vector $\bs{u}^t$ approximates the numbers of accepted future customers (of different types) under the optimal policy, and the vector $\bs{d}^t$ approximates the numbers of future customer arrivals (of different types). In each resolving period $t\in \mcT$, the decision maker updates the empirical estimations $\hat{\bs{p}}^t$ and solves the corresponding fluid LP based on the current inventory $\bs{b}^t$ and the estimation $\hat{\bs{p}}^t$. Then, we update the approximations as follows: $\bs{u}^t=\bs{y}^{t, *}$ and $\bs{d}^t=(T-t+1)\hat{\bs{p}}^t$. In each non-resolving time period $t\notin \mcT$, since the decision maker cannot access the optimal solution $\bs{y}^{t, *}$ for the current period, the decision is determined based on $\bs{u}^t$ and $\bs{d}^t$. At the end of each period, we update the approximations $\bs{u}^t$ and $\bs{d}^t$ by only two subtraction operations in a ``greedy'' fashion as described in the algorithm: Once a type-$j$ customer arrives at period $t$, we subtract the number of future type-$j$ arrivals by one, i.e., $\bs{d}^{t+1}=\bs{d}^{t}-\bs{e}_j$; once a type-$j$ customer is accepted at period $t$, we decrease the number of future accepted type-$j$ arrivals by one, i.e., $\bs{u}^{t+1}=\bs{u}^{t}-\bs{e}_j$. 

Given the approximations $\bs{u}^t$ and $\bs{d}^t$, we adopt the idea of the ``argmax'' policy in \cite{arlotto2019uniformly} and \cite{vera2021online} to make accept/reject decisions. At each period $t$, the decision maker observes the arrival type $j$ and accepts the request only if it is feasible to do so (i.e., $\boldsymbol{A}_j\le \boldsymbol{b}^t$), $u_j^t\ge d_j^t-u_j^t$, and at least one customer should be accepted ($u_j^t \ge 1$). Note that $\boldsymbol{u}_t$ approximates the number of customers that should be accepted and $\boldsymbol{d}_t$ approximates the future demand. Intuitively, $d_j^t-u_j^t$ represents the number of type $j$ requests that should be rejected and the decision maker accepts the type-$j$ request if and only if more should be accepted than rejected (i.e., $u_j^t\ge d_j^t-u_j^t$). In the following, we use $\mcA$ to denote the AIR policy in Algorithm~\ref{alg:period_resolving}. 

In the following, we specify a resolving schedule $\mathcal{T}$ with $|\mathcal{T}|=\mcO(\log \log T)$, and then prove the constant regret bound.

\subsection{Resolving Schedule}\label{subsec:resolve_schedule}
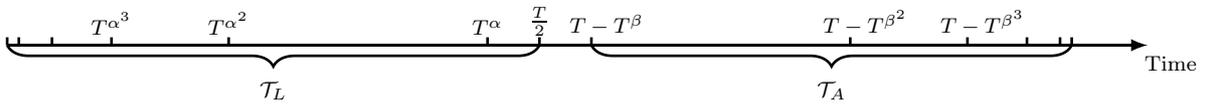
\begin{figure}[ht!]
    \centering
    \begin{tikzpicture}[%
        every node/.style={
            font=\footnotesize,
        },
        line width = 1pt
    ]
    \draw[line width=0.4mm, -latex] (0,0) -- (15,0)
    node[anchor=north, pos=1.02, align=center] {Time};
    \foreach \x in {0, 6.31688, 7.68312, 2.91283, 11.0872, 1.3715, 12.6285, 0.588723, 13.4113, 0.153104, 13.8469, 7, 14}{
        \draw (\x cm,3pt) -- (\x cm,0pt);
    }
    
    \node[anchor=south] at (7,0) {$\frac{T}{2}$};
    \node[anchor=south] at (6.31688,0) {$T^\alpha$};
    \node[anchor=south] at (7.68312+0.2,0) {$T-T^\beta$};
    \node[anchor=south] at (2.91283,0) {$T^{\alpha^2}$};
    \node[anchor=south] at (11.0872+0.2,0) {$T-T^{\beta^2}$};
    \node[anchor=south] at (1.3715,0) {$T^{\alpha^3}$};
    \node[anchor=south] at (12.6285+0.2,0) {$T-T^{\beta^3}$};
    
    \draw[decorate,decoration={brace,amplitude=8pt, mirror}] (0,0) -- (7,0)
        node[anchor=south,midway,below=10pt, align=center] {$\mcT_L$};
    \draw[decorate,decoration={brace,amplitude=8pt, mirror}] (7.68312,0) -- (14,0)
        node[anchor=south,midway,below=10pt, align=center] {$\mcT_A$};
    \end{tikzpicture}
    \caption{Illustration of resolving time set $\mcT=\mcT_L\cup\mcT_A$.}
    \label{fig:illustrate_resolving_time}
\end{figure}

To achieve the constant regret, we introduce the time set $\mcT = \mcT_L\cup \mcT_A$. Specifically, the first subset is called the ``learning'' time set, specified as 
\[\mcT_L=\left\{\left\lceil T^{\alpha^{K_L}}\right\rceil, \dots, \left\lceil T^{\alpha^3}\right\rceil, \left\lceil T^{\alpha^2}\right\rceil, \left\lceil T^\alpha\right\rceil \right\}\bigcup\left\{\left\lceil \frac{T}{2}\right\rceil\right\} \]
with $\alpha\in (0, 1)$ and $K_L = \lceil \log_{\frac{1}{\alpha}}\log_3 T \rceil$. The second subset is called the ``approximation'' time set, and is specified as 
\[\mcT_A=\left\{\left\lceil T-T^\beta \right\rceil, \left\lceil T-T^{\beta^2} \right\rceil, \left\lceil T-T^{\beta^3} \right\rceil, \dots, \left\lceil T-T^{\beta^{K_A}}\right\rceil\right\}\] with $\beta\in(\frac12, 1)$ and $K_A=\lceil \log_{\frac{1}{\beta}}\log_3 T \rceil$. To facilitate understanding, we illustrate the resolving times in Figure~\ref{fig:illustrate_resolving_time}. 

As Figure~\ref{fig:illustrate_resolving_time} shows, the learning time set $\mcT_L$ is concentrated at the beginning of the time horizon. This is similar to the schedules in \cite{agrawal2014dynamic} and \cite{gupta2014experts}, but our method requires solving fewer LPs, i.e., $\mcO(\log \log T)$ times compared to $\mcO(\log T)$ times in \cite{agrawal2014dynamic} and \cite{gupta2014experts}. Since empirical estimations fluctuate drastically at the beginning of the time horizon, the learning time set is designed to update the estimations $\hat{\bs{p}}^t$ promptly to avoid the accumulation of learning error. In contrast to the learning time set, the approximation time set $\mcT_A$ is concentrated at the end of the time horizon. It shares some similarities with the NRM literature, e.g., \cite{jasin2012re} and \cite{bumpensanti2020re}. At the end of the time horizon, the solution to $\phi(\bs{b}^t, (T-t+1)\hat{\bs{p}}^t)$ usually changes drastically because of the scarce inventory. Since the first-order computations may incur significant errors in this case, the approximation time set is thus designed to control the deviation of the approximated solution $\bs{u}^t$ from the true optimal fluid solution. The total number of resolvings is $|\mcT|= K_L+K_A+1=\mcO(\log\log T)$. Then, we present our main result in the following theorem. 

\begin{theorem}[Regret Bound]\label{thm:regret_bound}
    Given the resolving schedule $\mcT$ with $\alpha\in (0, 1)$ and $\beta\in (\frac12, 1)$, the regret of the AIR policy is $\mcO(1)$. 
\end{theorem}

According to Theorem~\ref{thm:regret_bound}, the AIR algorithm can achieve a constant regret by solving $\mcO(\log\log T)$ LPs. Before proceeding, we provide an overview of the proof idea of Theorem~\ref{thm:regret_bound} in four steps. First, we decompose the regret by periods, and identify events under which a revenue loss occurs. By doing so, we reduce the analysis to bounding the probability of such events. Second, we prove that $\bs{d}^t$ in Algorithm~\ref{alg:period_resolving} is a good approximation of future arrival numbers. Third, we identify that $\bs{u}^t$ is an optimal solution of a surrogate LP with high probability, and hence can show that $\bs{u}^t$ is a good approximation of the optimal solution of the fluid problem. Lastly, given that $\bs{u}^t$ and $\bs{d}^t$ are good approximations, we can prove that the event probability in the first step is small under the argmax decision, resulting in a constant regret. In the following subsection, we present a more detailed proof of Theorem~\ref{thm:regret_bound} step by step. Before that, we first provide remarks to compare our techniques with the literature reviewed in Section~\ref{subsec:literature} to highlight our contribution.

\begin{remark}[Comparison with Literature]
    First, \cite{bumpensanti2020re} consider the problem where the probabilities are known at the beginning, and provide a constant-regret algorithm with an $\mcO(\log\log T)$ resolving schedule similar to our approximation time set. They use the thresholding technique to adjust the acceptance probabilities in the PAC control: If the probability is lower (higher, resp.) than a threshold, then the probability is adjusted to 0 (1, resp.). The high-level philosophy of the thresholding policy is similar to the argmax policy, which can be seen as a thresholding action with $0.5$ as the threshold. However, unlike our policy, their policy between two resolving periods is static (i.e., the acceptance probabilities are not updated), and the induction proof in \cite{bumpensanti2020re} highly relies on the knowledge of arrival probabilities. Even if we complement the resolving time set with our learning time set, their proof cannot easily be adapted to the unknown-probability case. 

    Second, for the known-probability case, \cite{banerjee2024good} provide a lazy-resolving algorithm achieving a constant regret with $\mcO(\log\log T)$ \emph{expected} number of resolvings. Their algorithm sets lower confidence bounds for the numbers of accepted customers of different types based on the optimal solution of the fluid model, and updates the bounds when the number of accepted customers exceeds the corresponding bound. The proof cannot be easily extended to the unknown-probability case because the lower confidence bounds are sensitive to the estimation error of the arrival probabilities. 

    Third, for the unknown-probability case, \cite{xie2025benefits} achieve constant regret by implementing a similar policy with per-period resolving. However, the proof cannot be directly used for our infrequent resolving schedule because the probability estimation is infrequently updated and the optimal solution to the fluid model is not available in most periods. To deal with this problem, we provide a novel proof to identify a surrogate LP which admits the approximated solution as an optimal solution with high probability, such that the approximated solution can be compared with the optimal solution to the hindsight problem. \hfill \qed
\end{remark}

\subsection{Proof of Regret Bound}\label{subsec:regret_proof}
In this section, we provide a sketch of the proof of Theorem~\ref{thm:regret_bound}. In our discussions, without loss of generality, we assume the time periods in $\mcT$ are indexed in an ascending way, i.e., $T_1<T_2<\cdots <T_{|\mcT|}$. We then present some properties of the time set $\mcT$. Since $\min_{t\in \mcT} t\le 3$ and $\max_{t\in \mcT} t\ge T-3$, we focus on the properties for $t\in [3 , T-3]$, such that there exist $T_{k-1}$ and $T_k$ satisfying $t\in [T_{k-1}, T_k)$. Then, we have the following lemma depicting the relation between $t$ and $T_{k-1}$. 
\begin{lemma}\label{lem:t_relation}
    Given the resolving schedule $\mcT$ and any $t\in [T_{k-1}, T_k)$, we have $T_{k-1} \ge (t-1)^{\alpha}$ and $T-T_{k-1}\le (T-t+1)^{1/\beta}$. 
\end{lemma}
According to Lemma~\ref{lem:t_relation}, we can bound any time $t$ by the latest LP solving time $T_{k-1}$. Next we present the main steps in the proof. The proof can be decomposed into four steps as follows.

$ $\newline
\noindent\tb{Step 1: Regret decomposition.} Since $V^*(T)$ is hard to analyze, to bound the regret \eqref{eq:regret_define}, we start with a benchmark serving as an upper bound to the optimal value $V^*(T)$. Similar to \cite{vera2021online}, we consider the \emph{hindsight} benchmark that knows types of all future arrivals. Specifically, given full arrival information $\boldsymbol{Z}^t$ from time $t$ onward, the hindsight problem at time $t$ is defined as $\phi(\bs{b}^t, \bs{Z}^t)$ where $\bs{b}^t$ is the vector of the remaining inventory at time $t$ with $\bs{b}^1=T\bs{\rho}$ being the initial inventory, $Z_j^t=\sum_{\ell=t}^T Y_j^\ell$ is the number of future type-$j$ arrivals from period $t$ onward, and $\bs{Z}^t$ is the corresponding vector. Moreover, we have $\phi(\bs{b}^{T+1}, \bs{Z}^{T+1})=0$ because $\bs{Z}^{T+1}=\bs{0}$. Then, the hindsight benchmark is defined as $V^H(T):=\mE[\phi(\bs{b}^1, \bs{Z}^1)].$
Intuitively, the hindsight problem has access to future information and thus earns a higher reward than any nonanticipative online algorithm. The following result formalizes this intuition.
\begin{lemma}[Upper Bound] \label{lem:upper_bound}
    Given any $\bs{b}^1\ge \bs{0}$, we have $V^H(T) \ge V^*(T).$
\end{lemma}

According to Lemma~\ref{lem:upper_bound}, it is sufficient to bound the regret by analyzing the difference between the hindsight problem and the performance of our policy. 
Let $j^t$ denote the random type of the $t$-th arrival, and let $x^t_{\mcA, j^t}\in \{0, 1\}$ denote the decision of the AIR policy in period $t$. 
Then, we have 
\begin{equation}\label{eq:regret_decompose}
    \begin{split}
    \text{Reg}^{\mcA}(T) &\le V^H(T) - \mE\left[\sum_{t=1}^T r_{j^t} x^t_{\mcA, j^t}\right]\\
    &= \mE\left[\phi(\bs{b}^1, \bs{Z}^1) - \sum_{t=1}^T r_{j^t}  x^{t}_{\mcA, j^t}\right]\\
    & = \mE\left[\sum_{t=1}^T \left(\phi(\bs{b}^t_{\mcA}, \bs{Z}^t) - \mathbb{E}\left[\phi(\bs{b}^{t+1}_{\mcA}, \bs{Z}^{t+1})+ r_{j^t}  x^t_{\mcA, j^t} \right]\right) \right]\\
    & = \sum_{t=1}^T \mE\left[\underbrace{\phi(\bs{b}^t_{\mcA}, \bs{Z}^t) - \phi(\bs{b}^t_{\mcA} -x^t_{\mcA, j^t} \bs{A}_{j^t}, \bs{Z}^t - \bs{e}_{j^t}) - r_{j^t}  x^t_{\mcA, j^t}}_{\Delta(\bs{b}^t_{\mcA}, \bs{Z}^t, j^t, x^t_{\mcA, j^t})} \right],
    \end{split}
\end{equation}
where $\bs{b}^t_{\mcA}=T\bs{\rho} - \sum_{\ell=1}^{t-1} x^\ell_{\mcA, j^\ell}\bs{A}_{j^\ell}$ is the random inventory vector at time $t$ under the algorithm $\mathcal{A}$. Therefore, it is sufficient to bound $\sum_{t=1}^T\mE\left[\Delta(\bs{b}^t_{\mcA}, \bs{Z}^t, j^t, x_{\mcA, j^t}^t)\right]$.

\begin{remark}[Alternative Benchmarks]
    In addition to the hindsight benchmark $V^H(T)$, there are two other commonly used benchmarks. First, the closest benchmark is the dynamic programming benchmark with the knowledge of the arrival probabilities, which is denoted by $V^{D}(T)$. Although it is tighter than the hindsight benchmark (i.e., $V^D(T)\le V^H(T)$), it is intractable for analysis due to the complexity of Bellman equations. 
    Second, some researchers (e.g., \citealt{jasin2015performance}, \citealt{chen2024improved}) consider the fluid benchmark with the knowledge of the arrival probabilities, i.e., $V^F(T):= \phi(\bs{b}^1, T\bs{p})$, which is convenient to analyze. However, using this benchmark in the degenerate case, we cannot achieve an upper bound better than $\Theta(\sqrt{T})$ because $V^F(T)-V^*(T)= \Omega(\sqrt{T})$ (see \citealt{bumpensanti2020re}).
    The selection of the hindsight benchmark is due to the balance between bound tightness and analytical tractability because $V^F(T) \ge V^H(T) \ge V^D(T)\ge V^*(T) $ and analytical tractability decreases in this order. \hfill \qed
\end{remark}
\vspace{0.05in}

In the following, we introduce properties of the term $\Delta(\bs{b}^t_{\mcA}, \bs{Z}^t, j^t, x_{\mcA, j^t}^t)$.
\begin{proposition}\label{prop:delta_property}
    For the term $\Delta(\bs{b}, \bs{Z}, j, x)$, we have the following properties:
    \begin{enumerate}
        \item[\textnormal{(i)}] For any $(\bs{b}, \bs{Z}, j, x)$ with $\bs{b}\ge x\bs{A}_j$ and $\bs{Z}\ge \bs{e}_j$, we have $\Delta(\bs{b}, \bs{Z}, j, x) \le r_\phi$, where $r_\phi$ is a constant independent of $T$. 
        \item[\textnormal{(ii)}] If there exists an optimal solution $\bs{y}^*$ to $\phi(\bs{b}, \bs{Z})$ such that $y^*_j\ge 1$, then  $\Delta(\bs{b}, \bs{Z}, j, 1)=0$.
        \item[\textnormal{(iii)}] If there exists an optimal solution $\bs{y}^*$ to $\phi(\bs{b}, \bs{Z})$ such that $Z_j-y^*_j\ge 1$, then $\Delta(\bs{b}, \bs{Z}, j, 0)=0$. 
    \end{enumerate}
\end{proposition}

Proposition~\ref{prop:delta_property}(i) states that the term $\Delta(\bs{b}^t_{\mcA}, \bs{Z}^t, j^t, x_{\mcA, j^t}^t)$ is upper bounded by a constant $r_{\phi}$. More importantly, Proposition~\ref{prop:delta_property}(ii) and (iii) provide conditions under which the per-period optimality gap is zero. For example, if there exists an optimal solution $\boldsymbol{y}^*$ to $\phi(\boldsymbol{b},\boldsymbol{Z})$ such that $y_j^*\ge 1$, it is optimal for the clairvoyant to accept $y_j^*$ type-$j$ customers. Since the reward is independent of time, the clairvoyant can accept the request either now or later without loss of optimality. Therefore, the per-period optimality gap of taking the acceptance action equals zero. Leveraging this proposition, it holds that
\begin{align}\label{eq:regret_reduce}
    \text{Reg}^{\mcA}(T) \le \sum_{t=1}^T \mE\left[\Delta(\bs{b}^t_{\mcA}, \bs{Z}^t, j^t, x_{\mcA, j^t}^t)\right]\le \sum_{t=1}^T r_\phi \mathbb{P}\left(\Delta(\bs{b}^t_{\mcA}, \bs{Z}^t, j^t, x_{\mcA, j^t}^t)>0\right),
\end{align}
which reduces the problem to bounding the probability $\mathbb{P}\left(\Delta(\bs{b}^t_{\mcA}, \bs{Z}^t, j^t, x_{\mcA, j^t}^t)>0\right)$. We then show this probability is relatively small by arguing the sufficient conditions stated in Proposition~\ref{prop:delta_property}(ii) and (iii) happen almost all the time. Notably, those sufficient conditions depend on the optimal solution to the hindsight problem in each period which is not available to the decision maker because resolving only happens at certain time points. Also, the conditions depend on the exact demand information which is not available to the decision maker. Therefore, we aim to show that under our resolving time schedule, those factors are well approximated. In the following, since the context is clear, we omit the dependence on policy $\mcA$ for $\bs{b}$ and $x$ in the notation. 

$ $\newline
\noindent\tb{Step 2: Bound demand approximation error.} In this step, we prove that the approximated number of future arrivals $\bs{d}^t$ in Algorithm~\ref{alg:period_resolving} is close to the true demand $\bs{Z}^t$.

\begin{proposition}[Demand Approximation Error]\label{prop:d_diff_1_learning}
    Given a time $t\in [T_{k-1}, T_k)$, we have
    \begin{enumerate}
        \item[\textnormal{(i)}] With probability larger than $1-\frac{2}{(t-1)^2} - \frac{2}{(T-t+1)^2}$, it holds that
        \[|d_j^t - Z_j^t| \le (T-T_{k-1}+1) \sqrt{\frac{\log(t-1)}{T_{k-1}-1}} + \sqrt{(T-T_{k-1}+1)\log(T-t+1)}.\]
        \item[\textnormal{(ii)}] With probability larger than $1-\frac{4}{(T-t+1)^2}$, it holds that
        \[|d_j^t - Z_j^t|\le (T-T_{k-1}+1) \sqrt{\frac{\log(T-t+1)}{T_{k-1}-1}} + \sqrt{(T-T_{k-1}+1)\log(T-t+1)}.\]
    \end{enumerate} 
\end{proposition}

According to Proposition~\ref{prop:d_diff_1_learning}, the demand estimation error is relatively small with high probability. To explain the bounds in Proposition~\ref{prop:d_diff_1_learning}, we note that each bound consists of two terms: The first term is due to the gap between the empirically estimated probability and the underlying true probability; the second term is due to the deviation of the random future demand from the true expected future demand. Note that since the empirical probability is also infrequently updated, the demand estimation $\bs{d}^t$ consists of $\hat{\bs{p}}^{T_{k-1}}(T-T_{k-1}+1)$ and some first-order operations, and hence the concentration bounds in Proposition~\ref{prop:d_diff_1_learning} contain $T_{k-1}$. 

$ $\newline
\noindent\textbf{Step 3: Surrogate LP for $\bs{u}^t$.} In the literature on argmax policies (e.g., \citealt{vera2019bayesian} and \citealt{vera2021online}), the proposed policy makes decisions based on the optimal solution to the fluid problem in each period. This approach allows them to bridge the proposed policy with the hindsight policy using two LPs: the fluid LP and the hindsight LP. However, we solve the fluid problem only in a few selected periods, and approximate the optimal solution by $\bs{u}^t$ in other periods. In this case, $\bs{u}^t$ is not an optimal solution to the fluid problem $\phi(\bs{b}^t, (T-t+1)\hat{\bs{p}}^t)$ for most time periods. Consequently, the linkage between the proposed policy and the hindsight LP breaks down, and the proof technique in the literature cannot directly apply to our proof. To overcome this challenge, we prove that $\bs{u}^t$ is a good approximation of the optimal solution to the hindsight problem by introducing a surrogate LP. In the following, we show that $\bs{u}^t$ is an optimal solution to the surrogate LP with high probability. 
\begin{proposition}[Surrogate LP]\label{prop:surrogate_LP_learning}
    Given the AIR policy with the resolving schedule $\mcT$ with $\alpha\in (0, 1)$ and $\beta\in (\frac12, 1)$, there exist constants $c_1$ and $c_2$ independent of  $\; T$ such that when $t\in [c_1, T-c_2]$, with probability larger than $1-\frac{n}{(T-t+1)^2}-\frac{n}{(\min\{T-t+1, t-1\})^2}$, we have 
    \begin{enumerate}
        \item[\textnormal{(i)}] $\bs{d}^t\ge \frac{T-t+1}{2}\bs{p}\ge \bs{2}$.
        \item[\textnormal{(ii)}] $\bs{u}^t$ is an optimal solution of the LP $\phi(\bs{b}^t, \bs{d}^t)$.
    \end{enumerate}
\end{proposition}

The first part of Proposition~\ref{prop:surrogate_LP_learning} is proved by concentration inequalities. The second part is proved by induction: In the resolving period $T_{k-1}$, $\bs{u}^{T_{k-1}}$ is certainly the optimal solution to $\phi(\bs{b}^{T_{k-1}}, \bs{d}^{T_{k-1}})$. Then, due to the argmax policy and the designed first-order computations in Algorithm~\ref{alg:period_resolving}, we can show that the statement holds for any period $\ell\in (T_{k-1}, t]$ as long as $\bs{d}^t\ge \bs{2}$. Proposition~\ref{prop:surrogate_LP_learning} shows that the approximated solution $\bs{u}^t$ is optimal to a surrogate LP, $\phi(\bs{b}^t, \bs{d}^t)$, with high probability. Recall that we prove that the demand estimation $\bs{d}^t$ is not far away from the random demand $\bs{Z}^t$. Then, we prove that the optimal solution to the surrogate LP $\phi(\bs{b}^t, \bs{d}^t)$ is not far away from the optimal solution to the hindsight LP $\phi(\bs{b}^t, \bs{Z}^t)$. Thus, we can show that if $u_j^t\ge  d_j^t/2$, then there exists an optimal solution $\bs{y}^*$ of $\phi(\bs{b}^t, \bs{Z}^t)$ such that $y^*_j\ge 1$ with high probability, implying that $\Delta(\bs{b}^t, \bs{Z}^t, j, 1)=0$ according to Proposition~\ref{prop:delta_property}. The logic for the case with $u_j^t<  d_j^t/2$ is similar.

$ $\newline
\noindent\tb{Step 4: Bound the probability $\mathbb{P}\left(\Delta(\bs{b}^t_{\mcA}, \bs{Z}^t, j^t, x_{\mcA, j^t}^t)>0\right)$.}
We first define the ``\textit{good event}'' at period $t\in [c_1, T-c_2]$ to be that the conditions in both Propositions~\ref{prop:d_diff_1_learning} and \ref{prop:surrogate_LP_learning} hold for all $j$ in this period, and let the ``\textit{bad event}'' denote the opposite. The probability of the good event at period $t$ is at least $1-\frac{c_3}{(T-t+1)^2}-\frac{c_4}{(t-1)^2}$ where $c_3$ and $c_4$ are positive constants independent of $T$. In the following, we prove that $\Delta(\bs{b}^t, \bs{Z}^t, j^t, x_{j^t}^t)=0$ under the good event for most periods. 

According to Proposition~\ref{prop:delta_property}, in order to show $\Delta(\bs{b}^t, \bs{Z}^t, j^t, x_{j^t}^t)=0$, it suffices to show that there exists an optimal solution $\bs{y}^*$ of $\phi(\bs{b}^t, \bs{Z}^t)$ such that $y^*_{j^t}\ge 1$ if $x_{j^t}^{t}=1$ and $Z_{j^t}^t-y^*_{j^t}\ge 1$ otherwise. As Proposition~\ref{prop:surrogate_LP_learning} shows, we can bridge the approximated solution $\bs{u}^t$ in Algorithm~\ref{alg:period_resolving} with the optimal solution to the hindsight problem by two LPs, $\phi(\bs{b}^t, \bs{d}^t)$ and $\phi(\bs{b}^t, \bs{Z}^t)$. Specifically, let $\mathcal{Y}(\bs{b}, \bs{d})$ denote the set of optimal solutions to $\phi(\bs{b}, \bs{d})$ and 
\[\bar{\mfS}(\bs{b},\bs{d},j)=\max_{\bs{y}\in \mathcal{Y}(\bs{b}, \bs{d})} y_j,\]
which selects the largest value of $y_j$ among all optimal solutions to $\phi(\bs{b}, \bs{d})$.
Note that $\bar{\mfS}(\bs{b}^t,\bs{Z}^t,j^t)\ge 1$ implies that $\Delta(\bs{b}^t,\bs{Z}^t, j^t, 1)=0$ by Proposition~\ref{prop:delta_property}. Similarly, we define $\underline{\mfS}(\bs{b},\bs{d},j)=\min_{\bs{y}\in \mathcal{Y}(\bs{b}, \bs{d})} y_j$, and $Z_{j^t}-\underline{\mfS}(\bs{b}^t,\bs{Z}^t,j^t)\ge 1$ implies that $\Delta(\bs{b}^t,\bs{Z}^t, j^t, 0)=0$. We now show that  $\Delta(\bs{b}^t, \bs{Z}^t, j^t, x_{j^t}^{t})=0$ under the good event for most periods.
\begin{proposition}\label{prop:satisfying_prob}
    Given the AIR policy with the time set $\mcT$ with $\alpha\in (0, 1)$ and $\beta\in (\frac12, 1)$, there exist two constants $c_5$ and $c_6$ independent of $T$ such that when $t\in [c_5, T-c_6]$, under the good event, we have 
    \begin{enumerate}
        \item[\textnormal{(i)}] $\bar{\mfS}(\bs{b}^t, \bs{Z}^t, j^t)\ge 1$ if $x_{j^t}^{t}=1$ and $Z_{j^t}^t - \underline{\mfS}(\bs{b}^t, \bs{Z}^t, j^t)\ge 1$ if $x_{j^t}^{t}=0$.
        \item[\textnormal{(ii)}] $\Delta(\bs{b}^t, \bs{Z}^t, j^t, x_{j^t}^{t})=0$.
    \end{enumerate}
\end{proposition}

The proof idea of Proposition~\ref{prop:satisfying_prob} is as follows: Suppose a type-$j$ customer is accepted at period $t$ and the good event occurs. In this case, we have $\bs{d}^t\approx \bs{Z}^t$, $\bs{d}^t\ge \frac{T-t+1}{2}\bs{p}$ and that $\bs{u}^t$ is an optimal solution of $\phi(\bs{b}^t, \bs{d}^t)$. Since $\bs{d}^t\approx \bs{Z}^t$, we have $\bs{u}^t\approx \tilde{\bs{y}}^*$, where $\tilde{\bs{y}}^*$ denotes the optimal solution to the hindsight problem $\phi(\bs{b}^t, \bs{Z}^t)$. Then, when $T-t+1$ is greater than a constant, we have $\tilde{y}^*_j \approx u_j^t \ge d_j^t/2 \ge \frac{T-t+1}{4}p_j\ge 1$, resulting in the results in Proposition~\ref{prop:satisfying_prob}. 
According to Proposition~\ref{prop:satisfying_prob}, for most periods, under the good event, we have $\bar{\mfS}(\bs{b}^t, \bs{Z}^t, j^t)\ge 1$ if $x_{ j^t}^{t}=1$ and $Z_{j^t}^t - \underline{\mfS}(\bs{b}^t, \bs{Z}^t, j^t)\ge 1$ if $x_{j^t}^{t}=0$, resulting in $\Delta(\bs{b}^t, \bs{Z}^t, j^t, x_{j^t}^{t})=0$. Therefore, the probability $\mathbb{P}\left(\Delta(\bs{b}^t, \bs{Z}^t, j^t, x_{ j^t}^t)>0\right)$ is upper bounded by the bad event probability. 
Finally, with the above four steps, we are prepared to prove Theorem~\ref{thm:regret_bound} by bounding the right-hand side of \eqref{eq:regret_decompose}.
\begin{align*}
    \text{Reg}^{\mcA}(T) &\le \sum_{t=c_5}^{T-c_6} r_\phi \mathbb{P}\left(\Delta(\bs{b}^t, \bs{Z}^t, j^t, x_{ j^t}^{t})>0\right) + (c_5+c_6)r_\phi\\
    &\le \sum_{t=c_5}^{T-c_6} r_\phi \left(\frac{c_3}{(T-t+1)^2} + \frac{c_4}{(t-1)^2}\right) + (c_5+c_6)r_\phi \\
    &\le \left(\frac{\pi^2}{6}(c_3+c_4)+c_5+c_6\right)r_\phi,
\end{align*}
which is independent of $T$. Thus, Theorem~\ref{thm:regret_bound} is proved.

\begin{remark}[Proof Challenges under Infrequent Resolving]
    We would like to emphasize that the infrequent resolving requirement significantly complicates the problem even under the finite-support assumption. First, we discuss related literature. Under per-period resolving, the constant regret for the known-probability case has already been extended to the unknown-probability case (see \citealt{xie2025benefits}). In contrast, under infrequent resolving, the best known result \cite{jasin2015performance} extends the algorithm in \cite{jasin2012re} to handle the unknown-probability case, but the regret bound changes from $\mcOd(1)$ to $\mcOd(\log^2 T)$, implying the hardness of infrequent resolving. 
    
    Second, we discuss the technical details. Let $\hat{\bs{y}}^t$ and $\bar{\bs{y}}^{t}$ denote the optimal solutions to $\phi\left(\bs{b}^t, (T-t+1)\hat{\bs{p}}^t\right)$ and $\phi\left(\bs{b}^t, \bs{Z}^t\right)$, respectively. 
    If the fluid LP $\phi\left(\bs{b}^t, (T-t+1)\hat{\bs{p}}^t\right)$ is resolved per period, we can always utilize the optimal fluid solution $\hat{\bs{y}}^t$ rather than the approximation $\bs{u}^t$ to determine the argmax action. In this case, we can use the Lipschitz property of LP (see Theorem 2.4 in \citealt{mangasarian1987lipschitz}) to bound the difference $\Vert \bar{\bs{y}}^{t} - \hat{\bs{y}}^t\left\Vert \le c_{13} (T-t+1)\Vert \bs{p} - \hat{\bs{p}}^t\right\Vert$, where $c_{13}$ is a constant independent of $T$. Then, we can use concentration inequalities to prove that for some small constant $c_{14}$, $\left\Vert \bs{p} - \hat{\bs{p}}^t\right\Vert \le c_{14}$ with high probability. Subsequently, we can deduce that $\left\Vert \bar{\bs{y}}^{t} - \hat{\bs{y}}^t\right\Vert \le c_{13}c_{14} (T-t+1)$ with high probability and then prove the constant regret bound similar to \cite{vera2019bayesian}. Indeed, \cite{xie2025benefits} have already adopted a similar idea to derive a constant regret for the per-period resolving case. 

    However, in our work, we consider an \textit{infrequent resolving} schedule, and hence the optimal fluid solution $\hat{\bs{y}}^t$ is only accessible at resolving periods $t\in \mcT$, whose size is $\mcO(\log \log T)$ rather than $T$. At each non-resolving period $t\in [T_{k-1}, T_k)$, since $\hat{\bs{y}}^t$ is not accessible, we use an approximate solution $\bs{u}^t$, which equals the latest optimal fluid solution combined with some first-order computations (see Algorithm~\ref{alg:period_resolving}). Therefore, we need to bound the difference $\Vert\bar{\bs{y}}^{t}- \bs{u}^t\Vert$. Unlike $\hat{\bs{y}}^t$, $\bs{u}_t$ is not an optimal solution to the fluid LP, so we cannot directly use the Lipschitz property of LP to bound the difference. Without the help of the surrogate LP, we may need to trace back to the latest optimal fluid solution $\bs{u}^{T_{k-1}} = \hat{\bs{y}}^{T_{k-1}}$, which is an optimal solution to the fluid LP at period $T_{k-1}$. Specifically, the difference $\left\Vert\bar{\bs{y}}^{t}- \bs{u}^t\right\Vert$ can be bounded as follows:
    \begin{align}\label{eq:remark}
    \left\Vert\bar{\bs{y}}^{t}- \bs{u}^t\right\Vert \le\left\Vert\bar{\bs{y}}^{t}- \hat{\bs{y}}^{T_{k-1}}\right\Vert + \left\Vert\bs{u}^{T_{k-1}}- \bs{u}^t\right\Vert = \left\Vert\bar{\bs{y}}^{t}- \hat{\bs{y}}^{T_{k-1}}\right\Vert + \left\Vert \sum_{\ell=T_{k-1}}^{t-1} x^\ell_{\mcA, j^\ell} \cdot \bs{e}_{j^\ell} \right\Vert. 
    \end{align}
    If we consider $T_{k-1}=\lfloor T/2\rfloor$ and $T_{k}=\lfloor T-T^\beta\rfloor$, then the second term in the rightmost formula of \eqref{eq:remark} will be $\Theta(t-T_{k-1})=\Theta\left((T-t+1)^{1/\beta}\right)$ with high probability. Since $1/\beta>1$, the proof idea for the per-period resolving case cannot directly induce a constant regret. To address this challenge, we introduce the high-probability surrogate LP to bridge $\bs{u}_t$ and $\bs{y}^{t, *}$. By doing so, under the well-designed resolving schedule, we can deduce that $\Vert\bar{\bs{y}}^{t}- \bs{u}^t\Vert = \mcO\left((T-t+1)^{1/2\beta}\right) = o(T-t+1)$, which allows us to prove a constant regret bound. \hfill \qed
\end{remark}

\subsection{Finite Resolving}\label{subsec:finite_resolve_olp} 
In the above analysis, we establish the constant regret bound when the number of resolvings is $\mcO(\log\log T)$. Although the number $\mcO(\log\log T)$ is nearly a constant, it increases with the time horizon, which may still prevent its application to huge-size or time-sensitive problems. A natural question that follows is, what if we are only allowed to solve LPs a finite number of times? In this subsection, we unveil a more detailed relation between resolving frequency and algorithm performance by considering the case when the number of resolvings is a finite number $M$.

\begin{figure}[ht!]
    \centering
    \begin{tikzpicture}[%
        every node/.style={
            font=\footnotesize,
        },
        line width = 1pt
    ]
    \draw[line width=0.4mm, -latex] (0,0) -- (15,0)
    node[anchor=north, pos=1.02, align=center] {Time};
    \foreach \x in {0, 7.68312, 11.0872, 0.588723, 7, 14, 12.6285, 13.4113}{
        \draw (\x cm,3pt) -- (\x cm,0pt);
    }

    \node[anchor=south] at (7,0) {$\frac{T}{2}$};
    \node[anchor=south] at (7.68312+0.2,0) {$T-T^\beta$};
    \node[anchor=south] at (11.0872+0.2,0) {$T-T^{\beta^2}$};
    \node[anchor=south] at (0.588723+0.2,0) {$T^{\beta^{M-1}}$};
    
    \draw[decorate,decoration={brace,amplitude=8pt, mirror}] (0,0) -- (7,0)
        node[anchor=south,midway,below=10pt, align=center] {$\mcT^{F}_L(M)$};
    \draw[decorate,decoration={brace,amplitude=8pt, mirror}] (7.68312,0) -- (14,0)
        node[anchor=south,midway,below=10pt, align=center] {$\mcT^{F}_A(M)$};
    \end{tikzpicture}
    \caption{Illustration of finite-resolving time set $\mcT^{F}(M)=\mcT^{F}_L(M)\cup\mcT^{F}_A(M)$.}
    \label{fig:illustrate_resolving_time_finite}
\end{figure}
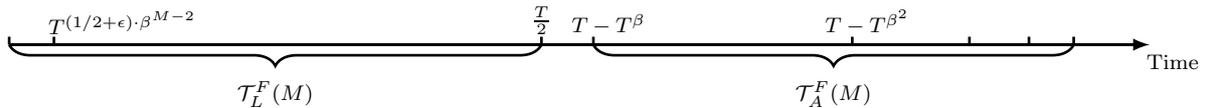

In this case, we need to adjust the resolving times for learning and approximation. Given the number $M\ge 2$ and $\beta\in\left(\frac12, 1\right)$, we define the finite-resolving schedule $\mcT^{F}(M) = \mcT^{F}_L(M)\cup \mcT^{F}_A(M)$ as $$\mcT^{F}_L(M)=\left\{ \left\lceil T^{\beta^{M-1}}\right\rceil, \left\lceil \frac{T}{2} \right\rceil\right\}$$ 
and 
$$\mcT^{F}_A(M)=\left\{ \left\lceil T-T^{\beta}\right\rceil, \left\lceil T-T^{\beta^2}\right\rceil, \dots, \left\lceil T-T^{\beta^{M-2}}\right\rceil \right\}.$$
To facilitate understanding, we illustrate the resolving schedule in Figure~\ref{fig:illustrate_resolving_time_finite}. As Figure~\ref{fig:illustrate_resolving_time_finite} shows, when the number of resolvings is restricted, we should invest more computational power to the approximation set, and the learning set only needs two time points. More specifically, we do not solve the fluid model until we get enough samples to derive a relatively accurate estimation of $\bs{p}$, i.e., at the first time point $\left\lceil T^{\beta^{M-1}} \right\rceil$. Then, at time point $\lceil T/2\rceil$, we solve the fluid model with a more accurate estimation based on the collected samples. The insight behind this choice is that, at early stages, there is plenty of inventory and wrong actions can be made up by the remaining periods. Specifically, if we accept too many type-$j$ customers at early periods, then we can reject more in the remaining periods to make up. However, at late periods, we cannot tolerate so many wrong actions. As for the approximation set, in order to correct the approximation error in time, the smaller time points (further away from $T$) are more crucial. Therefore, we keep smaller time points when the number of resolvings is limited.

We have the following theorem regarding the performance of the algorithm. 

\begin{theorem}\label{thm:olp_finite_resolve}
    For $M\ge 2$ and any $\epsilon\in(0, \frac12)$, given the finite-resolving schedule
    $\mcT^{F}(M)$ with $\beta=\frac12+\epsilon$, the regret of the AIR policy is $\mcO\left(T^{(1/2+\epsilon)^{M-1}}\right)$. 
\end{theorem}

The proof of Theorem~\ref{thm:olp_finite_resolve} is similar to Theorem~\ref{thm:regret_bound} and is given in Appendix~\ref{proof:olp_finite_resolving}. As we prove in Section~\ref{subsec:regret_proof}, for any period $t$, if there exist resolving times $T_{k-1}$ and $T_k$ (again we use $T_k$ to denote the $k$-th resolving time in $\mcT^F(M)$) such that $t\in [T_{k-1}, T_k)$, then we can prove that the revenue loss at period $t$ is $\mcO\left(\frac{1}{(t-1)^2} + \frac{1}{(T-t+1)^2}\right)$. Similarly, we can find that each period $t\in [T_M, \lceil T-T^{\beta^{M-1}}\rceil)$ has the same property, resulting in $\mcO(1)$ regret during $\left[\left\lceil T^{ \beta^{M-1}}\right\rceil, \left\lceil T-T^{\beta^{M-1}}\right\rceil\right)$. Then, due to the revenue loss during the remaining periods, the regret is $\mcO\left( T^{\beta^{M-1}}\right)$.

According to Theorem~\ref{thm:olp_finite_resolve}, the AIR policy can guarantee an $\mcO\left(T^{(1/2+\epsilon)^{M-1}}\right)$ regret by solving LPs $M$ times. For example, we can achieve an $\mcO(T^{(1/2+\epsilon)^2})\approx \mcO(T^{1/4})$ regret bound by solving LPs only $M=3$ times, which beats the existing  regret bound $\mcO(\sqrt{T})$ for LP-free algorithms without the non-degeneracy assumption. Therefore, Theorem~\ref{thm:olp_finite_resolve} indicates that the AIR policy can achieve outstanding performance with very limited resolvings.

\begin{remark}[Role of Finite-Support Assumption]
    As mentioned in the introduction, we assume that the underlying distribution $\mcP$ has finite supports. We would like to explain the role of the assumption in the above proof. First, the finite-support assumption enables us to aggregate customers by types and derive a tractable fluid model \eqref{eq:fluid_LP_intro} with finite variables. If the distribution $\mcP$ has infinite supports, then the fluid model becomes an intractable infinite linear program. In that case, the typical approach is to work on the dual problem instead of the primal (see, \citealt{li2022online}, \citealt{bray2025logarithmic} and \citealt{ma2025optimal}). However, in the degenerate case, the linkage between the optimal primal solution and the optimal dual solution becomes relatively weak, resulting in large regrets for the above algorithms. 

    Second, the finite-support assumption ensures that the arrival probability of each customer type is positive, which plays an important role in Proposition~\ref{prop:satisfying_prob}. As we explained after Proposition~\ref{prop:satisfying_prob}, one of the key parts is to deduce $\frac{T-t+1}{4}p_j \ge 1$ when $T-t+1$ is greater than a constant independent of $T$. However, for the continuous-support case, the probability measure of each type is zero and hence the proof cannot hold. 

    Indeed, as \cite{bray2025logarithmic} showed, under the continuous-support assumption, the regret of any policy is lower bounded by $\Omega(\log T)$, i.e., a constant-regret algorithm is impossible. \hfill \qed
\end{remark}

\section{Known Arrival Probabilities}\label{sec:nrm}
In this section, we consider a variation of the problem in which the arrival probabilities $\bs{p}$ are known at the beginning. Such cases are widely studied in the  network revenue management (NRM) literature. We show that the AIR policy can be easily adapted to this case and achieve a better performance guarantee due to the additional distribution information. 

In the following, we first propose a modified algorithm, which we call the Argmax with Infrequent Resolving and Known Probabilities (AIR-KP) algorithm. 

\vspace{0.15in}
\begin{breakablealgorithm}
    \caption{Argmax with infrequent resolving and known probabilities (AIR-KP) policy}\label{alg:period_resolving_nrm}
    \begin{algorithmic}
    \State Input: Time set $\mathcal{T}^{\mcK}= \{T_1, T_2, T_3, \dots, T_{|\mcT^{\mcK}|}\}$.
    \State Initialize $\bs{b}^1 \gets T\bs{\rho}$, $\bs{u}^{1} \gets \bs{0}$ and $\bs{d}^1 \gets \bs{0}$.
    \For{$t=1, 2, 3, \dots, T$} 
    \If{$t\in \mathcal{T}^{\mcK}$} \Comment{Infrequent resolving}
        \State Solve the fluid problem $\phi(\bs{b}^t, (T-t+1)\bs{p})$ and obtain its optimal solution $\bs{y}^{t, *}$. 
        \State Set $u^t_j\gets y^{t, *}_j$ and $d^t_j\gets p_j (T-t+1)$ for any $j$.
    \EndIf
        \State Observe arrival type $j$.
        \If{ $\bs{A}_j\le \bs{b}^t$, $u_j^t\ge 1$, and $u_j^t\ge d_j^t-u_j^t$} \Comment{Argmax between $u_j^t$ and $d_j^t-u_j^t$}
            \State Accept the request. 
            \State Set $\bs{b}^{t+1} \gets \bs{b}^t-\bs{A}_j$. \Comment{Update the remaining capacity}
            \State Set $\bs{u}^{t+1} \gets \bs{u}^t-\bs{e}_j$. \Comment{Approximate the optimal solution}
        \Else
        \State Reject the request and set $\bs{b}^{t+1} \gets \bs{b}^t$ and $\bs{u}^{t+1}\gets \bs{u}^t$.
        \EndIf
        \State Set $\bs{d}^{t+1} \gets \bs{d}^{t}-\bs{e}_j$.\Comment{Approximate the future demand}
    \EndFor
    \end{algorithmic}
    \end{breakablealgorithm}
\vspace{0.15in}

Algorithm~\ref{alg:period_resolving_nrm} is the same as Algorithm~\ref{alg:period_resolving} except that we replace the empirical estimation $\hat{\bs{p}}^t$ with the known arrival probabilities $\bs{p}$. Next, we present the resolving schedule. \vspace{0.1in}

\begin{figure}[ht!]
    \centering
    \begin{tikzpicture}[%
        every node/.style={
            font=\footnotesize,
        },
        line width = 1pt
    ]
    \draw[line width=0.4mm, -latex] (0,0) -- (15,0)
    node[anchor=north, pos=1.02, align=center] {Time};
    \foreach \x in {0, 7.68312, 11.0872,  12.6285,  13.4113,  13.8469,  14}{
        \draw (\x cm,3pt) -- (\x cm,0pt);
    }
    
    \node[anchor=south] at (0,0) {$1$};
    \node[anchor=south] at (7.68312+0.2,0) {$T-T^\beta$};
    \node[anchor=south] at (11.0872+0.2,0) {$T-T^{\beta^2}$};
    \node[anchor=south] at (12.6285+0.2,0) {$T-T^{\beta^3}$};

    \draw[decorate,decoration={brace,amplitude=8pt, mirror}] (0,0) -- (14,0)
        node[anchor=south,midway,below=10pt, align=center] {$\mcT^{\mcK}$};
    \end{tikzpicture}
    \caption{Illustration of resolving schedule $\mcT^{\mcK}$ for known-probability case.}
    \label{fig:illustrate_resolving_time_nrm}
\end{figure}
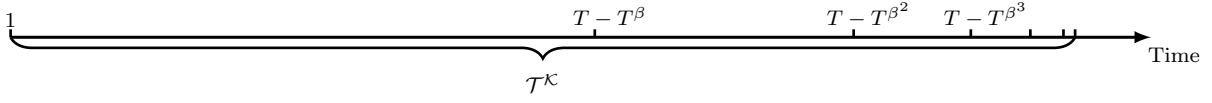

\tb{Resolving Schedule.} Given the arrival probabilities at the beginning, we can drop the learning time set $\mcT_L$ in $\mcT$ and add an initial solving in period 1. In this case, the resolving schedule for the known-probability case becomes
$$\mcT^{\mcK}=\{1\}\cup \left\{\left\lceil T-T^{\beta}\right\rceil, \left\lceil T-T^{\beta^2}\right\rceil, \dots, \left\lceil T-T^{\beta^{K_A}}\right\rceil\right\},$$ where $K_A=\lceil \log_{\frac{1}{\beta}}\log_3 T\rceil$ and $\beta\in (\frac12, 1)$. For ease of understanding, we illustrate the schedule in Figure~\ref{fig:illustrate_resolving_time_nrm}. 

We show the regret bound of the AIR-KP policy in the following theorem. 
\begin{theorem}[Regret Bound for NRM]\label{thm:regret_bound_nrm}
    Given the arrival probabilities and the resolving schedule $\mcT^{\mcK}$ with $\beta\in (\frac12, 1)$, the regret of the AIR-KP policy is $\mcO(1)$. 
\end{theorem}

The idea of the proof of Theorem~\ref{thm:regret_bound_nrm} is similar to that of Theorem~\ref{thm:regret_bound}, which is described in Section~\ref{subsec:regret_proof}. We present the detailed proof in Appendix~\ref{proof:nrm_regret}.  According to Theorem~\ref{thm:regret_bound_nrm}, for the known-probability case, the AIR-KP policy can guarantee a constant regret bound for the NRM problem under the resolving schedule $\mcT^{\mcK}$. Note that the set $\mcT^{\mcK}$ is smaller than $\mcT$ because the learning time set is dropped, but the order is still $O(\log\log T)$. In other words, the distribution information can reduce the number of resolvings, but the order regarding the time horizon $T$ stays the same. 

\subsection{Finite Resolving}\label{subsec:finite_resolve_nrm} 
Similar to the base model, in the following, we consider the known-probability case with a finite number of resolvings. Similar to the case with unknown probabilities, we keep the smaller time points in the approximation set, that is $$\mcT^{\mcK, F}(M) = \{1\} \cup \left\{ \left\lceil T-T^{\beta} \right\rceil, \left\lceil T-T^{\beta^2} \right\rceil, \dots, \left\lceil T-T^{\beta^{M-1}} \right\rceil \right\}.$$ 
In the following theorem, we provide a better regret bound than that in Section~\ref{subsec:finite_resolve_olp}. 
\begin{theorem}\label{thm:nrm_finite_resolve}
    For $M\ge 1$ and any $\epsilon \in (0, \frac12)$, given the arrival probabilities and the resolving time set $\mcT^{\mcK, F}(M)$ with $\beta=\frac12 +\epsilon$, the regret of the AIR-KP policy is $\mcO(T^{(1/2+\epsilon)^{M}})$. 
\end{theorem}

According to Theorem~\ref{thm:nrm_finite_resolve}, we can guarantee an $\mcO(T^{(1/2+\epsilon)^M})\approx \mcO(T^{(1/2)^M})$ regret by solving LPs $M$ times. Note that the result in Theorem~\ref{thm:nrm_finite_resolve} is stronger than that in Theorem~\ref{thm:olp_finite_resolve} because there is no need to learn $\bs{p}$ in the known-probability case. Specifically, in the unknown-probability case, in order to mitigate the accumulation of learning error, we need to resolve the LP at the midpoint of the time horizon, i.e., $\lceil T/2\rceil$ in $\mathcal{T}^F(M)$ (see Figure~\ref{fig:illustrate_resolving_time_finite}). In contrast, in the known-probability case, there is no learning error and hence such a resolving point is not necessary. As a result, one can have one more resolving point in the latter part of the time horizon ($\lceil T-T^{\beta^{M-1}}\rceil$), which leads to a smaller regret with the same number of resolvings.

The result is close to Theorem~6.1 in \cite{jasin2012re}, which shows that solving LPs $M$ times can induce an $\mcOd(T^{(1/2)^M})$ regret bound under the non-degeneracy assumption. However, different from \cite{jasin2012re}, our resolving schedule also works for the degenerate case. 
Moreover, in the following proposition, we provide a lower bound for the regret of the AIR-KP policy with any two resolving time points. 

\begin{proposition}\label{prop:lower_bound}
    Given any resolving schedule with no more than two resolvings, the regret of the AIR-KP policy is $\Omega(T^{1/4})$. 
\end{proposition}

According to Theorem~\ref{thm:nrm_finite_resolve} and Proposition~\ref{prop:lower_bound}, the regret bound $\mcO(T^{(1/2+\epsilon)^M})$ (with $\epsilon\to 0$) is nearly tight when $M=2$. Moreover, in Appendix~\ref{append:epsilon}, we discuss potential reasons why there is an additional $\epsilon$ compared to the regret bound $\mcOd(T^{(1/2)^M})$ in \cite{jasin2012re}.

\subsection{Discussion of Resolving Schedules} \label{subsec:NRM_schedule}
In this subsection, in order to show the power of our proof framework, we revisit several resolving schedules proposed in the literature for the known-probability case, i.e., \cite{jasin2012re} and \cite{bumpensanti2020re}, and provide corresponding modified schedules for the unknown-probability case. With similar proofs, we can show that the AIR-KP policy is no worse than the proposed policies in \cite{jasin2012re} and \cite{bumpensanti2020re}. 

\begin{enumerate}
    \item \tb{Periodic Resolving in \cite{jasin2012re}.} In \cite{jasin2012re}, the authors propose a resolving algorithm for the known-probability case with resolving schedule being
    $$\mathcal{T}^{\mcK, P}(\omega)=\left\{1, 1+\omega, 1+2\omega, \dots, 1+K_P \omega \right\},$$
    with $K_P=\left\lfloor \frac{T-1}{\omega} \right\rfloor$.
    In this case, for $t\in [T_{k-1}, T_k)$, we have $T_{k-1}\ge t-\omega$. Given the resolving schedule $\mathcal{T}^{\mcK, P}(\omega)$, the AIR-KP policy guarantees a constant regret bound for the known-probability case.

    \begin{lemma}\label{lem:regret_periodic_nrm}
        Given the arrival probabilities and the resolving time set $\mcT^{\mcK, P}(\omega)$, the regret of the AIR-KP policy is $\mcO(\sqrt{\omega}\log \omega)=\tilde{\mcO}(\sqrt{\omega})$, which is independent of $T$.  
    \end{lemma}
    
    According to Lemma~\ref{lem:regret_periodic_nrm}, as the resolving interval $\omega$ increases, the regret bound grows at an $\tilde{\mcO}(\sqrt{\omega})$ rate, which is close to the $\mcOd(\sqrt{\omega})$ regret bound in \cite{jasin2012re}. Moreover, in the following proposition, the periodic resolving schedule can be directly applied to the unknown-probability case. 
    
    \begin{proposition}\label{prop:regret_periodic}
        Given the resolving time set $\mcT^{\mcK, P}(\omega)$, the regret of the AIR policy is $\mcO(\omega)$, which is independent of $T$. 
    \end{proposition}

    The result in Proposition~\ref{prop:regret_periodic} with $\omega=1$ directly improves the regret in the degenerate case from $\tilde{\mcO}(\sqrt{T})$ in \cite{chen2024improved} to $\mcO(1)$. 

    \item \tb{Midpoint Resolving in \cite{jasin2012re}.} In \cite{jasin2012re}, the authors also consider the following resolving schedule:
    $$\mathcal{T}^{\mcK, M}=\{1\}\cup \left\{\left\lceil T-T/2\right\rceil, \left\lceil T-T/2^2\right\rceil, \dots, \left\lceil T-T/2^{K_M}\right\rceil\right\}$$ 
    with $K_M=\lceil \log_2 T \rceil $. In this case, for $t\in [T_{k-1}, T_k)$, we have $T-T_{k-1} \le 2(T-t+1)$. Given the resolving schedule $\mathcal{T}^{\mcK, M}$, our AIR policy guarantees the following regret bound. 
    
    \begin{lemma}\label{lem:regret_midpoint_nrm}
        Given the arrival probabilities and the resolving time set $\mcT^{
        \mcK, M}$, the regret of the AIR-KP policy is $\mcO(1)$.  
    \end{lemma}
    
    According to Lemma~\ref{lem:regret_midpoint_nrm}, given the resolving schedule $\mcT^{\mcK, M}$, the AIR-KP policy can guarantee a constant bound for the known-probability case. However, for the unknown-probability case, since the empirical estimation at the beginning is inaccurate and the second resolving time is $\lceil T/2\rceil$, the regret of the AIR policy will be $\mcO(T)$. In order to extend such a schedule to fit the unknown-probability case, we supplement the midpoint resolving schedule with a learning set similar to $\mcT_L$ in the base model. Specifically, we construct a midpoint resolving time set $$\mcT^{M} = \left\{\left\lceil T/2^2 \right\rceil, \left\lceil T/2^3 \right\rceil, \dots, \left\lceil T/2^{K_M}\right\rceil \right\} \bigcup \mcT^{\mcK, M}.$$ 
    In this case, we have $T-T_{k-1} \le 2(T-t+1)$ and $T_{k-1}\ge \frac{t-1}{2}$. Under this resolving schedule, our AIR policy can guarantee a constant regret bound. 

    \begin{proposition}\label{prop:regret_midpoint_olp}
        Given the resolving time set $\mcT^{M}$, the regret of the AIR policy is $\mcO(1)$.  
    \end{proposition}

    Note that \cite{jasin2015performance} proposes a midpoint-resolving algorithm which can achieve an $\mcOd(\log^2 T)$ regret under the non-degeneracy assumption. Our result in Proposition~\ref{prop:regret_midpoint_olp} provides a way to improve the regret bound even without the non-degeneracy assumption. 

    \item \tb{Infrequent Resolving in \cite{bumpensanti2020re}.} In \cite{bumpensanti2020re}, the authors propose a resolving algorithm for the known-probability case with the resolving schedule being the same as $\mcT^{\mcK}$ with $\beta= 5/6$. According to Theorem~\ref{thm:regret_bound_nrm}, our AIR-KP policy can always guarantee a constant regret bound with a more flexible schedule selection, i.e., $\beta\in (\frac12, 1)$. Moreover, in order to modify such schedules to fit the unknown-probability case, we need to add a learning time set $\mcT_L$ specified in Section~\ref{subsec:resolve_schedule}. 
\end{enumerate}

\subsection{Discussion of Arrival Processes} \label{subsec:arrival}
Before we close this section, we discuss some extensions regarding the arrival process. First, we consider the case in which the arrival process is a non-stationary process with arrival probabilities $\{\bs{p}^t\}_{t\in [T]}$. Similar to \cite{zhu2023assign}, given the scaling factor $\gamma$, we set the arrival probabilities in the asymptotic regime as $p_j^t(\gamma) = p_j^{\lceil t/\gamma \rceil}$ and $T(\gamma)=\gamma T$. If $\max\{t: p_j^t>0\}$ is different for different $j$'s, then \cite{zhu2023assign} show that the regret is lower bounded by $\Omega(\sqrt{\gamma})$. Otherwise, using similar techniques as in our analysis, we still have a constant bound for the regret. 

Second, we consider the case in which the arrival probabilities $\{\bs{p}^t\}_{t\in [T]}$ evolve according to an irreducible and aperiodic Markov process with $L$ states $\{\bs{\sigma}^i\}_{i=1}^L$ and stationary distribution $\bs{\pi}^*$. In this case, we can use $\sum_{i=1}^L \pi_{i}^* \sigma^i_j$ as the arrival probability of type-$j$ customers in the fluid model. Then, due to the exponential convergence of Markov chains (see, e.g., Theorem 15.0.1 in \citealt{meyn2012markov}), similar to Theorem~\ref{thm:regret_bound_nrm}, we can still have a constant regret bound as shown in the following proposition (the proof of Proposition~\ref{prop:markov} can be found in Appendix~\ref{proof:markov}).

\begin{proposition}\label{prop:markov}
    Suppose the arrival probability vector $\bs{p}^t$ evolves according to an irreducible, aperiodic and finite-state Markov chain. Given the resolving schedule $\mcT^K$ with $\beta\in \left(\frac12, 1\right)$, the regret of the AIR-KP policy is $\mcO(1)$. 
\end{proposition}

\section{Numerical Experiments}\label{sec:numerical}
In this section, we conduct numerical experiments to compare the performance of different policies. Since the exact problem \eqref{eq:exact_formula_finite_type} is intractable, we replace $V^*(T)$ in \eqref{eq:regret_define} by the hindsight benchmark $V^H(T)$ to evaluate policy performance in the following numerical experiments, consistent with the theoretical analysis of regret. We perform all numerical experiments on a machine with a 2.0 GHz Quad-Core Intel Core i5 CPU, using Gurobi 12.0 and MATLAB 2024b. 

\subsection{OLP Policy Comparison}\label{subsec:olp_compare}

In the following, we numerically compare the AIR policy with several OLP algorithms as follows (see Appendix~\ref{subsec:olp_alg} for the detailed descriptions of these algorithms):
\begin{enumerate}
    \item \tb{AIR Policy.} We set the parameters in the resolving schedule as $\alpha=\beta=0.7$.
    \item \tb{Argmax with Frequent Resolving (AFR) Algorithm.} We study the case when the LP is resolved per period, which is similar to Algorithm~3 in \cite{xie2025benefits}. 
    \item \tb{Adaptive Allocation (ADA) Policy.} We implement Algorithm 1 in \cite{chen2024improved}. 
    \item \tb{Simple and Fast (SFA) Policy.} We implement Algorithm~5 in \cite{li2020simple} with stepsize $\gamma_t=1/\sqrt{t}$. 
    \item \tb{Decoupling Learning and Decision (DLD) Policy. } We implement Algorithm~2 in \cite{gao2024decoupling} (the parameters are $T_e=\lfloor T^{2/3}\rfloor$, $\alpha_e=T^{-1/3}$ and $\alpha_p=T^{-2/3}$) with $\mcA_L$ being the subgradient algorithm with stepsize $\gamma_{L, t}=1/t$ and $\mcA_D$ being the subgradient algorithm with stepsize $\alpha_e$ before time $T_e$ and $\alpha_p$ after that. In order to satisfy the budget constraints, the modified algorithm will always reject a request if the request cannot be fulfilled, i.e., $\bs{A}_{j^t}\nleq \bs{b}^t$.
    \item \tb{Budget-Updating Fast (BUF) Policy. } We implement Algorithm~5 in \cite{ma2025optimal}. 
\end{enumerate}

In the following, we compare the performance of the above algorithms. For each parameter set, we run 2,000 simulations. To gain deeper insights into degeneracy, we first focus on the single-resource problem similar to the one studied in \cite{bumpensanti2020re}. Specifically, we consider two types of customers and one type of resource. The arrival probability of either customer type is $0.5$. The rewards of the two types of customers are $r_1=2$ and $r_2=1$, and each customer consumes one unit of resource. The budget per period is $\rho$.

\begin{figure}[ht]
    \begin{tikzpicture}
        \begin{axis}[
            width=15cm, height=7cm,     
            xmin = 0.2,     
            xmax = 0.8,    
            ymin = 0,     
            axis background/.style = {fill=white},
            ylabel = {$\text{Reg}^{\mcA}(T)$},
            xlabel = {$\rho$},
            legend pos= north west,
            tick align = outside,
            ]
    
            
            \addplot[red, mark=o,thick] table[x=T, y=argmax_regret, col sep=comma]{data_single_xi_50k.csv};
            \addlegendentry{AIR}
            \addplot[blue, mark=star,thick] table[x=T, y=xie_regret, col sep=comma]{data_single_xi_50k.csv};
            \addlegendentry{AFR}
            \addplot[green, mark=square,thick] table[x=T, y=chen_regret, col sep=comma]{data_single_xi_50k.csv};
            \addlegendentry{ADA}
            \addplot[magenta, mark=asterisk,thick] table[x=T, y=li_regret, col sep=comma]{data_single_xi_50k.csv};
            \addlegendentry{SFA}
            \addplot[cyan, mark=triangle,thick] table[x=T, y=gao_regret, col sep=comma]{data_single_xi_50k.csv};
            \addlegendentry{DLD}
            \addplot[orange, mark=diamond,thick] table[x=T, y=ma_regret, col sep=comma]{data_single_xi_50k.csv};
            \addlegendentry{BUF}
        \end{axis} 
    \end{tikzpicture}
    \caption{Regret under different policies as functions of $\rho$ when $m=1$, $n=2$, $r_1=2$, $r_2=1$, $p_1=p_2=0.5$, $T=50,000$ and $\alpha=\beta=0.7$. }
    \label{fig:regret_comp_single_resource_b}
    \end{figure}
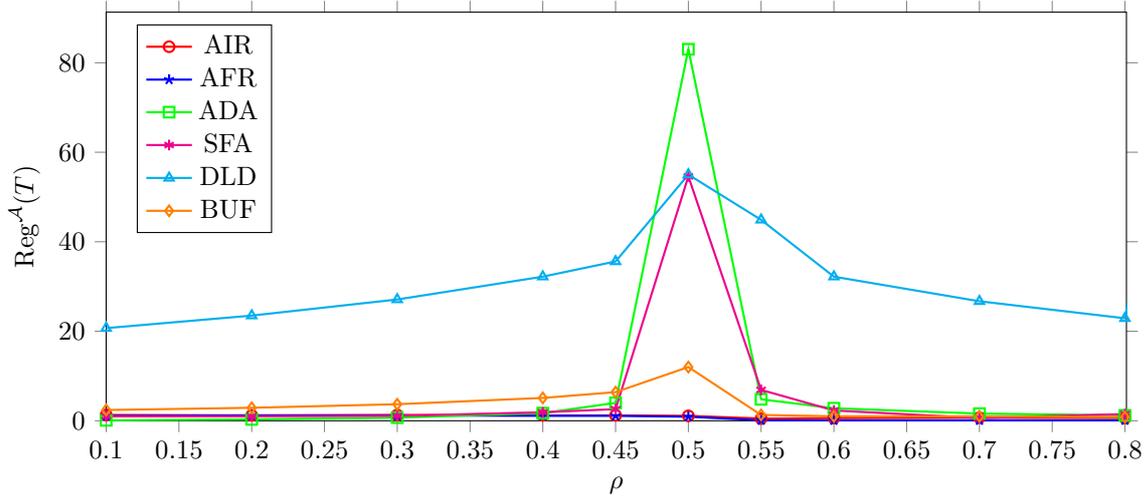

First, in Figure~\ref{fig:regret_comp_single_resource_b}, we fix the time horizon $T$ at $50,000$, and test the algorithms as the budget factor $\rho$ changes. Due to the setup of the single-resource problem, the fluid problem at time $1$ is degenerate when $\rho$ takes the value $0.5$ and hence the non-degeneracy measure $\delta$ equals $|\rho-0.5|$. According to Figure~\ref{fig:regret_comp_single_resource_b}, the regret of most policies is large when $\rho$ is close to 0.5 (i.e., $\delta\to 0$). Therefore, the degenerate case is the hardest case for most policies. However, both the AIR policy and the AFR policy guarantee a constant regret in this case. 

We then focus on the degenerate case and numerically test the above algorithms as the time horizon $T$ increases. Specifically, given $m=10$ and $n=2$, we randomly generate the parameters $\bs{A}$, $\bs{p}$ and $\bs{r}$, and construct $\bs{\rho}$ such that the optimal solution to the resulting fluid problem is degenerate (see Appendix~\ref{subsec:numerical_data} for detailed parameters). We run 200 simulations for each set of parameters and summarize the regret and the runtime of the above policies in Table~\ref{tab:regret_comp_10x5}. In addition, we present the detailed resolving schedule in Table~\ref{tab:resolving_schedule}.

\begin{table}[ht]
    \centering
    \caption{Regret and runtime of different policies when $m=10$, $n=2$ and $\alpha=\beta=0.7$. Superscript ``$*$'' marks LP-free policies. Some cases are marked as ``$-$'' due to the extremely long runtime.}
    \label{tab:regret_comp_10x5}
    \resizebox{\textwidth}{!}{
    \begin{tabular}{@{} *{16}{c} @{}}
    \toprule
       \multicolumn{1}{c}{\multirow{2}{*}{Time Horizon}} & \multicolumn{6}{c@{}}{Regret} & \multicolumn{6}{c@{}}{Runtime (seconds)} & \multicolumn{3}{c@{}}{\# of Resolvings}\\
       \cmidrule(l){2-7}  \cmidrule(l){8-13} \cmidrule(l){14-16}
         & AIR & AFR & ADA & SFA\textsuperscript{*} & DLD\textsuperscript{*} & BUF\textsuperscript{*} & AIR & AFR & ADA & SFA\textsuperscript{*} & DLD\textsuperscript{*}  & BUF\textsuperscript{*} & AIR & AFR & ADA\\
         \midrule
        2,500 & 2.5 & 1.5 & 7.7 & 45.6 & 62.3  & 48.3  & 0.062 & 13.3 & 13.3 & 0.002 & 0.003 & 0.002 & 13 & 2500 & 2500\\
        5,000 & 2.2 & 1.2 & 10.5 & 57.6 & 82.6  & 59.0  & 0.064 & 26.4 & 26.6 & 0.004 & 0.005 & 0.004 & 13 & 5000 & 5000\\
        7,500 & 2.2 & 1.6 & 12.0 & 66.6 & 96.4  & 65.7  & 0.065 & 39.4 & 39.3 & 0.005 & 0.007 & 0.005 & 13 & 7500 & 7500\\
        10,000 & 2.2 & 1.4 & 13.2 & 74.4 & 109.7 & 72.5  & 0.066 & 54.4 & 54.2 & 0.007 & 0.009 & 0.007 & 13 & 10,000 & 10,000\\
        12,500 & 2.1 & 1.2 & 14.3 & 80.9 & 118.8 & 76.0 & 0.077 & 66.6 & 66.5 & 0.009 & 0.012 & 0.009 & 15 & 12,500 & 12,500\\
        15,000 & 2.2 & 1.3 & 15.3 & 86.8 & 128.1 & 79.7 & 0.086 & 73.3 & 73.2 & 0.011 & 0.015 & 0.011 & 15 & 15,000 & 15,000\\
        17,500 & 2.2 & 1.1 & 16.6 & 92.1 & 136.0 & 82.9 & 0.083 & 84.4 & 84.3 & 0.012 & 0.016 & 0.012 & 15 & 17,500 & 17,500\\
        20,000 & 2.1 & 1.0 & 17.4 & 97.0 & 141.6 & 85.9 & 0.084 & 103.2 & 103.1 & 0.014 & 0.019 & 0.013 & 15 & 20,000 & 20,000\\
        100,000 & 2.2 & - & - & 192.0 & 260.1 & 126.6 & 0.141 & - & - & 0.080 & 0.105 & 0.078 & 15 & - & -\\
        200,000 & 2.1 & - & - & 260.2 & 330.9 & 151.6 & 0.204 & - & - & 0.169 & 0.219 & 0.163 & 15 & - & -\\
        300,000 & 2.1 & - & - & 313.8 & 379.2 & 166.1 & 0.231 & - & - & 0.220 & 0.291 & 0.213 & 15 & - & -\\
        \bottomrule
    \end{tabular}}
\end{table}

From Table~\ref{tab:regret_comp_10x5}, we observe several interesting phenomena regarding the regret and runtime. First, the regret of each LP-free policy (SFA, DLD or BUF) increases with the time horizon. In contrast, the regret of both the AFR policy and the AIR policy remains constant, highlighting the effectiveness of LP resolving. Second, the regret of the ADA policy, which also solves LPs per period, also increases with the time horizon. This implies that the choice of the base policy or the interpretation of the fluid model's solution is important. Third, the runtime of the AIR policy is almost negligible, while the performance of the AIR policy is close to that of the AFR policy, which solves LPs per period, and is much better than that of the LP-free policies. For example, when $T=20{,}000$, the AIR policy, which solves LPs only 15 times (see Table~\ref{tab:resolving_schedule} for the detailed resolving schedule), achieves a regret of the same order as the AFR policy with only about 0.1\% of the runtime. This suggests that an appropriate resolving schedule is important. 

\begin{table}[ht]
    \aboverulesep = 0mm \belowrulesep = 0mm 
    \centering
    \caption{Resolving schedule $\mcT$ with $\alpha=\beta=0.7$.}
    \label{tab:resolving_schedule}
    \resizebox{\textwidth}{!}{
    \begin{tabular}{c|rrrrrrrr|rrrrrrr}
    \toprule
        Time Horizon & \multicolumn{8}{c|}{$\mcT_L$} & \multicolumn{7}{c}{$\mcT_A$} \\
         \midrule
        2500 & & 3 & 4 & 7 & 15 & 47 & 240 & 1250 & 2261 & 2454 & 2486 & 2494 & 2497 & 2498 &  \\
        5000 & & 3 & 5 & 8 & 19 & 65 & 389 & 2500 & 4621 & 4936 & 4982 & 4993 & 4996 & 4998 & \\
        7500 & & 3 & 5 & 9 & 22 & 80 & 516 & 3750 & 6985 & 7421 & 7479 & 7492 & 7496 & 7498 & \\
        10,000 & & 3 & 5 & 10 & 24 & 92 & 631 & 5000 & 9370 & 9909 & 9977 & 9991 & 9996 & 9998 & \\
        12,500 & 3 & 4 & 5 & 10 & 26 & 102 & 738 & 6250 & 11763 & 12399 & 12475 & 12491 & 12496 & 12497 & 12498 \\
        15,000 & 3 & 4 & 6 & 11 & 28 & 112 & 839 & 7500 & 14162 & 14889 & 14973 & 14990 & 14995 & 14997 & 14998\\
        17,500 & 3 & 4 & 6 & 11 & 29 & 120 & 934 & 8750 & 16567 & 17381 & 17472 & 17490 & 17495 & 17497 & 17498 \\
        20,000 & 3 & 4 & 6 & 11 & 30 & 129 & 1025 & 10000 & 18976 & 19872 & 19971 & 19990 & 19995 & 19997 & 19998 \\
        100,000 & 3 & 4 & 7 & 16 & 52 & 282 & 3163 & 50000 & 96838 & 99719 & 99949 & 99985 & 99994 & 99997 & 99998 \\
        200,000 & 3 & 5 & 8 & 19 & 66 & 396 & 5138 & 100000 & 194863 & 199605 & 199935 & 199982 & 199993 & 199996 & 199998\\
        300,000 & 3 & 5 & 9 & 21 & 76 & 483 & 6824 & 150000 & 293177 & 299518 & 299925 & 299980 & 299992 & 299996 & 299998\\
        \bottomrule
    \end{tabular}}
\end{table}

\tb{Impact of $\alpha$ and $\beta$.} In Table~\ref{tab:alpha_beta}, we illustrate the impact of $\alpha$ and $\beta$ on the regret and the number of resolvings. According to Table~\ref{tab:alpha_beta}, as $\alpha$ or $\beta$ increases, the regret decreases with diminishing margins while the number of resolvings grows with increasing margins. Therefore, to balance the regret and the computational efficiency, we recommend choosing moderately large values of $\alpha$ and $\beta$ (between 0.7 and 0.9) in practice.

\begin{table}[ht]
    \centering
    \aboverulesep=0ex
    \belowrulesep=0ex 
    \caption{Regret and number of resolvings of AIR policy under different $\alpha$ and $\beta$ when $m=10$, $n=2$ and $T=20{,}000$. }
    \label{tab:alpha_beta}
    \subfloat[Regret]{\begin{tabular}{c|cccccccccc}
        \toprule
        \diagbox{$\alpha$}{$\beta$} & 0.60 & 0.70 & 0.80 & 0.90 & 0.95 \\
        \midrule 
        0.10 & 883.6 & 822.7 & 822.8 & 884.1 & 643.9 \\
        0.30 & 112.8 & 95.8 & 112.7 & 128.7 & 86.5 \\
        0.50 & 6.1 & 4.8 & 4.0 & 3.2 & 3.1\\
        0.70 & 4.8 & 2.2 & 1.6 & 1.6 & 1.5\\
        0.90 & 4.3 & 1.7 & 1.7 & 1.4 & 1.4\\
        0.95 & 3.0 & 1.7 & 1.5 & 1.4 & 1.4\\
        \bottomrule
    \end{tabular}}
    \qquad
    \subfloat[Number of resolvings]{\begin{tabular}{c|cccccccccc}
        \toprule
        \diagbox{$\alpha$}{$\beta$} & 0.60 & 0.70 & 0.80 & 0.90 & 0.95 \\
        \midrule 
        0.10 & 7 & 9 & 12 & 22 & 38 \\
        0.30 & 8 & 10 & 13 & 23 & 39 \\
        0.50 & 10 & 12 & 15 & 25 & 41\\
        0.70 & 13 & 15 & 18 & 28 & 44\\
        0.90 & 26 & 28 & 31 & 41 & 57\\
        0.95 & 42 & 44 & 47 & 57 & 73\\
        \bottomrule
    \end{tabular}}
\end{table}

\tb{Impact of $n$.} In order to identify the impact of $n$, we consider the case with $m=10$ types of resources and $T=20{,}000$ periods. Let $\mathcal{U}[a, b]$ denote the uniform distribution over the interval $[a, b]$. Given the value of $n$, we consider 20 instances with the parameters generated as follows: First, we generate $A_{ij}\sim \mathcal{U}[0, 1]$, $r_{j}\sim \mathcal{U}[0, 1]$ and $\rho_i\sim\mathcal{U}[0.3, 0.5]$; second, to generate the underlying arrival probabilities, we sample $n$ values $z_j$'s from $\mathcal{U}[0, 1]$, and then choose $\bs{p}=(p_1, p_2, \dots, p_n)$ with $p_j=z_j/\left(\sum_{j'=1}^n z_{j'}\right)$. In this case, we set $\alpha=\beta=0.9$ and the number of resolvings is 41. The reason is that as the number of customer types increases, the optimal solution of the fluid problem $\phi(\bs{b}^t, (T-t+1)\hat{\bs{p}})$ changes more significantly and frequently and hence the resolving frequency should be slightly increased. We define the relative loss as the ratio $\left(V^H(T) -V^\mcA(T)\right)/V^H(T)$. In Table~\ref{tab:different_n}, we summarize the performance of the AIR policy under different $n$. According to Table~\ref{tab:different_n}, as the number of customer types increases, the regret of the AIR policy increases with diminishing margins. 

\begin{table}[ht]
    \centering
    \aboverulesep=0ex
    \belowrulesep=0ex 
    \caption{Performance of AIR Policy under different $n$ when $m=10$, $\alpha=\beta=0.9$, and $T=20{,}000$.}
    \label{tab:different_n}
    \resizebox{\textwidth}{!}{
    \begin{tabular}{c|cccccccccc}
    \toprule
         \diagbox{Statistics}{$n$} & 10 & 20 & 30 & 40 & 50 & 60 & 70 & 80 & 90 & 100 \\
         \midrule
         Avg. Regret & 2.1 & 4.1 & 9.5 & 11.0 & 13.3 & 15.8 & 18.5 & 21.8 & 21.3 & 22.8\\
         Avg. Relative Loss (\%) & 0.03 & 0.05 & 0.13 & 0.13 & 0.15 & 0.17 & 0.22 & 0.24 & 0.25 & 0.27 \\
         Avg. Runtime (seconds) & 0.44 & 0.40 & 0.40 & 0.40 & 0.39 & 0.41 & 0.44 & 0.42 & 0.39 & 0.39\\
         \bottomrule
    \end{tabular}}
    
\end{table}

\textbf{Impact of $m$. } Lastly, we investigate the impact of $m$. Similar to Table~\ref{tab:different_n}, we consider the case with $n=50$ customer types and $T=20{,}000$ periods, and randomly generate 20 instances for each given $m$. In Table~\ref{tab:different_m}, we summarize the performance of the AIR policy under different $m$. According to Table~\ref{tab:different_m}, the number of resource types does not significantly affect the performance of the AIR policy.

\begin{table}[ht]
    \centering
    \aboverulesep=0ex
    \belowrulesep=0ex 
    \caption{Performance of AIR Policy under different $m$ when $n=50$, $\alpha=\beta=0.9$, and $T=20{,}000$.}
    \label{tab:different_m}
    \resizebox{\textwidth}{!}{
    \begin{tabular}{c|cccccccccc}
    \toprule
         \diagbox{Statistics}{$m$} & 10 & 20 & 30 & 40 & 50 & 60 & 70 & 80 & 90 & 100 \\
         \midrule
         Avg. Regret & 13.3 & 14.4 & 15.1 & 16.2 & 16.1 & 15.6 & 15.9 & 15.7& 15.1 & 15.9\\
         Avg. Relative Loss (\%) & 0.1 & 0.2 & 0.2 & 0.2 & 0.2 & 0.2 & 0.2 & 0.2 & 0.2 & 0.2 \\
         Avg. Runtime (seconds) & 0.45 & 0.46 & 0.43 & 0.49 & 0.49 & 0.49 & 0.45 & 0.46 & 0.49 & 0.46\\
         \bottomrule
    \end{tabular}}
    
\end{table}

\subsection{Finite Resolving}
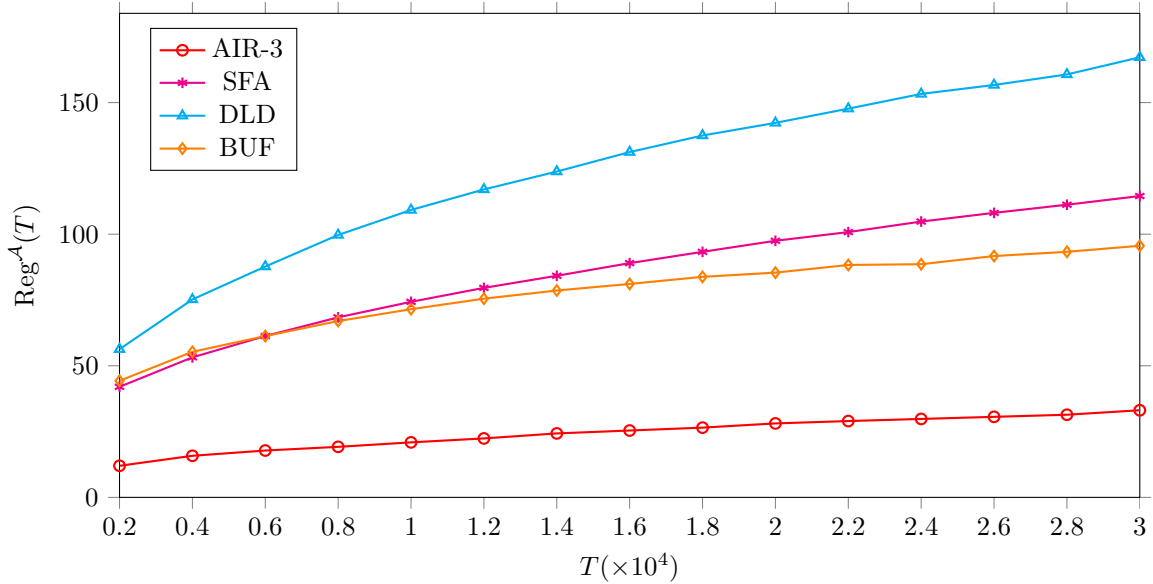
\begin{figure}[ht]
    \begin{tikzpicture}
        \begin{axis}[
            width=15cm, height=8cm,     
            xmin = 2000/10^4,     
            xmax = 30000/10^4,    
            ymin = 0,     
            axis background/.style = {fill=white},
            ylabel = {$\text{Reg}^{\mcA}(T)$},
            xlabel = {$T(\times 10^4)$},
            legend pos= north west,
            tick align = outside,
            ]
    
            
            \addplot[red, mark=o,thick] table[x expr=\thisrow{T}/10^4, y=argmax3_regret, col sep=comma]{data_finite_T_10x2.csv};
            \addlegendentry{AIR-3}
            \addplot[magenta, mark=asterisk,thick] table[x expr=\thisrow{T}/10^4, y=li_regret, col sep=comma]{data_finite_T_10x2.csv};
            \addlegendentry{SFA}
            \addplot[cyan, mark=triangle,thick] table[x expr=\thisrow{T}/10^4, y=gao_regret, col sep=comma]{data_finite_T_10x2.csv};
            \addlegendentry{DLD}
            \addplot[orange, mark=diamond,thick] table[x expr=\thisrow{T}/10^4, y=ma_regret, col sep=comma]{data_finite_T_10x2.csv};
            \addlegendentry{BUF}
        \end{axis} 
    \end{tikzpicture}
    \caption{Regret under different policies as functions of $T$ when $m=10$ and $n=2$. }
    \label{fig:regret_comp_finite_resolve_T}
    \end{figure}

In this case, we study the AIR policy with the resolving schedule $\mcT^F(3)$ with $\beta=0.7$, referred to as ``AIR-3'', and use the degenerate case with $m=10$ and $n=2$ in Section~\ref{subsec:olp_compare}. For each parameter set, we run 2,000 simulations.  
In Figure~\ref{fig:regret_comp_finite_resolve_T}, we compare the performance of the AIR-3 policy with LP-free policies (e.g., SFA, DLD, and BUF). According to Figure~\ref{fig:regret_comp_finite_resolve_T}, the AIR-3 policy can guarantee a low regret by solving LPs only three times, illustrating the power of LP resolving. 

\subsection{Known-Probability Case}\label{subsec:degeneracy_compare}

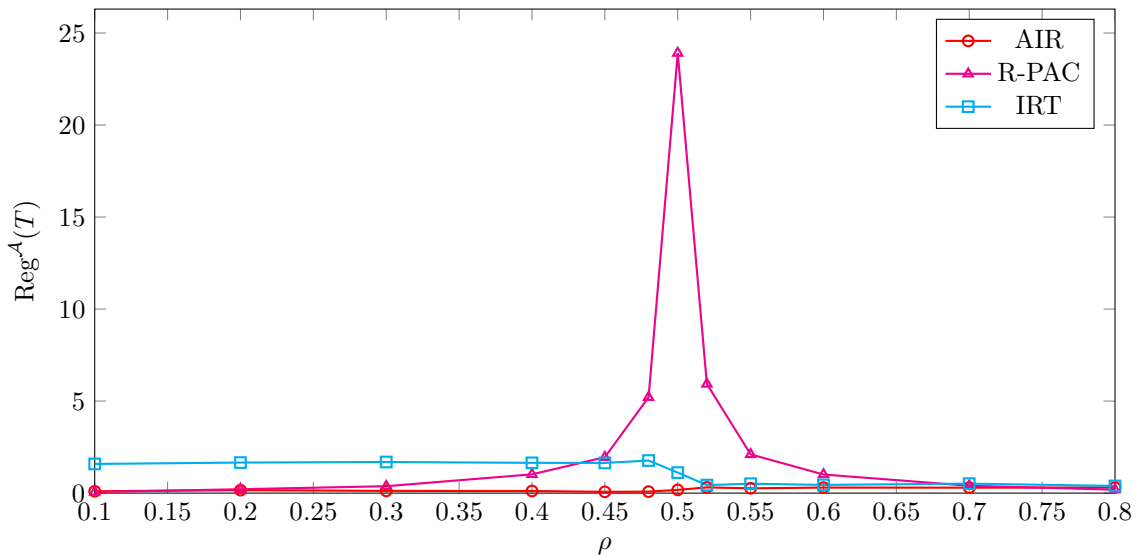
\begin{figure}[ht]
\begin{tikzpicture}
    \begin{axis}[
        width=15cm, height=6cm,     
        xmin = 0.1,     
        xmax = 0.9,    
        ymin = 0,     
        axis background/.style = {fill=white},
        ylabel = {$\text{Reg}^{\mcA}(T)$},
        xlabel = {$\rho$},
        ]

		
        \addplot[mark=o,red,thick] table[x=b, y=air_regret, col sep=comma]{data_KP.csv};
        \addlegendentry{AIR-KP}
        \addplot[mark=triangle,magenta,thick] table[x=b, y=pac_regret, col sep=comma]{data_KP.csv};
        \addlegendentry{R-PAC}
        \addplot[mark=square,cyan,thick] table[x=b, y=irt_regret, col sep=comma]{data_KP.csv};
        \addlegendentry{IRT}
	\end{axis} 
\end{tikzpicture}
\caption{Regret under different policies as functions of $\rho$ when $T=20{,}000$ and $\beta=5/6$; known-probability case. }
\label{fig:regret_comp_56}
\end{figure}

In this subsection, we consider the single-resource case in Section~\ref{subsec:olp_compare}. We compare three policies: R-PAC policy with per-period resolving in \cite{jasin2012re}, IRT policy in \cite{bumpensanti2020re}, and our AIR-KP policy with $\beta=5/6$. In this case, the resolving schedule of our policy is the same as that of the IRT policy. In Figure~\ref{fig:regret_comp_56}, we illustrate the regret $\text{Reg}^{\mcA}(T)$ as the budget per period $\rho$ changes. 

According to Figure~\ref{fig:regret_comp_56}, the regret of the R-PAC policy is large when the initial LP is degenerate ($\rho=0.5$), which agrees with the lower bound $\Omega(\sqrt{T})$ in \cite{bumpensanti2020re}. In contrast, both the IRT policy and our AIR-KP policy can deal with the degeneracy well and hence maintain constant regret. Note that both the IRT and the AIR-KP policies only solve LPs 13 times (with the same schedule) while the R-PAC policy solves LPs 20,000 times, which highlights the effectiveness of the resolving schedule.

\subsection{Markov-Modulated Arrival Probabilities. } Then, we investigate the performance of the AIR-KP policy when the arrival probabilities evolve according to a Markov process (see Section~\ref{subsec:arrival} for details).  Specifically, we consider the case with $m=10$, $n=50$, $L=10$, and $\alpha=\beta=5/6$. We consider 20 instances with the parameters generated as follows: First, we generate $A_{ij}\sim\mathcal{U}[0, 1]$, $r_j\sim\mathcal{U}[0, 1]$ and $\rho_i\sim\mathcal{U}[0.3, 0.5]$; second, to generate $\bs{\sigma}^\ell$, we sample $n$ values $z_j^\ell$'s from $\mathcal{U}[0, 1]$ and then select $\bs{\sigma}^\ell=(\sigma^\ell_1, \dots, \sigma^\ell_n)$ with $\sigma^\ell_j=z_j^\ell/\left(\sum_{j'=1}^n z_{j'}^\ell\right)$; third, to generate the transition matrix $\bs{Q}$, we sample $L^2$ values $w_{ij}$'s from $\mathcal{U}[0, 1]$ and then select $\bs{Q}=\{q_{ij}\}_{i\in [L], j\in [L]}$ with $q_{ij} = w_{ij}/\left(\sum_{j'=1}^L w_{ij'}\right)$; lastly, the initial state distribution is a uniform distribution over the $L$ states. The performance of the AIR-KP policy is reported in Table~\ref{tab:markov}. According to Table~\ref{tab:markov}, the proposed AIR-KP policy also works well in this case.

\begin{table}[ht]
    \centering
    \aboverulesep=0ex
    \belowrulesep=0ex 
    \caption{Regret and runtime of AIR-KP policy when $m=10$, $n=50$, $L=10$, and $\alpha=\beta=5/6$; known-probability case under Markov-modulated arrival process.}
    \label{tab:markov}
    \resizebox{\textwidth}{!}{
    \begin{tabular}{c|cccccccccc}
    \toprule
         \diagbox{Statistics}{$T$} & 2,500 & 5,000 & 7,500 & 10,000 & 12,500 & 15,000 & 17,500 & 20,000 \\
         \midrule
         Avg. Regret & 5.3 & 5.5 & 5.8 & 6.1 & 6.2 & 6.3 & 6.3 & 6.3 \\
         Avg. Relative Loss (\%) &  0.5 & 0.3 & 0.2 & 0.2 & 0.1 & 0.1 & 0.1 & 0.1\\
         Avg. Runtime (seconds) & 0.16 & 0.17 & 0.17 & 0.17 & 0.18 & 0.17 & 0.19 & 0.20\\
         \bottomrule
    \end{tabular}}
    
\end{table}

\section{Concluding Remarks}\label{sec:conclusion}
In this paper, we investigate the OLP problem under the random input assumption, assuming a finite-support underlying distribution. We propose an infrequent resolving algorithm that guarantees constant regret by solving LPs $\mcO(\log\log T)$ times. This algorithm strikes a superior balance between algorithm performance (i.e., regret) and computational efficiency (i.e., resolving frequency). Compared with LP-based algorithms, we significantly reduce the number of resolvings without significantly sacrificing algorithm performance; compared with LP-free algorithms, we substantially improve performance with only a slight increase in resolving frequency. Moreover, given a finite number $M$ of resolvings, we design a resolving schedule such that our policy guarantees an $\mcO\left(T^{(1/2+\epsilon)^{M-1}}\right)$ regret bound. Furthermore, when the arrival probabilities are known at the beginning, we can also guarantee a constant regret by solving LPs $\mcO(\log\log T)$ times, and guarantee an $\mcO\left(T^{(1/2+\epsilon)^{M}}\right)$ regret by solving LPs only $M$ times. 

From a methodological point of view, our work introduces a novel framework for proving regret bounds of infrequent resolving algorithms. The design of the resolving schedule highlights the importance of resolving at both the beginning and the end of the time horizon. These insights may be helpful in designing infrequent resolving algorithms for other problems. 

\newpage 
\bibliographystyle{informs2014}
\bibliography{reference}

\newpage
\begin{appendices}
\begin{center}
    \textbf{\Large Appendix}
\end{center}

\renewcommand{\theHsection}{A\arabic{section}}

\section{Concentration Inequality}

In this paper, we mainly use Hoeffding's inequality in \cite{hoeffding1994probability}. To be self-contained, we present the inequality in the following lemma. 
\begin{lemma}[Hoeffding's Inequality]\label{lem:hoeffding}
    Let $V_1, V_2, \dots, V_k$ be i.i.d. Bernoulli random variables with mean $\bar{v}$. Then, we have 
    \begin{align*}
        &\mathbb{P}\left(\sum_{\ell=1}^k V_\ell - k\bar{v}\ge \sqrt{k\log \xi}\right) \le \frac{1}{\xi^2}\\
        &\mathbb{P}\left(\sum_{\ell=1}^k V_\ell - k\bar{v}\le -\sqrt{k\log \xi}\right) \le \frac{1}{\xi^2}\\
        &\mathbb{P}\left(\left|\sum_{\ell=1}^k V_\ell - k\bar{v}\right|\ge \sqrt{k\log \xi}\right) \le \frac{2}{\xi^2}.
    \end{align*}
\end{lemma}

\section{Discussion of $\epsilon$}\label{append:epsilon}
    Consider a simple problem with two types of customers and one type of resource. The arrival probability of either customer type is $0.5$. The rewards of the two types of customers are $r_1=2$ and $r_2=1$, and each customer consumes one unit of resource. Given the time horizon $T$, the initial inventory is $\rho T$ with $\rho\ge 0.5$. Given the resolving schedule $\left\{1, T- T^{1/2+\epsilon}\right\}$, the time horizon is divided into two parts, $[1, T-T^{1/2+\epsilon})$ and $[T-T^{1/2+\epsilon}, T]$. Consider the event when all the following four conditions hold:
    \begin{align*}
        &\Lambda_1\left(1, T-T^{1/2+\epsilon}\right) \le \frac12\left(T-T^{1/2+\epsilon}\right) - 2\sqrt{T} \qquad \Lambda_1\left(T-T^{1/2+\epsilon}, T\right) \le \frac12 T^{1/2+\epsilon}\\
        &\Lambda_2\left(1, T-T^{1/2+\epsilon}\right) \ge \frac12\left(T-T^{1/2+\epsilon}\right) + 2\sqrt{T} \qquad \Lambda_2\left(T-T^{1/2+\epsilon}, T\right) \le \frac12 T^{1/2+\epsilon},
    \end{align*}
    where $\Lambda_i(t_1, t_2)$ is the random number of type-$i$ arrivals during the time interval $[t_1, t_2)$. Following the proof of Proposition~\ref{prop:lower_bound}, we can similarly prove that the probability of the above event is greater than a constant. 
    
    If we set $\epsilon=0$, under the above event, the proposed policy will accept all type-1 customers and at most $\left(\rho-\frac12\right)T + \frac12 \sqrt{T}$ type-2 customers. In contrast, the hindsight benchmark will accept all type-1 customers and at least $\left(\rho-\frac12\right)T + \frac32 \sqrt{T}$ type-2 customers, implying an $\Omega(\sqrt{T})$ regret. This is because the error in the first part cannot be fully made up by the second part, i.e., $\frac12  T^{1/2+\epsilon} \le 2\sqrt{T}$ when $\epsilon=0$. Therefore, a positive $\epsilon$ is necessary for our work even in the non-degenerate case.

    Then, a natural question is why the $\epsilon$ term does not appear in the finite-resolving bound (under the non-degeneracy assumption) in \cite{jasin2012re}. We highlight that the finite-resolving schedule in \cite{jasin2012re} is adaptively determined based on the realization of the arrival process. In contrast, our schedule is independent of the arrival process and is determined at the beginning of the time horizon. 

    Under an adaptive schedule, we resolve the LP when the number of arrivals significantly deviates from the expectation, and hence control the wrong actions more carefully. Thus, an adaptive schedule has the potential to improve the performance. However, the analysis of an adaptive schedule is complicated, such that the proof in \cite{jasin2012re} requires both the non-degeneracy assumption and a finite number of resolvings $M$ independent of $T$. Moreover, there are no existing results showing that in the degenerate case, solving LPs twice can lead to an $O(T^{\frac{1}{4}})$ regret bound.
    Therefore, we leave the analysis of adaptive schedules without the non-degeneracy assumption to future work.

\section{Omitted Proofs}
    In this section, we provide omitted proofs in the main text. 

\subsection{Proof of Lemma~\ref{lem:upper_bound}}
Let $\mu^*$ denote the optimal policy of $V^*(T)$. First, for any sample path $\omega$ with the demand $\bs{Z}^1(\omega)$, let $\bar{y}_j(\omega)=\sum_{t=1}^T x_{t, j}^{\mu^*}(\omega)$ for each $j$. Then, it can be verified that $\bar{\bs{y}}(\omega)$ is always a feasible solution to the problem $\phi(\bs{b}^1, \bs{Z}^1(\omega))$ because of the feasibility constraints in \eqref{eq:exact_formula_finite_type}. Then, we have $\mE\left[\mE[\phi(\bs{b}^1, \bs{Z}^1(\omega))]\right]\ge \bs{r}^{\mathsf{T}}\bar{\bs{y}}(\omega)$. Taking expectations on both sides, we can deduce that $V^H(T) \ge V^*(T)$. \hfill \qed

\subsection{Proof of Proposition~\ref{prop:delta_property}}
In this proof, we fix $(\bs{b}, \bs{Z}, j)$ with $\bs{b}\ge \bs{A}_j x$ and $\bs{Z}\ge \bs{e}_j$, and define $\bar{a}:=\max_{i, j} a_{ij}$ and $\bar{r}=\max_{j} r_j$. First, according to Theorem 2.4 in \cite{mangasarian1987lipschitz}, for any optimal solution $\bs{y}_1^*$ of $\phi(\bs{b}, \bs{Z})$, there exists an optimal solution $\bs{y}_2^*$ of $\phi(\bs{b}-x\bs{A}_j, \bs{Z}-\bs{e}_j)$ such that 
$$ \Vert \bs{y}_1^* - \bs{y}_2^*\Vert_\infty \le \kappa_1 \cdot \max\{\max_i a_{ij}, 1\} \le \kappa_1 \cdot \max\{\bar{a}, 1\},$$
where $\kappa_1$ only depends on the matrix $\bs{A}$. Then, we have 
\begin{align*}
    \Delta(\bs{b}, \bs{Z}, j, x) = \phi(\bs{b}, \bs{Z})  - \phi(\bs{b}-x\bs{A}_j, \bs{Z}-\bs{e}_j) - x r_j\le \sum_{j=1}^n \kappa_1 r_j \cdot \max\{\bar{a}, 1\}  \le n \kappa_1 \bar{r} \cdot \max\{\bar{a}, 1\}. 
\end{align*}
Therefore, we have $\Delta(\bs{b}, \bs{Z}, j, x) \le r_\phi := n \kappa_1 \bar{r} \cdot \max\{\bar{a}, 1\} $, which is independent of $T$. 

Let $\bs{y}^*$ be an optimal solution of $\phi(\bs{b}, \bs{Z})$ with $y_j^*\ge 1$. Then, we prove that $\bs{y}^*-\bs{e}_j$ is an optimal solution to $\phi(\bs{b}-\bs{A}_j, \bs{Z}-\bs{e}_j)$. First, it is obvious that $\bs{y}^*-\bs{e}_j$ is a feasible solution. Second, suppose there exists a feasible solution $\tilde{\bs{y}}$ of $\phi(\bs{b}-\bs{A}_j, \bs{Z}-\bs{e}_j)$ such that $\bs{r}^{\mathsf{T}} \tilde{\bs{y}} > \bs{r}^{\mathsf{T}} (\bs{y}^*-\bs{e}_j)$. Since $\tilde{\bs{y}}+\bs{e}_j$ is a feasible solution to $\phi(\bs{b}, \bs{Z})$, we have $\bs{r}^{\mathsf{T}}(\tilde{\bs{y}}+\bs{e}_j) > \bs{r}^{\mathsf{T}}\bs{y}^*$, which contradicts the optimality of $\bs{y}^*$. Therefore, we can deduce that $\bs{y}^*-\bs{e}_j$ is an optimal solution to $\phi(\bs{b}-\bs{A}_j, \bs{Z}-\bs{e}_j)$, implying that $\Delta(\bs{b}, \bs{Z}, j, 1) = 0$. Similarly, we can prove that $\Delta(\bs{b}, \bs{Z}, j, 0)=0$ if $Z_j-y_j^*\ge 1$. \hfill \qed

 \subsection{Proof of Proposition~\ref{prop:d_diff_1_learning}}
For $t\in [T_{k-1},T_k)$, we have 
\begin{align*}
    |d_j^t - Z_j^t| &= \left|\frac{(\sum_{\ell=1}^{T_{k-1}-1} Y_j^\ell)\cdot (T-T_{k-1}+1)}{T_{k-1}-1} - \sum_{\ell=T_{k-1}}^{t-1} Y_j^\ell - \sum_{\ell=t}^T Y_j^\ell\right| \\
    &= \left|\frac{(\sum_{\ell=1}^{T_{k-1}-1} Y_j^\ell)\cdot (T-T_{k-1}+1)}{T_{k-1}-1} - \sum_{\ell=T_{k-1}}^T Y_j^\ell\right| \\
    &\le (T-T_{k-1}+1) \left( \left|\frac{(\sum_{\ell=1}^{T_{k-1}-1} Y_j^\ell)}{T_{k-1}-1}-p_j \right|  + \left| p_j -\frac{  \sum_{\ell=T_{k-1}}^T Y_j^\ell}{T-T_{k-1}+1}\right|\right),
\end{align*}
where the first equality follows from the definition of $d_j^t$. Since the arrival process is i.i.d. across time, by Hoeffding's inequality, it holds that
\begin{align*}
    &\mathbb{P}\left(|d_j^t - Z_j^t|\ge (T-T_{k-1}+1) \sqrt{\frac{\log(t-1)}{T_{k-1}-1}} + \sqrt{(T-T_{k-1}+1)\log(T-t+1)}\right)\\
    \le &\mathbb{P}\left(\left|\frac{(\sum_{\ell=1}^{T_{k-1}-1} Y_j^\ell)}{T_{k-1}-1}-p_j \right|\ge \sqrt{\frac{\log(t-1)}{T_{k-1}-1}} \right)+\mathbb{P}\left(\left| p_j - \frac{\sum_{\ell=T_{k-1}}^T Y_j^\ell}{T-T_{k-1}+1}\right|\ge  \sqrt{\frac{\log(T-t+1)}{T-T_{k-1}+1}}\right)\\
    \le&\frac{2}{(t-1)^2}+\frac{2}{(T-t+1)^2}.
\end{align*} 
Similarly, we have $|d_j^t - Z_j^t|\le (T-T_{k-1}+1) \sqrt{\frac{\log(T-t+1)}{T_{k-1}-1}} + \sqrt{(T-T_{k-1}+1)\log(T-t+1)}$ with probability at least $1-\frac{4}{(T-t+1)^2}$. \hfill \qed

\subsection{Proof of Lemma~\ref{lem:t_relation}}
Given $t\in [T_{k-1}, T_k)$, there exist $n_1$ and $n_2$ such that $T_{k-1}\ge \max\{\lceil T^{\alpha^{n_1+1}}\rceil, \lceil T-T^{\beta^{n_2}}\rceil\}$ and $T_{k}\le \min\{\lceil T^{\alpha^{n_1}}\rceil, \lceil T-T^{\beta^{n_2+1}}\rceil\}$. Then, we have 
\begin{align*}
    &(t-1)^{\alpha}\le (T_{k}-1)^{\alpha} \le (\lceil T^{\alpha^{n_1}}\rceil-1)^{\alpha} \le (T^{\alpha^{n_1}})^\alpha = T^{\alpha^{n_1+1}} \le T_{k-1} \\
    &(T-t+1)^{1/\beta} \ge (T-T_k+1)^{1/\beta} \ge (T-\lceil T-T^{\beta^{n_2+1}}\rceil+1)^{1/\beta} \ge (T^{\beta^{n_2+1}})^{1/\beta} = T^{\beta^{n_2}} \ge T-T_{k-1}. \tag*{\qed}
\end{align*}

\subsection{Proof of Proposition~\ref{prop:surrogate_LP_learning}}
We prove the two statements in Proposition~\ref{prop:surrogate_LP_learning} one by one. Before proceeding, we first simplify $d_j^t-(T-t+1)p_j$ as follows:
\begin{align*}
    d_j^t - (T-t+1)p_j &= \frac{(\sum_{\ell=1}^{T_{k-1}-1} Y_j^\ell)\cdot (T-T_{k-1}+1)}{T_{k-1}-1} - \sum_{\ell=T_{k-1}}^{t-1} Y_j^\ell - (T-t+1)p_j\\
    &= (T-T_{k-1}+1)\left(\frac{\sum_{\ell=1}^{T_{k-1}-1} Y_j^\ell}{T_{k-1}-1} - p_j\right) + (t-T_{k-1})\left(p_j -\frac{\sum_{\ell=T_{k-1}}^{t-1} Y_j^\ell}{t-T_{k-1}}\right). 
\end{align*}

\vspace{0.1in}
\noindent\textnormal{(i)}
Consider the time period $t \in [3, T-3]$ such that there exist $T_{k-1}$ and $T_k$ satisfying $t\in[T_{k-1}, T_{k})$. There are two possible cases:
\begin{enumerate}
    \item[(a)] When $T_{k-1}< \lceil \frac{T}{2}\rceil$, similar to the proof of Proposition \ref{prop:d_diff_1_learning}, by Hoeffding's inequality,  we have 
    \begin{align*}
        &\mathbb{P}\left(d_j^t-(T-t+1)p_j\le -(T-T_{k-1}+1)\sqrt{\frac{\log(t-1)}{T_{k-1}-1}} - \sqrt{(t-T_{k-1})\log(T-t+1)}\right)\\
        \le& \mathbb{P}\left(\frac{(\sum_{\ell=1}^{T_{k-1}-1} Y_j^\ell)}{T_{k-1}-1}-p_j \le -\sqrt{\frac{\log(t-1)}{T_{k-1}-1}} \right)+\mathbb{P}\left( p_j -\frac{\sum_{\ell=T_{k-1}}^{t-1} Y_j^\ell}{t-T_{k-1}} \le  -\sqrt{\frac{\log(T-t+1)}{t-T_{k-1}}}\right)\\
        \le& \frac{1}{(t-1)^2}+\frac{1}{(T-t+1)^2}.
    \end{align*}
    Since $\lceil \frac{T}{2} \rceil\in \mcT$ and $T_{k-1}<\lceil \frac{T}{2}\rceil$,  we have $t-T_{k-1}\le T-T_{k-1}+1\le T \le 2(T-T_k+1)\le 2(T-t+1)$. Then, according to Lemma~\ref{lem:t_relation}, with probability at least $1-\frac{1}{(t-1)^2}-\frac{1}{(T-t+1)^2}$, it holds 
    $$d_j^t \ge (T-t+1)p_j - 2(T-t+1)\sqrt{\frac{\log(t-1)}{(t-1)^{\alpha}-1}} - \sqrt{2(T-t+1)\log(T-t+1)}. $$ 
    
    Since $\alpha>0$, there exists a constant $\eta_j$ such that when $t\ge \eta_j$, we have $2\sqrt{\frac{\log(t-1)}{(t-1)^\alpha-1}}\le \frac{p_j}{4}$. Moreover, there exists a constant $\eta'_j$ such that when $t\le T-\eta'_j$, we have \[\frac{p_j }{4}(T-t+1) - \sqrt{2(T-t+1)\log(T-t+1)}\ge 0 \text{ and } \frac{p_j (T-t+1)}{2}\ge 2.\] Therefore, when $t\in [\eta_j, \min\{T-\eta'_j, \lceil \frac{T}{2}\rceil\}]$, we have 
    \[d_j^{t}\ge \frac{3p_j (T-t+1)}{4} - \sqrt{2(T-t+1)\log(T-t+1)}\ge \frac{p_j (T-t+1)}{2}\ge 2\] with probability at least $1-\frac{1}{(t-1)^2}-\frac{1}{(T-t+1)^2}$.

    \item[(b)] When $T_{k-1} \ge \lceil \frac{T}{2} \rceil$, according to the Hoeffding's inequality, we have 
    $$d_j^{t} \ge (T-t+1)p_j - (T-T_{k-1}+1)\sqrt{\frac{\log(T-t+1)}{T_{k-1}-1}} - \sqrt{(t-T_{k-1})\log(T-t+1)},$$ 
    with probability at least $1-\frac{2}{(T-t+1)^2}$. Since $T_{k-1}\ge \lceil\frac{T}{2}\rceil$, we have $T_{k-1}-1 \ge \frac{T-T_{k-1}+1}{2}$. According to Lemma~\ref{lem:t_relation}, we have 
    \begin{align*}
    d_j^{t}&\ge (T-t+1)p_j - (T-T_{k-1}+1) \sqrt{\frac{2\log(T-t+1)}{T-T_{k-1}+1}} - \sqrt{(T-T_{k-1}+1) \log (T-t+1)} \\
    &= (T-t+1)p_j - (\sqrt{2}+1)\sqrt{(T-T_{k-1}+1) \log (T-t+1)}\\
    &\ge (T-t+1)p_j - (\sqrt{2}+2) (T-t+1)^{\frac{1}{2\beta}}\sqrt{\log (T-t+1)}.    
    \end{align*}
    Since $\frac{1}{2\beta}<1$, there exists a constant $\eta''_j$ such that when $t\le T-\eta''_j$, we have $d_j^{t}\ge \frac{p_j(T-t+1)}{2} \ge 2$ with probability $1-\frac{2}{(T-t+1)^2}$. 
\end{enumerate}
Let $c_1 = \max\{\max_j \eta_j$, 3\} and $c_2 = \max\{\max_j \eta'_j, \max_j \eta''_j, 3\}$, which are independent of $T$. The above proof implies that, when $t\in [c_1, T-c_2]$, we have $\bs{d}^t\ge \frac{p_j(T-t+1)}{2}\ge \bs{2}$ with probability $1-\frac{n}{(T-t+1)^2}-\frac{n}{(\min\{t-1, T-t+1\})^2}$.

\vspace{0.2in}
\noindent \textnormal{(ii)} According to the proof of (i), when $t\in [c_1, T-c_2]\cap[T_{k-1}, T_k)$, we have $\bs{d}^{t} \ge \bs{2}$ with high probability. Then, it suffices to prove that if $\bs{d}^{t}\ge \bs{2}$, then $\bs{u}^t$ is an optimal solution to the surrogate LP $\phi(\bs{b}^t,\bs{d}^t)$. Given that $\bs{d}^{t}\ge \bs{2}$, we have $\bs{d}^\ell\ge \bs{2}$ when $\ell\in [T_{k-1}, t-1]$ due to the monotonicity of $\bs{d}^t$. When $t=T_{k-1}$, the solution $\bs{u}^t$ is an optimal solution solved from $\phi(\bs{b}^t, \bs{d}^t)$. Then, we prove the statement by induction. 

Suppose $\bs{u}^\tau$ is an optimal solution to $\phi(\bs{b}^\tau, \bs{d}^\tau)$ with $\tau\in [T_{k-1}, t)$. Since $\bs{d}^{\tau}\ge\bs{2}$, if the AIR algorithm accepts the arriving customer of type $j^t$, then we have $u^{\tau}_{j^t}\ge \frac12 d^{\tau}_{j^t} \ge 1$. Thus, we have $\bs{u}^{\tau+1}=\bs{u}^\tau-\bs{e}_{j^t}\ge \bs{0}$ is a feasible solution of $\phi(\bs{b}^{\tau+1}, \bs{d}^{\tau+1})$. If there exists a feasible solution $\tilde{\bs{u}}$ of $\phi(\bs{b}^{\tau+1}, \bs{d}^{\tau+1})$ such that $\bs{r}^{\mathsf{T}}\tilde{\bs{u}} > \bs{r}^{\mathsf{T}}\bs{u}^{\tau+1}$, then we have $\tilde{\bs{u}}+\bs{e}_{j^t}$ feasible to $\phi(\bs{b}^{\tau}, \bs{d}^{\tau})$ and $\bs{r}^{\mathsf{T}}(\tilde{\bs{u}}+\bs{e}_{j^t})> \bs{r}^{\mathsf{T}}\bs{u}^{\tau}$, which contradicts the optimality of $\bs{u}^{\tau}$. Then, we can deduce that $\bs{u}^{\tau+1}$ is an optimal solution to $\phi(\bs{b}^{\tau+1}, \bs{d}^{\tau+1})$. If the AIR algorithm rejects the arriving customer, we can similarly prove the optimality of $\bs{u}^{\tau+1}$. Therefore, we can prove that $\bs{u}^t$ is optimal to $\phi(\bs{b}^t, \bs{d}^t)$. \hfill \qed

\subsection{Proof of Proposition~\ref{prop:satisfying_prob}}
In the following, we always consider the good event.
First, we bound the difference $|\phi(\bs{b}^t, \bs{Z}^t)-\phi(\bs{b}^t, \bs{d}^t)|$. According to Theorem~2.4 in \cite{mangasarian1987lipschitz}, for any optimal solution $\bs{y}_1$ to $\phi(\bs{b}^t, \bs{Z}^t)$, there exists an optimal solution $\bs{y}_2$ to $\phi(\bs{b}^t, \bs{d}^t)$ such that $\Vert \bs{y}_1 -\bs{y}_2 \Vert_\infty \le \kappa_2\Vert \bs{d}^t-\bs{Z}^t\Vert_\infty$, where $\kappa_2$ is a constant independent of $T$. Therefore, we have
$$|\phi(\bs{b}^t, \bs{Z}^t)-\phi(\bs{b}^t, \bs{d}^t)| = |\bs{r}^{\mathsf{T}} (\bs{y}_1-\bs{y}_2)| \le \left(\sum_{j=1}^n r_j\right) \Vert \bs{y}_1-\bs{y}_2\Vert_\infty \le \kappa_2\left(\sum_j r_j\right) \Vert \bs{d}^t-\bs{Z}^t\Vert_\infty.$$
Note that $\bar{\mfS}(\bs{b},\bs{d},j)$ can be formulated as an LP:
\begin{align*}
\max_{\bs{y}\ge \bs{0}} \quad & y_j\\
    \text{s.t.}\quad & \bs{r}^{\mathsf{T}} \bs{y} \ge \phi(\bs{b}, \bs{d}),\\
    & \bs{A} \bs{y} \le \bs{b},\\
    &\bs{y} \le \bs{d}.
\end{align*}
According to Theorem~2.4 in \cite{mangasarian1987lipschitz}, we have
$$\Vert \bar{\mfS}(\bs{b}^t, \bs{Z}^t, j) - \bar{\mfS}(\bs{b}^t, \bs{d}^t, j)\Vert_\infty \le \kappa_3 (\Vert \bs{d}^t-\bs{Z}^t\Vert_\infty + |\phi(\bs{b}^t, \bs{Z}^t)-\phi(\bs{b}^t, \bs{d}^t)|) \le \kappa_4 \Vert \bs{d}^t-\bs{Z}^t\Vert_\infty,$$
where $\kappa_4=\kappa_3+\kappa_3\kappa_2\sum_j r_j$.

Consider the case when a type-$j$ customer arrives at time $t$ and the AIR policy accepts this request. We then have 
\begin{align*}
    \bar{\mfS}(\bs{b}^t, \bs{Z}^t, j)&\ge \bar{\mfS}(\bs{b}^t, \bs{d}^t, j) -\kappa_4\Vert \bs{d}^t-\bs{Z}^t\Vert_\infty \ge u_j^t -\kappa_4 \Vert \bs{d}^t-\bs{Z}^t\Vert_\infty\\
    &\ge \frac{d_j^t}{2} -\kappa_4 \Vert \bs{d}^t-\bs{Z}^t\Vert_\infty \ge \frac{p_j (T-t+1)}{4} - \kappa_4 \Vert \bs{d}^t-\bs{Z}^t\Vert_\infty,
\end{align*}
where the second inequality holds because $\bs{u}^t$ is an optimal solution to $\phi(\bs{b}^t, \bs{d}^t)$, the third inequality is due to the argmax operation in Algorithm~\ref{alg:period_resolving}, and the last inequality is due to Proposition~\ref{prop:surrogate_LP_learning}. 

Given $t\in [T_{k-1}, T_k)$, we discuss two possible cases:
\begin{enumerate}
    \item[\textnormal{(a)}] When $T_{k-1}< \lceil \frac{T}{2} \rceil$, since $\lceil \frac{T}{2} \rceil\in \mcT$ and $T_{k-1}<\lceil \frac{T}{2} \rceil$, we have $t\le \lceil \frac{T}{2} \rceil$ and $T-T_{k-1}+1 \le T \le 2(T-t+1)$. According to Proposition~\ref{prop:d_diff_1_learning}, we have 
    \begin{align*}
        &\bar{\mfS}(\bs{b}^t, \bs{Z}^t, j) \\
        \ge& \frac{p_j (T-t+1)}{4} - \kappa_4 (T-T_{k-1}+1)\sqrt{\frac{\log(t-1)}{T_{k-1}-1}} - \kappa_4 \sqrt{(T-T_{k-1}+1)\log(T-t+1)}\\
        \ge& \frac{p_j (T-t+1)}{4} - 2\kappa_4(T-t+1)\sqrt{\frac{\log(t-1)}{(t-1)^\alpha-1}} - \kappa_4 \sqrt{2(T-t+1)\log(T-t+1)}.
    \end{align*}
    where the last inequality is because of Lemma~\ref{lem:t_relation}.
    Since $\alpha>0$, there exists a constant $\theta_j$ such that when $t\ge \theta_j$, we have $2\kappa_4\sqrt{\frac{\log(t-1)}{(t-1)^\alpha-1}}\le \frac{p_j}{8}$. Moreover, there exists a constant $\theta'_j$ such that when $t\le T-\theta'_j$, we have $\frac{(T-t+1)p_j}{8}-\kappa_4\sqrt{2(T-t+1)\log(T-t+1)}\ge 1.$ Therefore, when $t\in [\theta_j, \min\{\lceil \frac{T}{2}\rceil, T-\theta'_j\}]$, we have $\bar{\mfS}(\bs{b}^t, \bs{Z}^t, j) \ge \frac{(T-t+1)p_j}{8}-\kappa_4\sqrt{2(T-t+1)\log(T-t+1)}\ge 1$. 
    \item[\textnormal{(b)}] When $T_{k-1} \ge \lceil \frac{T}{2} \rceil$, we have $T_{k-1}-1 \ge \frac{T-T_{k-1}+1}{2}$. According to Proposition~\ref{prop:d_diff_1_learning}, we have
    \begin{align*}
        &\bar{\mfS}(\bs{b}^t, \bs{Z}^t, j) \\
        \ge& \frac{p_j (T-t+1)}{4} - \kappa_4 (T-T_{k-1}+1)\sqrt{\frac{\log(T-t+1)}{T_{k-1}-1}} -\kappa_4 \sqrt{(T-T_{k-1}+1)\log(T-t+1)}\\
        \ge& \frac{p_j (T-t+1)}{4} -\kappa_4(T-T_{k-1}+1)\sqrt{\frac{2\log(T-t+1)}{T-T_{k-1}+1}} - \kappa_4 \sqrt{(T-T_{k-1}+1)\log(T-t+1)}\\
        =& \frac{p_j (T-t+1)}{4} -\kappa_4(\sqrt{2}+1)\sqrt{(T-T_{k-1}+1)\log(T-t+1)}\\
        \ge& \frac{p_j (T-t+1)}{4} - \kappa_4 (2+\sqrt{2}) (T-t+1)^{\frac{1}{2\beta}}\sqrt{\log(T-t+1)}.
    \end{align*}
    where the last inequality is because of Lemma~\ref{lem:t_relation}. Similar to the above discussions, since $\beta>\frac12$, there exists a constant $\theta''_j$ such that $\bar{\mfS}(\bs{b}^t, \bs{Z}^t, j)\ge 1$ when $\lceil\frac{T}{2}\rceil \le t\le T-\theta''_j$. 
\end{enumerate}

Let $\kappa_5^A = \max_j \theta_j$ and $\kappa_6^A=\max\{ \max_j\theta'_j, \max_j \theta''_j\}$, which are independent of $T$. When $t\in [\kappa_5^A, T-\kappa_6^A] \cap [c_1, T-c_2]$, if the algorithm accepts a type-$j$ customer at time $t$, then we have $\bar{\mfS}(\bs{b}^t, \bs{Z}^t, j)\ge 1$. Similarly, for the rejection action, we can derive $\kappa_5^R$ and $\kappa_6^R$. By setting $c_5=\max\{\kappa_5^A, \kappa_5^R, c_1\}$ and $c_6=\max\{\kappa^A_6, \kappa^R_6, c_2\}$, we have $\bar{\mfS}(\bs{b}^t, \bs{Z}^t, j^t)\ge 1$ if $x_{ j^t}^t=1$ and $Z_{j^t}^t - \underline{\mfS}(\bs{b}^t, \bs{Z}^t, j^t)\ge 1$ if $x_{j^t}^t=0$. 
Finally, according to Proposition~\ref{prop:delta_property}, we have $\Delta(\bs{b}^t, \bs{Z}^t, j^t, x_{j^t}^t)=0$. \hfill \qed

\subsection{Proof of Theorem~\ref{thm:olp_finite_resolve}}\label{proof:olp_finite_resolving}
We discuss different time intervals.
\begin{enumerate}
    \item For the time interval $\left[1, \left\lceil T^{\beta^{M-1}} \right\rceil\right)$, Algorithm~\ref{alg:period_resolving} rejects all customers and incurs at most $T^{\beta^{M-1}} \cdot r_{\phi}$ regret.
    \item For the time interval $\left[\left\lceil T^{\beta^{M-1}} \right\rceil, \left\lceil T- T^{\beta^{M-1}} \right\rceil \right]$, the proofs of Propositions~\ref{prop:d_diff_1_learning}, \ref{prop:surrogate_LP_learning} and \ref{prop:satisfying_prob} still hold, and hence the regret at time $t$ can be bounded by $\left(\frac{c_3}{(T-t+1)^2} + \frac{c_4}{(t-1)^2}\right)r_\phi$. 
    \item For the time interval $\left[\left\lceil T- T^{\beta^{M-1}}\right\rceil+1, T\right]$, the regret is no greater than $ T^{\beta^{M-1}}r_\phi$. 
\end{enumerate}

To summarize, given the resolving time set $\mcT^{F}(M)$, the regret can be bounded as 
\begin{align*}
    \text{Reg}^{\mcA}(T) &\le T^{\beta^{M-1}} \cdot r_{\phi} + \sum_{t=\left\lceil T^{\beta^{M-1}}\right\rceil}^{\left\lceil T- T^{\beta^{M-1}}\right\rceil}\left(\frac{c_3}{(T-t+1)^2} + \frac{c_4}{(t-1)^2}\right)r_\phi + T^{\beta^{M-1}}r_\phi\\
    &\le 2T^{\beta^{M-1}}r_\phi + \frac{\pi^2}{6} (c_3+c_4)r_\phi =\mcO(T^{\beta^{M-1}}). \tag*{\qed}
\end{align*}

\subsection{Proof of Theorem~\ref{thm:regret_bound_nrm}}\label{proof:nrm_regret}

Similar to Lemma~\ref{lem:t_relation}, we have $T-T_{k-1}\le (T-t+1)^{1/\beta}$ for any $t\in [T_{k-1}, T_k)$. In this case, the regret decomposition is the same as \eqref{eq:regret_decompose} but the proof is simpler than the unknown-probability case. In the following, we present the main steps of the proof. 
First, we bound the approximation error, i.e., the difference between $\bs{d}^t$ and $\bs{Z}^t$. 
\begin{lemma}\label{lem:d_diff_1_learning_nrm}
    Consider the resolving schedule $\mcT^{\mcK}$. Given a time $t\in [T_{k-1}, T_k)$, we have $|d_j^t-Z_j^t|\le \sqrt{(T-T_{k-1}+1)\log(T-t+1)}$ with probability larger than $1-\frac{2}{(T-t+1)^2}$.
\end{lemma}

Second, we prove the relationship between $\bs{u}^t$ and a surrogate LP.
\begin{lemma}\label{lem:surrogate_LP_learning_nrm}
   Consider the known-probability case. For the AIR-KP policy with the resolving schedule $\mcT^{\mcK}$ with $\beta\in (1/2, 1)$, there exists a constant $c_7$ independent of $T$ such that when $t\le T-c_7$, with probability larger than $1-\frac{n}{(T-t+1)^2}$, we have 
    \begin{enumerate}
        \item $d_j^t\ge \frac{p_j(T-t+1)}{2}\ge 2$ for any $j$.
        \item $\bs{u}^t$ is an optimal solution to the surrogate LP $\phi(\bs{b}^t, \bs{d}^t)$.
    \end{enumerate}
\end{lemma}

Third, we bound the difference between $\bs{u}^t$ and the hindsight solution, and then prove the regret bound. The \textit{good event} is defined as the case when conditions in both Lemmas~\ref{lem:d_diff_1_learning_nrm} and \ref{lem:surrogate_LP_learning_nrm} are satisfied, and the probability is no less than $1-\frac{c_8}{(T-t+1)^2}$. 
\begin{lemma}\label{lem:satisfying_prob_nrm}
    Consider the known-probability case. For the AIR-KP policy with the resolving schedule $\mcT^{\mcK}$ with $\beta\in (\frac12, 1)$, there exists a constant $c_{10}$ independent of $T$ such that when $t\le T-c_{10}$, under the good event, we have 
    \begin{enumerate}
        \item $\bar{\mfS}(\bs{b}^t, \bs{Z}^t, j^t)\ge 1$ if $x_{t, j^t}^{\mcA}=1$ and $Z_{j^t}^t - \underline{\mfS}(\bs{b}^t, \bs{Z}^t, j^t)\ge 1$ if $x_{t, j^t}^{\mcA}=0$.
        \item $\Delta(\bs{b}^t, \bs{Z}^t, j^t, x_{t, j^t}^{\mcA})=0$.
    \end{enumerate}
\end{lemma}

Lastly, we can prove the constant regret bound.
\begin{align*}
    \text{Reg}^{\mcA}(T) &\le \sum_{t=1}^{T-c_{10}} r_\phi \mathbb{P}\left(\Delta(\bs{b}^t, \bs{Z}^t, j^t, x_{t, j^t}^{\mcA})>0\right) + c_{10} r_\phi \\
    &\le \sum_{t=1}^{T-c_{10}} r_\phi \frac{c_8}{(T-t+1)^2} + c_{10} r_\phi \le \left(\frac{\pi^2}{6}c_8+c_{10}\right)r_\phi.\tag*{\qed}
\end{align*}

 \subsection{Proof of Lemma~\ref{lem:d_diff_1_learning_nrm}}
In this case, for $t\in [T_{k-1}, T_k)$, by Hoeffding's inequality, it holds that 
\begin{align*}
    &\mathbb{P}\left(|d_j^t - Z_j^t|\ge \sqrt{(T-T_{k-1}+1)\log(T-t+1)}\right) \\
    =& \mathbb{P}\left(\left|p_j(T-T_{k-1}+1) - \sum_{\ell=T_{k-1}}^T Y_j^\ell\right|\ge \sqrt{(T-T_{k-1}+1)\log(T-t+1)}\right) \\
    \le& \frac{2}{(T-t+1)^2}.\tag*{\qed}
\end{align*}

\subsection{Proof of Lemma~\ref{lem:surrogate_LP_learning_nrm}}

First, for $t\in [T_{k-1}, T_k)$, by Hoeffding's inequality, we have
\begin{align*}
&\mathbb{P}\left(d_j^t \le (T-t+1)p_j - \sqrt{(t-T_{k-1})\log(T-t+1)} \right) \\
=& \mathbb{P}\left((t-T_{k-1})p_j - \sum_{\ell=T_{k-1}}^{t-1} Y^\ell_j \le - \sqrt{(t-T_{k-1})\log(T-t+1)} \right)\\
\le & \frac{1}{(T-t+1)^2}.
\end{align*}

Therefore, with probability greater than $1-\frac{1}{(T-t+1)^2}$, we have $d_j^t\ge (T-t+1)p_j - \sqrt{(t-T_{k-1})\log(T-t+1)} \ge (T-t+1)p_j - \sqrt{(T-t+1)^{1/\beta}\log(T-t+1)}$. 
Since $\frac{1}{2\beta}<1$, there exists a constant $c_7$ such that when $t\le T-c_7$, $\bs{d}^t \ge \frac{T-t+1}{2}\bs{p} \ge \bs{2}$. 
Similar to Proposition~\ref{prop:surrogate_LP_learning}, we can prove that $\bs{u}^t$ is an optimal solution to $\phi(\bs{b}^t, \bs{d}^t)$ when $\bs{d}^t\ge \bs{2}$. \hfill \qed

\subsection{Proof of Lemma~\ref{lem:satisfying_prob_nrm}}
In the following, we always consider the good event. Consider the case when a type-$j$ customer arrives at time $t$ and the AIR-KP policy accepts the request. Similar to Proposition~\ref{prop:satisfying_prob}, given $t\in [T_{k-1}, T_k)$ and $t\le T-c_7$, we have 
\begin{align*}
    \bar{\mfS}(\bs{b}^t, \bs{Z}^t, j)&\ge  \frac{p_j (T-t+1)}{4} - \kappa_7 \Vert \bs{d}^t-\bs{Z}^t\Vert_\infty\ge \frac{p_j (T-t+1)}{4} - \kappa_7 \sqrt{(T-T_{k-1}+1)\log(T-t+1)}\\
    &\ge \frac{p_j (T-t+1)}{4} - \kappa_7 \sqrt{2(T-t+1)^{1/\beta}\log(T-t+1)}.
\end{align*}

Since $\frac{1}{2\beta}<1$, there exists a constant $c_9$ such that when $t\le T-c_9$, we have $\bar{\mfS}(\bs{b}^t, \bs{Z}^t, j^t)\ge 1$ if $x_{t, j^t}^{\mcA}=1$ and $Z_{j^t}^t-\underline{\mfS}(\bs{b}^t, \bs{Z}^t, j^t)\ge 1$ otherwise. Then, we set $c_{10}=\max\{c_7, c_9, 3\}$, and finish the proof. \hfill \qed

\subsection{Proof of Theorem~\ref{thm:nrm_finite_resolve}}
We discuss different time intervals.
\begin{enumerate}
    \item For the time interval $\left[1, \left\lceil T- T^{\beta^{M}}\right\rceil\right)$, the proofs of Lemmas~\ref{lem:d_diff_1_learning_nrm}, \ref{lem:surrogate_LP_learning_nrm} and \ref{lem:satisfying_prob_nrm} still hold, and hence the regret at time $t$ can be bounded by $\frac{c_8}{(T-t+1)^2}r_\phi$. 
    \item For the time interval $\left[\left\lceil T- T^{\beta^{M}}\right\rceil, T\right]$, the regret is no greater than $ T^{\beta^{M}}r_\phi$. 
\end{enumerate}

To summarize, given the resolving time set $\mcT^{\mcK}$, the regret can be bounded as 
\begin{align*}
    \text{Reg}^{\mcA}(T) &\le  \sum_{t=1}^{\left\lceil T- T^{\beta^{M}}\right\rceil} \frac{c_8}{(T-t+1)^2} r_\phi + T^{ \beta^{M}} \cdot r_{\phi} \le \frac{c_8 \pi^2}{6} r_\phi + T^{ \beta^{M}}r_\phi =\mcO(T^{\beta^{M}}).\tag*{\qed}
\end{align*}

\subsection{Proof of Proposition~\ref{prop:lower_bound}}
According to Theorem~\ref{thm:regret_bound_nrm}, we have $\mE\left[\phi(T\bs{\rho}, \bs{Z}^1)\right] - V^*(T) = O(1)$, i.e., the difference between the hindsight problem and the optimal problem is upper bounded by a constant. Thus, in the following, we compare the policy with the hindsight problem. 

Let $T_1$ and $T_2$ (with $T_1<T_2$) denote the two resolving time points. Similar to \cite{bumpensanti2020re}, in this proof, we consider two types of customers and one type of resource. The arrival probability of either customer type is $0.5$. The rewards of the two types of customers are $r_1=2$ and $r_2=1$, and each customer consumes one unit of resource. The budget per period is $0.5$, resulting in a degenerate case. 

We start with the case where $T_1 = \omega(T^{1/4})$, which means that $\lim_{T\to\infty} \frac{T_1}{T^{1/4}} = \infty$. In this case, due to the initialization in Algorithm~\ref{alg:period_resolving_nrm}, the algorithm rejects all customers during the time interval $[1, T_1)$. Let $\Lambda_j(t_1, t_2)$ denote the random number of type-$j$ arrivals during the time interval $[t_1, t_2]$. We consider the event where $\frac12 (T_1-1) - \sqrt{T_1-1}\le \Lambda_1(1, T_1-1)\le \frac12 (T_1-1) $ and $\Lambda_1(T_1, T)\le \frac12 (T-T_1+1)$. In this case, we have $\Lambda_1(1, T) = \Lambda_1(1, T_1-1) + \Lambda_1(T_1, T) \le \frac12 T$ and hence the hindsight problem will accept all type-1 customers. However, the algorithm rejects $\Lambda_1(1, T_1-1)$ type-1 customers, and hence the corresponding revenue loss is at least $\Lambda_1(1, T_1-1) \ge \frac12 (T_1-1) -\sqrt{T_1-1} = \Omega(T^{1/4})$. Then, similar to \cite{bumpensanti2020re}, according to the Berry-Esseen theorem (see \citealt{shevtsova2011absolute}), we bound the probability of the event as follows:
\begin{align*}
    &\mP\left(\frac12 (T_1-1) -\sqrt{T_1-1}\le \Lambda_1(1, T_1-1)\le \frac12 (T_1-1) \; \& \; \Lambda_1(T_1, T)\le \frac12 (T-T_1+1) \right)\\
    =&\mP\left(-\frac{\sqrt{T_1-1}}{\frac12 \sqrt{T_1-1}}\le  \frac{\Lambda_1(1, T_1-1) - \frac12 (T_1-1)}{\frac12 \sqrt{T_1-1}} \le 0\right)\cdot \mP\left(\frac{\Lambda_1(T_1, T) - \frac12 (T-T_1+1)}{\frac12 \sqrt{T-T_1+1}} \le 0\right)\\
    \ge& \left(0.477 - 2\times \frac{0.4748}{\frac18 \sqrt{T_1-1}}\right) \cdot \left(0.5 - \frac{0.4748}{\frac18 \sqrt{T-T_1+1}} \right),
\end{align*}
which is greater than a constant when $T$ is no less than a threshold. Therefore, the regret is $\Omega(T^{1/4})$. Then, we consider the case where $T_1 = O(T^{1/4})$, and study different choices of $T_2$. 

\noindent\textbf{Case I: There exists a constant $c_{12}<1$ such that $T_1\le T_2\le c_{12}\cdot T$.} 

In this case, we consider the event when $\frac12 (T_2-1) \le \Lambda_1(1, T_2-1) \le \frac12 (T_2-1) + \sqrt{T_2-1}$ and $\Lambda_1(T_2, T) \le \frac12(T-T_2+1) - \sqrt{T}$. Under this event, the total number of type-1 arrivals is $\Lambda_1(1, T_2-1) + \Lambda_1(T_2, T) \le \frac12 T + \sqrt{T_2-1} - \sqrt{T}\le \frac12 T -(1-\sqrt{c_{12}})\sqrt{T}$, and hence the hindsight optimal policy accepts at least $(1-\sqrt{c_{12}})\sqrt{T}$ type-2 customers. Then, we analyze the performance of the AIR-KP policy with the resolving schedule $\{T_1, T_2\}$. During the time interval $[1, T_1)$, the policy rejects all customers. Then, the algorithm resolves the fluid model and derives the solution $\bs{u}^{T_1} = \left(\frac12(T-T_1), \frac12 T_1 \right)$ and the demand estimation $\bs{d}^{T_1} = \left(\frac12(T-T_1), \frac12(T-T_1) \right)$. Thus, during the interval $[T_1, T_2)$, the algorithm accepts all type-1 customers and no greater than $\frac12 T_1$ type-2 customers, resulting in $\nu\ge \Lambda_1(1, T_2-1)-T_1$ resource consumption. When the algorithm resolves the fluid model at time $T_2$, we can derive the solution $\bs{u}^{T_2}$ with $u_2^{T_2}= \max\left\{\frac12 T - \nu - \frac12(T-T_2+1), 0\right\}\le \max\left\{\frac12 (T_2-1) - \frac12 (T_2-1)-T_1, 0 \right\} \le T_1$. Therefore, the algorithm accepts at most $\frac32 T_1$ type-2 customers and induces at least $(1-\sqrt{c_{12}})\sqrt{T} - \frac32 T_1 = \Omega(\sqrt{T})$ revenue loss under this event. Then, we provide a lower bound for the event probability as follows (Let $\Phi(\cdot)$ denote the c.d.f of a standard normal distribution):
\begin{align*}
    &\mP\left(\frac12 (T_2-1) \le \Lambda_1(1, T_2-1) \le \frac12 (T_2-1) + \sqrt{T_2-1} \; \& \; \Lambda_1(T_2, T) \le \frac12(T-T_2+1) - \sqrt{T}\right)\\
    =& \mP\left(0 \le \frac{\Lambda_1(1, T_2-1) -\frac12 (T_2-1)}{\frac12 \sqrt{T_2-1}}  \le 2\right) \cdot \mP\left(\frac{\Lambda_1(T_2, T) - \frac12(T-T_2+1)}{\frac12 \sqrt{T-T_2+1}} \le - \frac{\sqrt{T}}{\frac12 \sqrt{T-T_2+1}} \right)\\
    \ge& \mP\left(0 \le \frac{\Lambda_1(1, T_2-1) -\frac12 (T_2-1)}{\frac12 \sqrt{T_2-1}}  \le 2\right) \cdot \mP\left(\frac{\Lambda_1(T_2, T) - \frac12(T-T_2+1)}{\frac12 \sqrt{T-T_2+1}} \le -  \frac{2}{\sqrt{1-c_{12}}} \right)\\
    \ge & \left(0.477 - 2\times \frac{0.4748}{\frac18 \sqrt{T_2-1}} \right) \cdot \left(\Phi\left(-  \frac{2}{\sqrt{1-c_{12}}}\right) - \frac{0.4748}{\frac18 \sqrt{T-T_2+1}} \right),
\end{align*}
which is greater than a constant when $T$ is above a threshold. Therefore, the regret of the algorithm is at least $\Omega(\sqrt{T})$. 

\noindent\textbf{Case II: $T-T_2+1 = \Omega(\sqrt{T})$ and $T-T_2+1 = o(T)$.} 

In this case, we consider the event when $\Lambda_1(1, T_2-1) \le \frac12 (T_2-1) - \sqrt{T-T_2+1}$ and $\frac12(T-T_2+1) + \frac12 \sqrt{T-T_2+1} \le \Lambda_1(T_2, T) \le \frac12(T-T_2+1) + \sqrt{T-T_2+1}$. Under this event, we have $\Lambda_1(1, T) = \Lambda_1(1, T_2-1) + \Lambda_1(T_2, T) \le \frac12 T$, and hence the hindsight optimal policy accepts all type-1 customers. Then, we analyze the performance of the AIR-KP policy with the resolving schedule $\{T_1, T_2\}$. Similar to Case I, the algorithm rejects all customers during the interval $[1, T_1)$. During the interval $[T_1, T_2)$, the algorithm accepts all type-1 customers and at most $\frac12 T_1$ type-2 customers. Then, the algorithm resolves the fluid model and obtains the solution $\bs{u}^{T_2}$ with $u_1^{T_2}= \min\left\{\frac12 T - \nu , \frac12 (T-T_2+1)\right\} \le \frac12(T-T_2+1)$. However, since $\Lambda_1(T_2, T)\ge \frac12(T-T_2+1)+\frac12\sqrt{T-T_2+1}$, the algorithm rejects at least $\frac12\sqrt{T-T_2+1}$ type-1 customers, resulting in $\Omega(T^{1/4})$ revenue loss. Then, given a sufficiently large $T$, we provide a lower bound for the event probability as follows:
\begin{align*}
    &\mP\left(\Lambda_1(1, T_2-1) \le \frac12 (T_2-1) - \sqrt{T-T_2+1} \; \right. \\
    &\quad\quad \left.\& \;  \frac12(T-T_2+1) + \frac12 \sqrt{T-T_2+1} \le \Lambda_1(T_2, T) \le \frac12(T-T_2+1) + \sqrt{T-T_2+1}\right)\\
    =& \mP\left(\frac{\Lambda_1(1, T_2-1) -\frac12 (T_2-1)}{\frac12 \sqrt{T_2-1}} \le - \frac{\sqrt{T-T_2+1}}{\frac12 \sqrt{T_2-1}} \right) \cdot \mP\left( 1 \le \frac{\Lambda_1(T_2, T)-\frac12(T-T_2+1)}{\frac12 \sqrt{T-T_2+1}} \le 2\right)\\
    \ge& \mP\left(\frac{\Lambda_1(1, T_2-1) -\frac12 (T_2-1)}{\frac12 \sqrt{T_2-1}} \le -2 \right) \cdot \mP\left( 1 \le \frac{\Lambda_1(T_2, T)-\frac12(T-T_2+1)}{\frac12 \sqrt{T-T_2+1}} \le 2\right)\\
    \ge& \left(0.022 - \frac{0.4748}{\frac18\sqrt{T_2-1}} \right)\cdot \left(0.135 - 2\times\frac{0.4748}{\frac18 \sqrt{T-T_2+1}} \right),
\end{align*}
which is greater than a constant when $T$ is above a threshold. Therefore, the regret of the algorithm is at least $\Omega(T^{1/4})$. 

\noindent\textbf{Case III: $T-T_2+1 = o(\sqrt{T})$.} 

In this case, we consider the event when $\Lambda_1(1, T_2-1) \le \frac12(T_2-1) - \sqrt{T_2-1}$. Under this event, given a sufficiently large $T$, we have $\Lambda_1(1, T) = \Lambda_1(1, T_2-1) + \Lambda_1(T_2, T) \le \frac12 (T_2-1) - \sqrt{T_2-1} + (T-T_2+1) \le \frac12 (T_2-1) - \frac12 \sqrt{T_2-1}$, and hence the hindsight optimal policy accepts at least $\frac12 \sqrt{T_2-1}$ type-2 customers. Then, similar to the analysis in Case I, the algorithm rejects all customers during the interval $[1, T_1)$. During the interval $[T_1, T_2)$, the algorithm accepts all type-1 customers and at most $\frac12 T_1$ type-2 customers. During the interval $[T_2, T]$, the algorithm accepts at most $T-T_2+1$ type-2 customers. Thus, the algorithm accepts at most $\frac12 T_1 + T-T_2+1$ type-2 customers, resulting in $\frac12 \sqrt{T_2-1} - \frac12 T_1 - (T-T_2+1) = \Omega(\sqrt{T}) - \mcO(T^{1/4}) - o(\sqrt{T}) = \Omega(\sqrt{T})$. Then, we provide a lower bound for the event probability as follows:
\begin{align*}
    &\mP\left(\Lambda_1(1, T_2-1) \le \frac12(T_2-1) - \sqrt{T_2-1}\right) =  \mP\left(\frac{\Lambda_1(1, T_2-1)-\frac12(T_2-1)}{\frac12 \sqrt{T_2-1}} \le - 2\right)  \ge  \left(0.022 - \frac{0.4748}{\frac18 \sqrt{T_2-1}}\right),
\end{align*}
which is greater than a constant when $T$ is greater than a threshold. Therefore, the regret is $\Omega(\sqrt{T})$. 

To summarize, given two resolving time points, the regret of the AIR-KP policy is $\Omega(T^{1/4})$. \hfill \qed

\subsection{Proof of Lemma~\ref{lem:regret_periodic_nrm}}
For the known-probability problem, following the proof of Theorem~\ref{thm:regret_bound_nrm}, we need to prove similar results in Lemmas~\ref{lem:surrogate_LP_learning_nrm} and \ref{lem:satisfying_prob_nrm}. Similar to Lemma~\ref{lem:surrogate_LP_learning_nrm}, we have 
$$d_j^t \ge (T-t+1)p_j -\sqrt{(t-T_{k-1})\log(T-t+1)} \ge (T-t+1)p_j -\sqrt{\omega \log(T-t+1)},$$
with probability $1-\frac{1}{(T-t+1)^2}$. There exists a constant $\tilde{\zeta}^P$ such that when $t\le T-\tilde{\zeta}^P$, $\bs{d}^t\ge \frac{(T-t+1)}{2}\bs{p}\ge \bs{2}$. Moreover, $\tilde{\zeta}^P$ is the minimal positive integer $x$ satisfying $\frac{(x+1)\min_j p_j}{2} - \sqrt{\omega \log(x+1)}\ge 0$. Therefore, we can deduce that $\tilde{\zeta}^P = \mcO(\sqrt{\omega}\log \omega) = \tilde{\mcO}(\sqrt{\omega})$. 

Similar to Lemma~\ref{lem:satisfying_prob_nrm}, under the good event, we have 
\begin{align*}
    \bar{\mfS}(\bs{b}^t, \bs{Z}^t, j)&\ge \frac{p_j (T-t+1)}{4} - \kappa_7\sqrt{(T-T_{k-1}+1)\log(T-t+1)}\\
    &\ge \frac{p_j (T-t+1)}{4} - \kappa_7\sqrt{(T-t+1+\omega)\log(T-t+1)}.
\end{align*}
Then, there exists a constant $\tilde{\theta}^P$ such that when $t\le T-\tilde{\theta}^P$, we have $\bar{\mfS}(\bs{b}^t, \bs{Z}^t, j^t)\ge 1$ if $x_{t, j^t}^{\mcA}=1$ and $Z_j^t-\underline{\mfS}(\bs{b}^t, \bs{Z}^t, j^t)\ge 1$ if $x_{t, j^t}^{\mcA}=0$. Moreover, $\tilde{\theta}^P$ is the minimal integer $x$ satisfying $\frac{(x+1)\min_j p_j}{4} - \sqrt{(x+1+\omega)\log(x+1)}\ge 1$, implying that $\tilde{\theta}^P = \mcO(\sqrt{\omega}\log\omega)=\tilde{\mcO}(\sqrt{\omega})$. 
Then, the regret bound can be bounded as $\text{Reg}^{\mcA}(T) \le \left(\frac{\pi^2}{6}c_8 + \tilde{\theta}^P\right)r_\phi = \tilde{\mcO}(\sqrt{\omega})$. \hfill \qed

\subsection{Proof of Proposition~\ref{prop:regret_periodic}}
For the unknown-probability case, following the proof of Theorem~\ref{thm:regret_bound}, we only need to prove similar results in Propositions~\ref{prop:surrogate_LP_learning} and \ref{prop:satisfying_prob}. 

\begin{lemma}\label{lem:surrogate_LP_learning_periodic}
    For the resolving schedule $\mcT^{\mcK, P}(\omega)$, there exist two constants $\eta^P=\mcO(\omega)$ and $\zeta^P=\mcO(\omega\log\omega)$ independent of $T$. 
    When $t\in [\eta^P, T-\zeta^P]$, with probability $1-\frac{c_{10}}{(T-t+1)^2}-\frac{c_{11}}{(\min\{T-t+1, t-1\})^2}$, we have 
    \begin{enumerate}
        \item $d_j^t\ge \frac{p_j(T-t+1)}{2}\ge 2$ for any $j$.
        \item $\bs{u}^t$ is an optimal solution to the surrogate LP $\phi(\bs{b}^t, \bs{d}^t)$. 
    \end{enumerate}
\end{lemma}

\begin{lemma}\label{lem:satisfying_prob_periodic}
    For the resolving schedule $\mcT^{\mcK, P}(\omega)$, there exist two constants $\iota^P=\mcO(\omega)$ and $\theta^P=\mcO(\omega\log \omega)$ independent of  $T$ such that when $t\in [\iota^P, T-\theta^P]$, under the good event, we have 
    \begin{enumerate}
        \item $\bar{\mfS}(\bs{b}^t, \bs{Z}^t, j^t)\ge 1$ if $x_{t, j^t}^{\mcA}=1$ and $Z_{j^t}^t - \underline{\mfS}(\bs{b}^t, \bs{Z}^t, j^t)\ge 1$ if $x_{t, j^t}^{\mcA}=0$.
        \item $\Delta(\bs{b}^t, \bs{Z}^t, j^t, x_{t, j^t}^{\mcA})=0$.
    \end{enumerate}
\end{lemma}
Then, the regret bound can be bounded as 
$\text{Reg}^{\mcA}(T) \le \left(\frac{\pi^2}{6} (c_{10}+c_{11})+\iota^P+\theta^P\right)r_\phi = \mcO(\omega)$. 

\vspace{0.1in}

\textit{Proof of Lemma~\ref{lem:surrogate_LP_learning_periodic}.}
To derive a similar result in Proposition~\ref{prop:surrogate_LP_learning}, we only need to show that $\bs{d}^{t}\ge \frac{T-t+1}{2} \bs{p}\ge \bs{2}$ with high probability. Consider the time period $t\in [T_{k-1}, T_k)$.
\begin{enumerate}
    \item When $T_{k-1}<\lceil \frac{T}{2}\rceil$, we have
    \begin{align*}
        d_j^t &\ge (T-t+1)p_j - (T-T_{k-1}+1)\sqrt{\frac{\log(t-1)}{T_{k-1}-1}} - \sqrt{(t-T_{k-1})\log(T-t+1)}\\
        &\ge (T-t+1)p_j - (T-t+\omega+1)\sqrt{\frac{\log(t-1)}{t-\omega-1}} - \sqrt{\omega\log(T-t+1)},
    \end{align*}
    with probability $1-\frac{1}{(t-1)^2}-\frac{1}{(T-t+1)^2}$. There exist  constants $\eta^P_j$ and $\zeta^P_j$ such that when $t\in [\eta^P_j, \min\{T-\zeta^P_j, \lceil\frac{T}{2}\rceil+\omega\}]$, we have $ \sqrt{\frac{\log(t-1)}{t-\omega-1}}\le \frac{p_j}{4}$, $\frac{p_j}{4}(T-t+1)-\frac{p_j \omega}{4}-\sqrt{\omega \log(T-t+1)}\ge 0$, and hence $d_j^t\ge \frac{p_j(T-t+1)}{2}\ge 2$. Moreover, we have $\eta^P_j=\mcO(\omega)$ and $\zeta_j^P=\mcO(\omega\log\omega)$. 
    \item When $T_{k-1}\ge \lceil \frac{T}{2}\rceil$, we have 
    \begin{align*}
        d_j^t &\ge (T-t+1)p_j - \sqrt{2(T-T_{k-1}+1)\log(T-t+1)} - \sqrt{(t-T_{k-1})\log(T-t+1)}\\
        &\ge (T-t+1)p_j -\sqrt{2(T-t+\omega+1)\log(T-t+1)} - \sqrt{\omega\log(T-t+1)}\\
        &\ge (T-t+1)p_j - \sqrt{2(T-t+1)\log(T-t+1)} - \sqrt{2 \omega\log(T-t+1)} - \sqrt{\omega \log(T-t+1)}, 
    \end{align*}
    with probability $1-\frac{2}{(T-t+1)^2}$. Then, there exists a constant $\bar{\zeta}^P_j$ such that when $t\le T-\bar{\zeta}^P_j$, we have $d_j^t\ge \frac{p_j (T-t+1)}{2}\ge 2$. Moreover, we have $\bar{\zeta}^P_j=\mcO(\omega\log \omega)$. 
\end{enumerate}
To summarize, by defining $\eta^P=\max_j \eta^P_j = \mcO(\omega)$ and $\zeta^P=\max\{\max_j \zeta^P_j, \max_j \bar{\zeta}^P_j\}=\mcO(\omega\log\omega)$, we finish the proof. 

\vspace{0.1in}

\textit{Proof of Lemma~\ref{lem:satisfying_prob_periodic}.}
Consider the case when a type-$j$ customer arrives at time $t\in [T_{k-1}, T_k)$ and the AIR policy accepts the request. Similar to Proposition~\ref{prop:satisfying_prob}, we discuss two cases.
\begin{enumerate}
    \item When $T_{k-1}< \lceil \frac{T}{2} \rceil$, we have
    \begin{align*}
        &\bar{\mfS}(\bs{b}^t, \bs{Z}^t, j) \\
        \ge& \frac{p_j (T-t+1)}{4} - \kappa_4 (T-T_{k-1}+1)\sqrt{\frac{\log(t-1)}{T_{k-1}-1}} - \kappa_4 \sqrt{(T-T_{k-1}+1)\log(T-t+1)}\\
        \ge& \frac{p_j (T-t+1)}{4} - \kappa_4(T-t+1+\omega)\sqrt{\frac{\log(t-1)}{t-\omega-1}} - \kappa_4\sqrt{(T-t+\omega+1)\log(T-t+1)}.
    \end{align*}
    There exists a constant $\iota^P_j$ such that when $t\ge \iota^P_j$, we have $\kappa_4\sqrt{\frac{\log(t-1)}{t-\omega-1}}\le \frac{p_j}{8}$. Moreover, there exists a constant $\theta^P_j$ such that when $t\le T-\theta^P_j$, we have $\frac{(T-t+1)p_j}{8}-\frac{\omega p_j}{8}-\kappa_4\sqrt{(T-t+\omega+1)\log(T-t+1)}\ge 1.$ Therefore, when $t\in [\iota^P_j, \min\{\lceil \frac{T}{2}\rceil+\omega, T-\theta^P_j\}]$, we have $\bar{\mfS}(\bs{b}^t, \bs{Z}^t, j) \ge  1$.  Moreover, we have $\iota_j^P=\mcO(\omega)$ and $\theta^P_j=\mcO(\omega\log \omega)$. 
    \item When $T_{k-1} \ge \lceil \frac{T}{2} \rceil$, we have
    \begin{align*}
        \bar{\mfS}(\bs{b}^t, \bs{Z}^t, j) 
        \ge& \frac{p_j (T-t+1)}{4} -\kappa_4(\sqrt{2}+1)\sqrt{(T-T_{k-1}+1)\log(T-t+1)}\\
        =& \frac{p_j (T-t+1)}{4} - \kappa_4 (\sqrt{2}+1) \sqrt{(T-t+\omega+1)\log(T-t+1)}.
    \end{align*}
    There exists a constant $\bar{\theta}^P_j$ such that when $\lceil\frac{T}{2}\rceil \le t\le T-\bar{\theta}^P_j$, we have $\bar{\mfS}(\bs{b}^t, \bs{Z}^t, j)\ge 1$. Moreover, we have $\bar{\theta}^P_j=\mcO(\omega\log \omega)$. 
    Then, similar to Proposition~\ref{prop:satisfying_prob}, we can finish the proof. 
\end{enumerate}

\hfill \qed

\subsection{Proof of Lemma~\ref{lem:regret_midpoint_nrm}}
Similar to Lemma~\ref{lem:regret_periodic_nrm}, we only need the following inequalities:
\begin{align*}
    d_j^t \ge (T-t+1)p_j -\sqrt{(t-T_{k-1})\log(T-t+1)} \ge (T-t+1)p_j -\sqrt{2(T-t+1)\log(T-t+1)}
\end{align*}
\begin{align*}
    \bar{\mfS}(\bs{b}^t, \bs{Z}^t, j)&\ge \frac{p_j (T-t+1)}{4} - \kappa_7\sqrt{(T-T_{k-1}+1)\log(T-t+1)}\\
    &\ge \frac{p_j (T-t+1)}{4} - \kappa_7\sqrt{\left(2(T-t+1)+1\right)\log(T-t+1)}.
\end{align*}

Then, following a similar proof, we can derive the constant bound $\mcO(1)$. 
\hfill \qed

\subsection{Proof of Proposition~\ref{prop:regret_midpoint_olp}}
Similar to Proposition~\ref{prop:regret_periodic}, we only need the following inequalities:
\begin{enumerate}
    \item When $T_{k-1}<\lceil \frac{T}{2} \rceil$, we have 
    \begin{align*}
        d_j^t & \ge (T-t+1)p_j - (T-T_{k-1}+1)\sqrt{\frac{\log(t-1)}{T_{k-1}-1}} - \sqrt{(t-T_{k-1})\log(T-t+1)}\\
        &\ge (T-t+1)p_j - \left(2(T-t+1)+1\right) \sqrt{\frac{2\log(t-1)}{t+1}} - \sqrt{\left(2(T-t+1)+1\right)\log(T-t+1)},
    \end{align*}
    and
    \begin{align*}
        &\bar{\mfS}(\bs{b}^t, \bs{Z}^t, j)\\
        \ge & \frac{p_j (T-t+1)}{4} - \kappa_4(T-T_{k-1}+1)\sqrt{\frac{\log(t-1)}{T_{k-1}-1}} - \kappa_4\sqrt{(T-T_{k-1}+1)\log(T-t+1)}\\
        \ge & \frac{p_j (T-t+1)}{4} - \kappa_4\left(2(T-t+1)+1\right)\sqrt{\frac{2\log(t-1)}{t+1}} - \kappa_4\sqrt{\left(2(T-t+1)+1\right)\log(T-t+1)}. 
    \end{align*}
    \item When $T_{k-1}\ge \lceil \frac{T}{2}\rceil$, we have 
    \begin{align*}
        d_j^t & \ge (T-t+1)p_j - (T-T_{k-1}+1)\sqrt{\frac{\log(T-t+1)}{T_{k-1}-1}} - \sqrt{(t-T_{k-1})\log(T-t+1)}\\
        &\ge (T-t+1)p_j - \left(2(T-t+1)+1\right) \sqrt{\frac{\log(T-t+1)}{\left(2(T-t+1)+1\right)}} - \sqrt{\left(2(T-t+1)+1\right)\log(T-t+1)},
    \end{align*}
    and 
    \begin{align*}
        &\bar{\mfS}(\bs{b}^t, \bs{Z}^t, j)\\
        \ge & \frac{p_j (T-t+1)}{4} - \kappa_4(T-T_{k-1}+1)\sqrt{\frac{\log(t-1)}{T_{k-1}-1}} - \kappa_4\sqrt{(T-T_{k-1}+1)\log(T-t+1)}\\
        \ge & \frac{p_j (T-t+1)}{4} - \kappa_4\left(2(T-t+1)+1\right)\sqrt{\frac{\log(t-1)}{\left(2(T-t+1)+1\right)}} - \kappa_4\sqrt{\left(2(T-t+1)+1\right)\log(T-t+1)}.
    \end{align*}
\end{enumerate}
Then, following a similar proof, we can derive the constant bound $\mcO(1)$. 
\hfill \qed

\subsection{Proof of Proposition~\ref{prop:markov}}\label{proof:markov}
In order to derive the same result as Theorem~\ref{thm:regret_bound_nrm}, we need to prove similar results in Lemmas~\ref{lem:d_diff_1_learning_nrm} and \ref{lem:surrogate_LP_learning_nrm}. Let $\bs{\pi}^\ell$ denote the state distribution at period $\ell$. First, for $t\in[T_{k-1}, T_k)$, we have 
\begin{align*}
    |d_j^t - Z_j^t| = \left|d_j^{T_{k-1}} - Z_j^{T_{k-1}} \right| &= \left|\sum_{i=1}^L \pi^*_i \cdot \sigma^i_j \cdot (T-T_{k-1}+1) - \sum_{\ell=T_{k-1}}^TY_j^\ell\right|\\
    &\le \sum_{\ell=T_{k-1}}^T \sum_{i=1}^L \left|\pi^*_i-\pi_i^\ell\right| \cdot \sigma^i_j +  \left| \sum_{\ell=T_{k-1}}^TY_j^\ell - \sum_{\ell=T_{k-1}}^T \sum_{i=1}^L \pi_i^\ell \cdot \sigma^i_j\right|.
\end{align*}
According to Theorem~15.0.1 in \cite{meyn2012markov}, there exist constants $c_{15}>0$ and $\psi\in (0, 1)$ such that $\max_i \Vert Q^k(i, \cdot)-\bs{\pi}^*\Vert_1\le c_{15} \psi^k$ for all $k\ge 0$, where $Q$ is the transition matrix. Consequently, for any initial distribution, $\Vert\bs{\pi}^\ell - \bs{\pi}^*\Vert_\infty \le c_{15} \cdot \psi^\ell$.  Thus, we have 
\begin{align*}
    \sum_{\ell=T_{k-1}}^T \sum_{i=1}^L \left|\pi^*_i-\pi_i^\ell\right| \cdot \sigma^i_j \le \sum_{\ell=T_{k-1}}^T L\cdot \Vert\bs{\pi}^\ell - \bs{\pi}^*\Vert_\infty \le \sum_{\ell=T_{k-1}}^T L\cdot c_{15} \cdot \psi^\ell \le \frac{L\cdot c_{15}}{1-\psi}. 
\end{align*}

Let $a = T_{k-1}$, $N = T - a + 1$ and $S = \sum_{\ell=a}^{T} Y_j^{\ell}$, so that the second term in the above decomposition equals $\left| S - \mE[S] \right|$. Consider the filtration $\{\mathcal{F}_n\}$ generated by $\left\{s^{a},\dots,s^{n},Y^{a},\dots,Y^{n}\right\}$
and the Doob martingale $M_n = \mE\left[S \mid \mathcal{F}_n\right]$ for $n = a-1, a, \dots, T$, where $\mathcal{F}_{a-1}$ is the trivial $\sigma$-field, so that $M_{a-1} = \mE[S]$ and $M_T = S$. Since $Y^{\ell}$ depends on the past only through $s^{\ell}$, we have
\begin{align*}
  M_n = \sum_{\ell=a}^{n} Y_j^{\ell} + g_n\left(s^{n}\right), \qquad
  g_n(s) := \sum_{h=1}^{T-n} \sum_{i=1}^{L} Q^{h}(s,i)\sigma_j^{i},
\end{align*}
where $Q$ is the transition matrix. Since $\left| \sum_{i=1}^{L} Q^{h}(s,i) \sigma_j^{i} - \sum_{i=1}^{L} \pi_i^{*} \sigma_j^{i} \right|
 \le \left\| Q^{h}(s,\cdot) - \pi^{*} \right\|_1 \le c_{15}\psi^{h}$, it holds for any two states $s$ and $s'$ that
\begin{align*}
  \left| g_n(s) - g_n(s') \right|
  \;\le\; \sum_{h=1}^{\infty} 2 c_{15}\,\psi^{h}
  \;=\; \frac{2 c_{15}\,\psi}{1-\psi}.
\end{align*}
Hence, conditional on $\mathcal{F}_{n-1}$, the random variable
$Y_j^{n} + g_n(s^{n})$ has range at most $c_{\psi} := 1 + 2 c_{15}\psi/(1-\psi)$, which implies $\left| M_n - M_{n-1} \right| \le c_{\psi}$. By Azuma's inequality (e.g., Theorem 7.2.1 in \citealt{alon2016probabilistic}), we can deduce that
\begin{align*}
    \mP\left(\left| \sum_{\ell=T_{k-1}}^TY_j^\ell - \sum_{\ell=T_{k-1}}^T \sum_{i=1}^L \pi_i^\ell \cdot \sigma^i_j\right| \ge 2c_{\psi}\sqrt{(T-T_{k-1}+1)\log(T-t+1)}\right) \le \frac{2}{(T-t+1)^2}.
\end{align*}
Therefore, for $t\in[T_{k-1}, T_k)$, with probability at least $1-\frac{2}{(T-t+1)^2}$, we have 
\begin{align*}
    |d_j^t - Z_j^t| \le \frac{L\cdot c_{15}}{1-\psi} + 2c_{\psi} \sqrt{(T-T_{k-1}+1)\log(T-t+1)},
\end{align*}
which is similar to Lemma~\ref{lem:d_diff_1_learning_nrm} except for additional constants $2c_\psi$ and $\frac{L\cdot c_{15}}{1-\psi}$. 

Similarly, for $t\in [T_{k-1}, T_k)$, by the same construction applied to the window $[T_{k-1}, t-1]$, we have 
\begin{align*}
    \mP\left(d_j^t \le \sum_{i=1}^L \pi^*_i \cdot \sigma^i_j \cdot (T-t+1) - 2c_\psi \sqrt{(t-T_{k-1})\log(T-t+1)} - \frac{L\cdot c_{15}}{1-\psi} \right) \le \frac{1}{(T-t+1)^2},
\end{align*}
which is similar to Lemma~\ref{lem:surrogate_LP_learning_nrm} except for additional constants $2c_\psi$ and $\frac{L\cdot c_{15}}{1-\psi}$.

Since the constant term $\frac{L\cdot c_{15}}{1-\psi}$ and the factor $c_\psi$ are independent of $T$, we can still derive a constant regret bound following the proof idea of Theorem~\ref{thm:regret_bound_nrm}. \hfill \qed

\section{Numerical Details}
In this section, we provide the omitted details in Section~\ref{sec:numerical}.

\subsection{OLP Algorithms}\label{subsec:olp_alg}

\vspace{0.05in}
\begin{breakablealgorithm}
    \caption{Argmax with frequent resolving (AFR) policy}\label{alg:afr}
    \begin{algorithmic}
    \State Initialize $\bs{b}^1 \gets T\bs{\rho}$ and $\bs{N}^1 \gets \bs{0}$.
    \For{$t=1, 2, 3, \dots, T$} 
        \State Compute the empirical estimations $\hat{p}_j^t \gets N_j^t/(t-1)$ for each $j$.  
        \State Solve the fluid problem $\phi(\bs{b}^t, (T-t+1)\hat{\bs{p}}^t)$ and obtain its optimal solution $\bs{y}^*$. 
        \State Observe arrival type $j$ and set $\bs{N}^{t+1} \gets \bs{N}^t +\bs{e}_j$.
        \If{$\bs{A}_j\le \bs{b}^t$ and $y^*_j\ge (T-t+1)\hat{p}_j^t-y^*_j$} \Comment{Argmax between $y^*_j$ and $(T-t+1)\hat{p}_j^t-y^*_j$}
            \State Accept the request. 
            \State Set $\bs{b}^{t+1} \gets \bs{b}^t-\bs{A}_j$. 
        \Else
        \State Reject the request and set $\bs{b}^{t+1} \gets \bs{b}^t$.
        \EndIf
    \EndFor
    \end{algorithmic}
    \end{breakablealgorithm}

\vspace{0.1in}

\begin{breakablealgorithm}
    \caption{Adaptive allocation (ADA) policy}\label{alg:ada}
    \begin{algorithmic}
    \State Initialize $\bs{b}^1 \gets T\bs{\rho}$ and $\bs{N}^1 \gets \bs{0}$.
    \For{$t=1, 2, 3, \dots, T$} 
        \State Compute the empirical estimations $\hat{p}_j^t \gets N_j^t/(t-1)$ for each $j$.  
        \State Solve the fluid problem $\phi(\bs{b}^t, (T-t+1)\hat{\bs{p}}^t)$ and obtain its optimal solution $\bs{y}^*$. 
        \State Observe arrival type $j$ and set $\bs{N}^{t+1} \gets \bs{N}^t +\bs{e}_j$.
        \If{$\bs{A}_j\le \bs{b}^t$}
            \State Accept the request with probability $y_j^*/\left((T-t+1)\hat{p}_j^t\right)$. \Comment{Probabilistic Allocation.}
            \State If accepted, set $\bs{b}^{t+1} \gets \bs{b}^t-\bs{A}_j$. 
            \State If rejected, set $\bs{b}^{t+1} \gets \bs{b}^t$.
        \Else  
            \State Reject the request and set $\bs{b}^{t+1} \gets \bs{b}^t$.
        \EndIf
    \EndFor
    \end{algorithmic}
    \end{breakablealgorithm}

\vspace{0.1in}

\begin{breakablealgorithm}
    \caption{Simple and fast (SFA) policy}\label{alg:sfa}
    \begin{algorithmic}
    \State Initialize $\bs{b}^1 \gets T\bs{\rho}$ and $\bs{q}^1=\bs{0}$.
    \For{$t=1, 2, 3, \dots, T$} 
        \State Observe arrival type $j$.
        \State Set $\tilde{x}^t\gets 1$ if $r_j>\bs{A}_j\cdot \bs{q}^t$ and $\tilde{x}^t\gets 0$ otherwise.
        \State Compute $\bs{q}^{t+1} \gets \bs{q}^t + \frac{1}{\sqrt{t}} (\bs{A}_j \tilde{x}^t - \bs{\rho})$.
        \State Compute $\bs{q}^{t+1} \gets \max\{\bs{q}^{t+1}, \bs{0}\}$.
        \If{$\bs{A}_j\le \bs{b}^t$ and $\tilde{x}_t=1$}
            \State Accept the request and set $\bs{b}^{t+1} \gets \bs{b}^t-\bs{A}_j$.
        \Else 
            \State Reject the request and set $\bs{b}^{t+1} \gets \bs{b}^t$.
        \EndIf
    \EndFor
    \end{algorithmic}
    \end{breakablealgorithm}

\vspace{0.1in}

\begin{breakablealgorithm}
    \caption{Decoupling learning and decision (DLD) policy}\label{alg:dld}
    \begin{algorithmic}
    \State Input: $T_e=\lfloor T^{2/3}\rfloor$, $\alpha_e=T^{-1/3}$ and $\alpha_p=T^{-2/3}$. 
    \State Initialize $\bs{b}^1 \gets T\bs{\rho}$, $\bs{q}^1_D\gets \bs{0}$ and $\bs{q}^1_L\gets\bs{0}$.
    \For{$t=1, 2, 3, \dots, T_e$} 
        \State Observe arrival type $j$.
            \State Set $\tilde{x}^t_D\gets 1$ if $r_j>\bs{A}_j\cdot \bs{q}^t_D$ and $\tilde{x}^t_D\gets 0$ otherwise.
            \State Compute $\bs{q}^{t+1}_D \gets \bs{q}^t_D + \alpha_e (\bs{A}_j \tilde{x}^t_D - \bs{\rho})$.
            \State Compute $\bs{q}^{t+1}_D \gets \max\{\bs{q}^{t+1}_D, \bs{0}\}$.
            \If{$\bs{A}_j\le \bs{b}^t$ and $\tilde{x}_D^t=1$}
                \State Accept the request and set $\bs{b}^{t+1} \gets \bs{b}^t-\bs{A}_j$.
            \Else 
                \State Reject the request and set $\bs{b}^{t+1} \gets \bs{b}^t$. 
            \EndIf
            \State Set $\tilde{x}^t_L\gets 1$ if $r_j>\bs{A}_j\cdot \bs{q}^t_L$ and $\tilde{x}^t_L\gets 0$ otherwise.
            \State Compute $\bs{q}^{t+1}_L \gets \bs{q}^t_L + \frac{1}{t} (\bs{A}_j \tilde{x}^t_L - \bs{\rho})$.
            \State Compute $\bs{q}^{t+1}_L \gets \max\{\bs{q}^{t+1}_L, \bs{0}\}$.
    \EndFor
    \State Set $\bs{q}^{T_e+1}_D\gets \bs{q}^{T_e+1}_L$.
    \For{$t=T_e+1, T_e+2, \dots, T$} 
        \State Observe arrival type $j$.
            \State Set $\tilde{x}^t_D\gets 1$ if $r_j>\bs{A}_j\cdot \bs{q}^t_D$ and $\tilde{x}^t_D\gets 0$ otherwise.
            \State Compute $\bs{q}^{t+1}_D \gets \bs{q}^t_D + \alpha_p (\bs{A}_j \tilde{x}^t_D - \bs{\rho})$.
            \State Compute $\bs{q}^{t+1}_D \gets \max\{\bs{q}^{t+1}_D, \bs{0}\}$.
            \If{$\bs{A}_j\le \bs{b}^t$ and $\tilde{x}_D^t=1$}
                \State Accept the request and set $\bs{b}^{t+1} \gets \bs{b}^t-\bs{A}_j$.
            \Else  
                \State Reject the request and set $\bs{b}^{t+1} \gets \bs{b}^t$.
            \EndIf
    \EndFor
    \end{algorithmic}
    \end{breakablealgorithm}

\vspace{0.1in}

\begin{breakablealgorithm}
    \caption{Budget-updating fast (BUF) policy}\label{alg:buf}
    \begin{algorithmic}
    \State Input: Time set $\mcT=\{T-\lceil \frac{T}{2^k} \rceil: k=1, 2, \dots, \lceil \log_2 T\rceil\}$.
    \State Initialize $\bs{b}^1 \gets T\bs{\rho}$, $\bs{d}^1=\bs{\rho}$, $l=0$ and $\bs{q}^1=\bs{0}$.
    \For{$t=1, 2, 3, \dots, T$} 
        \State Observe arrival type $j$.
        \State Set $\tilde{x}^t\gets 1$ if $r_j>\bs{A}_j\cdot \bs{q}^t$ and $\tilde{x}^t\gets 0$ otherwise.
        \If{$\bs{b}^t-\bs{A}_j\ge \bs{0}$ and $\tilde{x}^t=1$}
            \State Accept the request and set $\bs{b}^{t+1} \gets \bs{b}^t-\bs{A}_j$.
        \Else
            \State Reject the request and set $\bs{b}^{t+1} \gets \bs{b}^t$. 
        \EndIf
        \If{$t+1 \in \mcT$}
            \State Set $l\gets t+1$ and $\bs{d}^{t+1} = \frac{\bs{b}^{t+1}}{T-t}$.
        \Else
            \State Set $\bs{d}^{t+1} \gets \bs{d}^t$.
        \EndIf
        \State Set $\bs{q}^{t+1} = \max\{\bs{q}^{t} + \frac{1}{t-l+2} (\bs{A}_j \tilde{x}^t  - \bs{d}^{t+1}), \bs{0}\}$.
    \EndFor
    \end{algorithmic}
\end{breakablealgorithm}

\subsection{Numerical Setup of the Multi-resource Case}\label{subsec:numerical_data}
In this section, we present the parameters randomly generated for the multi-resource case with $m=10$ and $n=2$. 
\begin{align*}
    \bs{A} = \begin{bmatrix}
    0.226 & 0.146\\
    0.957 & 0.916\\
    0.005 & 0.876\\
    0.457 & 0.790\\
    0.285 & 0.960\\
    0.572 & 0.736\\
    0.701 & 0.206\\
    0.093 & 0.642\\
    0.903 & 0.923\\
    0.743 & 0.789\\
    \end{bmatrix}\quad 
    \bs{\rho} = \begin{bmatrix}
        0.128\\ 0.805 \\ 0.770 \\ 0.695\\ 0.844\\ 0.647\\ 0.181\\ 0.564\\ 0.812\\ 0.694\\
    \end{bmatrix}
    \quad \bs{p} = \begin{bmatrix}
        0.121\\ 0.879\\
    \end{bmatrix}\quad \bs{r} = \begin{bmatrix}
        0.689 \\ 0.710\\
    \end{bmatrix}
\end{align*}

\end{appendices}

		
		


		
	\end{document}